\documentclass[german,oneside]{itwmreptsg}

\usepackage[utf8]{inputenc}
\usepackage[T1]{fontenc}
\usepackage{hyperref}
\usepackage{subfigure}
\usepackage{xcolor}
\usepackage{array}
\usepackage{amssymb}

\definecolor{fhg00}{RGB}{23, 156, 125}
\definecolor{fhg01}{RGB}{235, 106, 10}
\definecolor{fhg02}{RGB}{0, 110, 145}

\hypersetup{
    colorlinks=true,
    linkcolor=fhg00,
    filecolor=fhg00,      
    urlcolor=fhg00,
    citecolor=fhg00
}
\title{Starker Effekt von Schnelltests}
\shorttitle{Starker Effekt von Schnelltests} 
\subtitle{Eine Analyse mit Hilfe der EpideMSE--Software}
\innertitle{Starker Effekt von Schnelltests}

\thirdfootfield{Projekte EpideMSE, SEEvacs}

\SpecificText{
\textbf{
Jan Mohring,
Michael	Burger,
Robert 	Feßler,
Jochen	Fiedler,
Neele Leithäuser,
Johanna	Schneider,
Michael Speckert,
Jaroslaw Wlazlo 
}

Kontakt:\\ 
Dr. Jan Mohring\\
jan.mohring@itwm.fraunhofer.de\\
Tel.: +49 631 31600-4393\\

Fraunhofer ITWM\\
Fraunhofer-Platz 1\\
67663 Kaiserslautern\\

Version 2.0 vom 28. Juni 2021
}

\graphicspath{{pics/}}
\usepackage{enumitem}
\usepackage{blindtext}
\usepackage{url}
\usepackage{siunitx}
\pdfmapfile{+pfr.map}

\usepackage[backgroundcolor=lightgray]{todonotes}

\newcounter{mycomment}

\usepackage{wrapfig}
\usepackage{marginnote}

\begin{document}

\maketitle
\tableofcontents
\newpage


\section{Zusammenfassung}
\label{sec:zusammenfassung}
Seit dem 27.04.2021 sinkt in Deutschland die Zahl der Neuinfektionen mit dem Coronavirus deutlich. Die dritte Welle ist gebrochen. In dieser Arbeit schlüsseln wir auf, welchen Beitrag
hierzu die drei Maßnahmenpakete Impfung, Kontaktbeschränkung und Testung geliefert haben. Dies erscheint uns wichtig, damit sich im kollektiven Gedächtnis die richtigen Lehren für zukünftige Ausbrüche festsetzen. Hier das Ergebnis unserer Datenanalyse: 

Alle drei Maßnahmen trugen in ähnlicher Größenordnung bei. Den stärksten Effekt aber -- und das mag verwundern -- hatten Tests. Dies ist umso bemerkenswerter, als eine systematische Strategie des Testens, Nachverfolgens und Isolierens deutlich geringere gesellschaftliche Kosten verursacht als Kontaktbeschränkungen. 
In einem \href{https://www.itwm.fraunhofer.de/content/dam/itwm/de/documents/PressemitteilungenPDF/2021/Paper_Einfluss_von_Tests_auf_die_Reproduktionsrate.pdf}{Bericht vom März 2021} hatten wir bereits das Potential anlassloser Massentests als Mittel zur Senkung der Reproduktionsrate von Covid-19 abgeschätzt \cite{mohring21tests}. In Abschnitt~\ref{sec:beiträge} belegen wir nun retrospektiv mit Zahlen und Grafiken, welch tragende Rolle das Testen tatsächlich beim Brechen der dritten Welle gespielt hat. 

Erst im Verbund mit Nachverfolgung und Isolation entfalten Tests ihre infektionshemmende Wirkung: Nach Isolation stecken die Entdeckten selbst niemanden mehr an und per Nachverfolgung können weitere Infizierte aufgespürt werden. Unsere Ergebnisse legen nahe, dass die Möglichkeiten der Nachverfolgung bisher kaum ausgeschöpft werden, d.h. das bereits wirksame Maßnahmenbündel Testung hat noch weiteres Potential, s. Abschnitt~\ref{sec:krankheitsverlauf}. 

Grundlage unser Analyse ist ein epidemiologisches Modell, das hinreichend komplex ist, um den gemessenen Verlauf der Fallzahlen detailliert zu reproduzieren, dabei aber mit so wenigen Parametern auskommt, dass fast alle Parameter eindeutig an die Messungen angepasst werden können. Variieren wir nun die noch offenen Parameter innerhalb plausibler Grenzen, ergibt sich beständig die obige Aussage.

Insbesondere passen wir pro Woche zwei {\em Kontaktraten} und eine {\em Entdeckungsrate} an. Die Kontaktrate misst kritische Kontakte pro Person und Tag, also Kontakte, die zur Infektion führen, wenn eine Person infiziert und die andere noch nicht immunisiert ist. 
Die Entdeckungsrate bezeichnet den Anteil neu Infizierter, die erkannt, per PCR-Test bestätigt und dem RKI gemeldet werden.

Die Anpassung aller Raten wird erst möglich, weil wir eine weit verbreitete Grund\-annahme in Frage stellen. Üblicherweise werden die starken Schwankungen der Fallzahlen im Wochenverlauf damit \href{https://www.rki.de/SharedDocs/FAQ/NCOV2019/FAQ_Liste_Fallzahlen_Meldungen.html}{erklärt}, dass Menschen am Wochenende seltener zum Arzt gehen und  Gesundheitsämter verzögert melden. Die gängige Vorstellung ist also, dass lediglich die gemeldeten Fallzahlen mäandern, während sich die Infektion selbst deutlich gleichmäßiger in der Gesellschaft ausbreitet. In dieser Vorstellung liefern gemessene Tageswerte also keinen Informationsgewinn gegenüber wöchentlichen Mittelwerten. Unsere Analyse zeigt jedoch, dass sich die Schwankungen im Wochenverlauf gut durch verminderte Kontakte am Wochenende erklären lassen, also einem realen Geschehen folgen, s. Abbildung~\ref{fig:deutschland:faelle}. Nur weil nun tatsächlich sieben mal mehr nutzbare Messungen zur Verfügung stehen, können wir überhaupt Kontakt- und Entdeckungsraten gleichzeitig anpassen. 
Vergleichen wir die rekonstruierten Entdeckungsraten unterschiedlicher Bundesländer, so zeigt sich länderübergreifend ein monotoner Anstieg ab Anfang März mit einem Einbruch über Ostern. Daneben fällt aber jeweils noch eine Phase mit besonders steilem Anstieg auf, die zwischen den Bundesländern zeitversetzt auftritt. Diese Anstiege korrelieren deutlich mit der jeweiligen Einführung von Schnelltests an Schulen, s. Abschnitt~\ref{sec:schlüsselrolle}. Es ist also anzunehmen, dass  innerhalb des Maßnahmenpakets Testung die Schultests nochmals besonders wirksam waren. Dies ist nachvollziehbar, da die Kombination von Schulpflicht und Testpflicht einen ungefilterten Blick in alle Bevölkerungsgruppen erlaubt. Insbesondere decken sich unsere Ergebnisse mit den statistischen Erhebungen von \href{https://epub.ub.uni-muenchen.de/76005/1/tr000.pdf}{Kauermann} für bayerische Schulen \cite{berger21schools}. 

Schließlich beleuchten wir noch einmal das Test-Inzidenz-Dilemma, s. Abschnitt \ref{sec:dilemma}. Um Maßnahmen für die Bevölkerung nachvollziehbar und planbar zu machen, sollten sie an Kennwerte wie die Inzidenz gekoppelt sein. Andererseits kann diese Kopplung aber dazu führen, dass eigentlich hilfreiche Maßnahmen bestraft werden. 
So senken Massentests die realen Neuinfektionen durch Isolation und Nachverfolgung von Beginn an. Da das Dunkelfeld aufgehellt wird, steigt die Inzidenz aber erst einmal und kann dann grundlos Kontaktbeschränkungen auslösen. Dies führt z.B. zu Schulschließungen, was die Entdeckungsrate wieder einbrechen lässt.     


Wir verstehen diese Arbeit auch ein wenig als Beitrag zur Ehrenrettung der Zunft der Corona-Modellierer, deren Ansehen zuletzt etwas litt, weil der starke Rückgang der Fälle ab Ende April nicht vorhergesagt wurde. Dazu zeigen wir, welche analytische Kraft zumindest retrospektiven  Modellanpassungen innewohnt. Um in Zukunft auch die Vorhersagekraft zu steigern, arbeiten wir am ITWM daran, Mutationen, klimatische Effekte und die Rückkopplung der Bedrohungslage auf Kontaktverhalten und politische Maßnahmen in unser Modell zu integrieren.  

\section{Aufgabenstellung, Methode und Datenbasis}
\label{sec:aufgabenstellung}
In dieser Studie versuchen wir zu quantifizieren, welchen Anteil die Maßnahmenbündel {\em Testen}, {\em Impfen} und {\em Kontaktbeschränkung} daran hatten, die dritte Coronawelle in Deutschland und ausgewählten Bundesländern einzudämmen. Während z.B. die Sterberate recht eindeutig durch die Impfung besonders Gefährdeter gesenkt wurde, sind bei den Neuinfektionen die Haupteinflussfaktoren weniger offensichtlich. Um diese geht es hier. Uns ist klar, dass Maßnahmen nicht unabhängig voneinander wirken. So fördern Maßnahmen, die Tests erzwingen, um gewisse Kontakte wieder zu ermöglichen, natürlich auch enorm die Testbereitschaft. Wenn wir hier von Maßnahmenbündeln sprechen, dann fassen wir darin alles zusammen, was entweder die Kontakt-, die Impf- oder die Entdeckungsrate ändert. Eine einzelne Maßnahme kann dabei zu mehreren Bündeln beitragen.  

Wer aus welchen Gründen von Corona verschont bleibt, lässt sich nicht direkt messen. Deshalb formulieren wir die Fragestellung als Parameteridentifikation. 
Dazu setzen wir ein Ausbreitungsmodell auf, das Impf-, Kontakt- und Entdeck\-ungsraten zu verschiedenen Zeiten als Parameter enthält. Setzen wir diese und weitere Parameter auf bestimmte Werte und führen eine Simulation durch, ergeben sich Vorhersagen für messbare Größen: Fall- und Sterbezahlen. Diese Vorhersagen können zunächst deutlich von den tatsächlich erhobenen Werten abweichen. Deshalb werden die Parameter nun so lange verändert, bis Vorhersage und Messung möglichst gut übereinstimmen. Die Hypothese ist, dass in diesem Fall auch die gewählten Impf-, Kontakt- und Entdeckungsraten den realen ähnlich sind. Damit ist die Parameteridentifikation abgeschlossen. 

Die Raten zu kennen sagt aber noch nichts darüber aus, welcher Effekt dominiert. 
Um diese Frage zu klären, werden drei weitere Szenarien durchgerechnet. In jedem Szenario wird nun eine der drei Raten so eingefroren, wie es dem Wegfall der zugehörigen Maßnahme während der dritten Welle entspräche. Anschließend wird geprüft, wie stark sich die Fallzahlen gegenüber dem realen Szenario erhöhen. Die Maßnahme, deren Wegfall zur größten Erhöhung führt, gilt uns dann als die mit der größten Wirkung.   

Der erste Schritt -- die Parameteridentifikation -- erfolgt natürlich automatisiert. Um die zugrunde liegende Optimierung zu beschleunigen, berechnen wir nicht nur, welche Parameter zu welchen Fallzahlen führen, sondern auch, wie sich die Fallzahlen bei kleinen Parameterschwankungen verändern. Dazu wird in allen Rechnungen {\em automatisches Differenzieren} \cite{griewank1989automatic} eingesetzt.

Neben den drei Raten enthält das Modell viele weitere Parameter, z.B. solche, die den individuellen Krankheitsverlauf oder den Impferfolg beschreiben. Zu diesen Parametern finden sich in der Literatur oft ungenaue oder widersprüchliche Angaben. Deshalb verwenden wir außer den vom \href{https://www.rki.de/SharedDocs/FAQ/NCOV2019/FAQ_Liste_Fallzahlen_Meldungen.html}{Robert Koch Institut} erfassten und vom \href{https://kitmetricslab.github.io/forecasthub/forecast}{Karlsruher Institut für Technologie} aufbereiteten \href{https://raw.githubusercontent.com/KITmetricslab/covid19-forecast-hub-de/master/data-truth/RKI/by_age/truth_RKI-Cumulative Cases by Age_Germany.csv}{Fall-}, 
\href{https://raw.githubusercontent.com/KITmetricslab/covid19-forecast-hub-de/master/data-truth/RKI/by_age/truth_RKI-Cumulative Deaths by Age_Germany.csv}{Sterbe-} und \href{https://www.rki.de/DE/Content/InfAZ/N/Neuartiges_Coronavirus/Daten/Impfquotenmonitoring.xlsx?__blob=publicationFile}{Impfzahlen}  keine weiteren Literaturwerte. Vielmehr identifizieren wir die meisten dieser Parameter gemeinsam mit den Kontakt- und Entdeckungsraten alleine aus den Meldedaten des RKI. Nur bei zwei Parametern gelingt uns dies nicht: der mittleren Wirksamkeit und dem mittleren Wirkbeginn von Impfungen. Diese variieren wir deshalb innerhalb plausibler Grenzen und passen dann jeweils alle übrigen Parameter an die Meldedaten an.   

\section{Epidemiologisches Modell}
\label{sec:modell}
In der Literatur stößt man im Wesentlichen auf zwei Modellklassen zur Beschreibung einer Epidemie: kohorten- und agentenbasierte. Übersichtsartikel finden sich z.B. in \cite{gonccalves2010comparing} oder 
\cite{siettos2013mathematical}. Agentenbasierte Verfahren spielen die Interaktion zwischen Individuen nach. Dadurch lassen sich Maßnahmen wie Wechselunterricht direkt abbilden. Allerdings übersteigt die Zahl freier Parameter bei Weitem die Zahl erfasster Messwerte, so dass sich die Modelle nur mit Hilfe vieler ungesicherter Hypothesen an eine reale Situation anpassen lassen. Die erheblichen Rechenzeiten verschärfen dieses Problem noch.    

Kohortenbasierte Modelle teilen die Bevölkerung in Gruppen ein, die ihren aktuellen Gesundheitszustand widerspiegeln. Das erste derartige Modell wurde bereits 1766 von Bernoulli erstellt, um die Ausbreitung der Pocken zu beschreiben \cite{bernoulli1760essai}. In ihrer heutigen Form gehen diese Modelle auf das so genannte SIR-Modell von Kermack und McKendrick \cite{Kermack27} aus dem Jahr 1927 zurück. Sie unterteilen die Bevölkerung in Anfällige (S), Infektiöse (I) und Erholte (R). Die Übergänge zwischen den Kohorten werden mit Hilfe von Ratengleichungen modelliert, was zu ODE-Systemen führt. Eine Verbesserung stellen die SEIR-Modelle dar, die auch die Latenz zwischen Infektion und eigener Infektiösität berücksichtigen. Hierzu wird die Zwischenkohorte der Exponierten (E) eingeführt. 
Solche Modelle, einschließlich vieler Variationen, sind gut etabliert und ihre Namen spiegeln die Reihenfolge der Übergänge wider. Einen guten Überblick liefert \cite{hethcote2000mathematics}. 

Um modellierte Infektionszahlen an Meldedaten anpassen oder die Wirkung von Teststrategien beurteilen zu können, muss auch abgebildet werden, mit welcher Rate und welchem Verzug Infizierte durch Tests sichtbar gemacht werden \cite{abbott2020epiforecasts, bortz20epidemiologie}. 
Bereits im Herbst 2020 legte die Gruppe um Priesemann ein Modell vor, das die Reproduktionsrate auch mit anlasslosen Tests und Nachverfolgung verknüpft \cite{contreras2021challenges}. In der politischen Diskussion wurde allerdings nur eine der möglichen Konsequenzen betont: Damit Nachverfolgung wieder greife, müsse die Inzidenz zunächst durch strenge Kontaktbeschränkungen reduziert werden \cite{linden2020uberschreitung}.

Ein zweiter Hebel von Tests wurde im öffentlichen Diskurs dagegen weitgehend ignoriert. Wird die Zeitspanne zwischen Infektion und Isolation reduziert, steckt die infizierte Person weniger weitere an. Darauf wiesen Kreck und Scholz in einer weniger beachteten Arbeit hin \cite{kreck2021studying}. Grundlage bildet ein Modell, das wie das unsere auf dem ursprünglichen Integralkern-Ansatz von Kermack und McKendrick basiert \cite{Kermack27}. Der Zeitpunkt der Isolation wurde möglicherweise deshalb kaum thematisiert, weil die Annahme vorherrschte, fast alle Infizierten mit Symptomen würden sich sofort selbst isolieren. Unsere Ergebnisse legen nahe, dass diese Annahme wohl falsch ist und dass die schnelle Isolation nach Tests die dritte Welle maßgeblich mit gebrochen hat, s. Abschnitt~\ref{sec:krankheitsverlauf}. 

Unser Modell stellt eine Verallgemeinerung der SEIR-Modelle dar. Es kennt nicht nur drei oder vier Kohorten. Vielmehr wird der Krankheitsverlauf in gleich lange Abschnitte aufgeteilt, die je nach Auflösung einen Tag oder auch nur ein paar Stunden umfassen können. Die Kohorten werden nun von Infizieren gebildet, die sich gemeinsam vor entsprechend langer Zeit angesteckt haben. In seiner einfachsten Form -- ohne Tests, Impfungen und Aufspaltung in Altersgruppen -- handelt es sich um ein Integralgleichungsmodell mit kastenförmigem Integralkern, wie es das erste Mal ebenfalls von Kermack und McKendrick beschrieben wurde \cite{Kermack27}, aber im Schatten des einfacheren SIR-Modells blieb \cite{breda2012formulation}.

Bezeichnen wir mit $N(t)$ die Zahl aller jemals Infizierten zur Zeit $t$, dann ergeben sich die aktuellen Neuinfektionen aus früheren Neuinfektionen gemäß 
\begin{align}
\dot{N}(t) &= \int\limits_{-\infty}^{\infty} K(t,t') \, \dot{N}(t') \, dt' \; . 
\label{eq:integralgleichung}
\end{align}
Für den Integralkern $K$ wird nun folgende Gestalt postuliert:
\begin{align}
    K(t,t') &= \left( 1 - \frac{N(t)}{G} \right) 
    \kappa(t) \; \omega(t-t') \; . 
    \label{eq:integralkern}
\end{align}
$\kappa(t)$ ist die Kontaktrate. Sie gibt an, wie viele Personen von einem voll Infektiösen im Mittel pro Tag angesteckt würden, wenn noch niemand immunisiert wäre. Die Klammer vor $\kappa$ liefert die Wahrscheinlichkeit, tatsächlich auf eine noch nicht immunisierte Person zu treffen. $G$ ist dabei die Größe der Gesamtpopulation. $\omega(\tau)$ gibt schließlich an, wie infektiös eine Person $\tau$ Tage nach der eigenen Ansteckung ist. In der Realität folgt $\omega$ der Virenlast während des individuellen Krankheitsverlaufs. In unserem Modell nehmen wir hingegen an, dass $\omega$ eine Kastengestalt hat:
\begin{align} 
    \omega(\tau) = \chi_{\left[ \tau^s, \tau^e \right]}(\tau) = \left\{ \begin{array}{ll}
    1 & \mbox{falls} \; \tau^s \le \tau \le \tau^e \\
    0 & \mbox{sonst}
    \end{array}\right. \; ,
    \label{eq:omega}
\end{align}
d.h. es gibt einen festen Beginn $\tau^s$ und ein festes Ende $\tau^e$ der infektiösen Phase. Vorher und nachher wird niemand angesteckt. Dazwischen ist die Person voll infektiös.
Damit vereinfacht sich Gleichung~\eqref{eq:integralgleichung} zu
\begin{align}
    \dot{N}(t) &= \left( 1 - \frac{N(t)}{G} \right) \kappa(t) \, 
    \left[ N(t-\tau^s) - N(t-\tau^e) \right] \; .
    \label{eq:simple}
\end{align}
Analysen und ein \href{https://www.itwm.fraunhofer.de/content/dam/itwm/de/documents/anwendungsfelder/20201201_fessler_IntegralEquationModelOfEpidemics.pdf}{Vergleich} mit dem klassischen SIR-Modell finden sich bei \href{https://www.itwm.fraunhofer.de/content/dam/itwm/de/documents/anwendungsfelder/20201201_fessler_IntegralEquationModelOfEpidemics.pdf}{Feßler} \cite{fessler21preprint}.
In unseren Beiträgen zum \href{https://covid19forecasthub.eu}{European Covid-19 Forecast Hub} und in älteren \href{https://www.itwm.fraunhofer.de/content/dam/itwm/de/documents/anwendungsfelder/EpideMSE_Whitepaper.pdf}{Berichten}  \cite{bortz20epidemiologie} firmiert das Modell als {\em dSEIR}-Modell, wobei {\em d} für {\em delayed} steht. Der Name {\em Boxkern-Modell} wäre wahrscheinlich treffender gewesen. Der Vorteil des Boxkern-Modells besteht darin, dass eine Latenzzeit ohne jegliche Weitergabe und ein striktes Ende der infektiösen Phase modelliert werden können. Ferner lassen sich auch überlappende Teilpopulationen wie Hospitalisierte und intensiv zu Betreuende einfach aus der Grundgröße $N$ ableiten.   

Sollen Effekte wie räumliche Ausbreitung oder unterschiedliche Sterberaten erfasst werden, teilen wir die Bevölkerung zusätzlich in Gruppen ein, die sich z.B. nach Wohnort oder Alter unterscheiden. Jede Gruppe wird anschließend wieder nach Krankheitsphasen unterteilt. An die Stelle einer skalaren Kontaktrate tritt in diesem Fall eine Kontaktmatrix, die auch Ansteckungen zwischen den Gruppen berücksichtigt. 
  
Um zusätzlich Impfungen und Tests abbilden zu können, wird jede Gruppe nicht nur durch die Infizierten in jeder Krankheitsphase beschrieben, sondern zusätzlich durch die Entdeckten und die per Impfung Immunisierten. 
Für die hier durchgeführten Analysen reichen glücklicherweise Eingruppenmodelle aus, d.h. innerhalb eines Modells wird zwischen Infizierten, Entdeckten und Geimpften unterschieden, nicht aber nach Ort oder Alter. 

Eine besondere Rolle kommt der Bestimmung der Dunkelziffer zu. Bisher haben wir
diese aus der Überlegung abgeleitet, dass das Verhältnis der Fälle zwischen jüngeren und älteren Altersgruppen eigentlich höher sein müsste als beobachtet \cite{fiedler21dunkelziffer}. Grundlage bildeten Kontaktmatrizen aus soziologischen Studien. Da die  Über-Achzigjährigen wegen der Impfungen aber kaum noch am Infektionsgeschehen teilnehmen, fallen sie mittlerweile als Referenz aus. In dieser Arbeit gelingt es uns, das Gegenstück zur Dunkelziffer --  die Entdeckungsrate -- auch direkt an Meldedaten anzupassen.

\subsection{Parameter}
\label{sec:parameter}
Die entscheidende Modellannahme besteht darin, dass wir die Wirkung der individuell variierenden Parameter gut reproduzieren können, wenn wir für alle Mitglieder einer Gruppe einen gemeinsamen Mittelwert verwenden. 

\begin{SCtable}[0.95\textwidth][!b]
	\centering
	\begin{tabular}{cp{0.6\textwidth}rc}
		\hline \noalign{\smallskip}
		Symbol & Bedeutung & \multicolumn{1}{l}{Wert} & Einheit \\
		\hline \noalign{\smallskip}
		$t_s$ & Beginn der Simulation & $25.01.2021$ & d \\
		$t_e$ & Ende der Simulation & $17.05.2021$ & d \\
		$t_f$ & Datum ab dem Sterbezahlen angepasst werden & $22.03.2021$ & d \\
		$g$ & Zahl der Gruppen & $1$ & \\
		$G_k$ & Zahl der Mitglieder in Gruppe $k$ & $83,2$ & $10^6$p\\
		$G$ & Zahl der Mitglieder aller Gruppen & $83,2$ & $10^6$p\\
		$\kappa_{kl}(t)$ & {\em Kontaktrate}: kritische Kontakte einer Person aus Gruppe $k$ mit Personen aus Gruppe $l$ pro Tag & \textcolor{fhg00}{$0,06$-$0,40$} & p/d \\ 
		$\tau_k^s$ & Beginn der infektiösen Phase in Gruppe  $k$ nach Infektion & \textcolor{fhg00}{$4,3\pm 0,6$} & d \\
		$\tau_k^e$ & Ende der infektiösen Phase in Gruppe $k$ nach Infektion & 
		\textcolor{fhg00}{$9,6 \pm 1,2$} & d \\
		$\tau^e_k-\tau^s_k$ & Dauer der infektiösen Phase in Gruppe $k$ & \textcolor{fhg00}{$5,3\pm1,0$} & d\\
        $\tau_k^d(t)$ & Zeit zwischen Ansteckung und Entdeckung eines Entdeckten & \textcolor{fhg00}{$6,6 \pm 0,3$} & d \\
        $\tau_k^r(t)$ & Zeit zwischen Ansteckung und Meldung eines Entdeckten
        & \textcolor{fhg00}{$7,6 \pm 0,2$} & d \\
		$\lambda_k(t)$ & {\em Entdeckungsrate}: Anteil der Infizierten aus Gruppe $k$, die später
		entdeckt werden & \textcolor{fhg00}{$23$-$71$} &  \%\\ 
        $\tau^m_k$ & Zeit nach Infektion, nach der Opfer aus Gruppe $k$ versterben & \textcolor{fhg00}{$27,7\pm0,3$} & d \\
        $\rho^m_k$ & Anteil Infizierter in Gruppe $k$, die versterben & \textcolor{fhg00}{$0,60\pm0,21$} & \% \\
        $V_k(t)$ & Zahl der Personen in Gruppe $k$, die bis zur Zeit $t$ eine Erstimpfung erhielten & $1$-$31$ & $10^6$p \\
        $\varepsilon_k(t)$ & Mittlere Wirksamkeit der Impfstoffe, die Gruppe $k$ zur Zeit $t$ verabreicht werden & \textcolor{fhg01}{$91$} & \%\\
        $\tau^p_k$ & Zeit nach Erstimpfung, ab der Impfschutz in Gruppe $k$ einsetzt & \textcolor{fhg01}{$14$} & d \\
        $N^0_k$ & Infizierte in Gruppe $k$ bei Simulationsbeginn & 
        \textcolor{fhg00}{$5,1\pm 5,3$} & $10^6$p \\
        $\alpha^N_k$ & Relative Steigung der Infizierten in Gruppe $k$ bei Simulationsbeginn & 
        \textcolor{fhg00}{$5,7\pm 2,6$} & $10^{-3}$/d \\
        $P^0_k$ & Erfolgreich Geimpfte in Gruppe $k$ bei Simulationsbeginn & 
        $0,27$ & $10^6$p \\
        $\alpha^P_k$ & Relative Steigung der erfolgreich Geimpften in Gruppe $k$ bei Simulationsbeginn & $5,7$ & 1/d \\
        $M^0_k$ & Verstorbene in Gruppe $k$ bei Simulationsbeginn & 
        \textcolor{fhg00}{$66\pm 2$} & $10^3$p \\
		\noalign{\smallskip}
		\hline
	\end{tabular}
	\caption{Modellparameter für Deutschland. Orange: angenommen. Grün: identifiziert.}
	\label{tab:parameter}
\end{SCtable}

Die Parameter unseres Modells sind in Tabelle~\ref{tab:parameter} aufgelistet. Obwohl wir in dieser Studie nur eine Gruppe verwenden, wollen wir hier das Modell doch in seiner ganzen Allgemeinheit darstellen. Deshalb tragen die meisten Parameter einen Gruppenindex. Die beispielhaft angegebenen Werte beziehen sich auf ein Modell, das an die \href{https://github.com/KITmetricslab/covid19-forecast-hub-de/tree/master/data-truth/RKI/by_age}{RKI-Meldedaten für Deutschland} angepasst wurde. Bei den schwarz gedruckten Werten handelt es sich um unzweifelhafte Konstanten. Die orangenen Impfparameter wurden aus Literaturwerten geschätzt und sind unsicher, s. 
Abschnitt~\ref{sec:anpassung}. Die grün gedruckten Werte sind angepasst. 
Die Fehlerbereiche entsprechen 99,7\% Konfidenzintervallen angenommener Normalverteilungen (3$\sigma$). 

Ein kritischer Kontakt ist definiert als Kontakt zwischen zwei Personen, der so eng ist, dass er zur Infektion führen würde, wenn die eine Person gerade infektiös und die andere noch nicht immun ist. $\kappa$ beschreibt also einerseits das Kontaktverhalten, aber auch -- und das ist nicht zu vergessen -- wie ansteckend die vorherrschende Virusvariante ist oder wie die Übertragbarkeit vom Wetter abhängt.  

Ein kritischer Kontakt kann z.B. auch zwischen zwei Immunen stattfinden. Es geht bei der Definition um die Art der Begegnung, nicht um den Status der sich Treffenden. Die Zahl kritischer Kontakte korrespondiert also mit politischen Maßnahmen wie Kontaktbeschränkung, aber nicht mit dem aktuellen Grad der Durchseuchung und kann also als Regelgröße verwendet werden. $\kappa$ wird andernorts auch Transmissions-, Übertragungs- oder Kontaktrate genannt.

Der individuelle Krankheitsverlauf wird sehr einfach und für jede Person in Gruppe $k$ gleich modelliert. Zwischen $\tau_k^s$ und $\tau_k^e$ Tagen nach der Infektion ist ein Infizierter gleichmäßig infektiös, davor und danach gar nicht.  
Die in Tabelle~\ref{tab:parameter} angegebenen Parameter entstammen keinen klinischen Studien, sondern wurden alleine durch Anpassung an den Epidemieverlauf im Frühjahr 2021 gewonnen. Es ist noch einmal wichtig zu betonen, dass sich die Zeiten auf ein kastenförmiges Infektionsprofil beziehen. In der Realität ist man bereits etwas früher und auch noch etwas später ansteckend, aber dafür schwächer. Für Einzelheiten s. Abschnitt~\ref{sec:krankheitsverlauf}.

Die Wirkung von Tests wird durch zwei Regelgrößen modelliert: die Wahrscheinlichkeit $\lambda_k(t)$, mit der Infizierte entdeckt werden, und die mittlere Zeit $\tau_k^d(t)$ nach Infektion, zu der eine solche Entdeckung erfolgt. Man beachte die Bedeutung der zweiten Größe. Werden Infizierte erst am Ende der infektiösen Phase aufgespürt, dann bringt auch eine hohe Entdeckungsrate wenig. Gelingt es hingegen, Personen noch während der Latenzzeit in Quarantäne zu nehmen, dann entfaltet ein Test seine maximale Wirkung. Werden Infizierte erst nach Ausbruch von Symptomen entdeckt, dann können sie bereits andere angesteckt haben. Werden letztere aber alle rechtzeitig aufgespürt und in Quarantäne genommen, dann ist die Wirkung auf die Ausbreitung des Virus so, als wäre die Ausgangsperson schon früher entdeckt worden. Mit Hilfe der Entdeckungszeit wird also auch die Effizienz der Nachverfolgung modelliert. 

Zusätzlich zur Entdeckungszeit führen wir noch die Meldezeit $\tau^r_k$ ein. $r$ steht für englisch {\em report}. Sie wirkt sich nicht auf das Infektionsgeschehen aus, ist aber nötig, um den Bezug zwischen simulierten und gemeldeten Fällen herzustellen.  

Die Impfung einer Gruppe $k$ beschreiben wir durch die mittlere Wirksamkeit $\varepsilon_k$ und die Zeit $\tau^p_k$ nach Erstimpfung, nach der der Impfschutz einsetzt. $p$ steht für {\em protection}. Auch hier verwenden wir ein recht simples Modell. Bis zu diesem Zeitpunkt gibt es keinerlei Impfschutz, danach beim Anteil $\varepsilon_k$ sofort den vollen und beim Rest gar keinen. Für erfolgreich Geimpfte wie für Genesene nehmen wir an, dass sie im Simulationszeitraum weder infiziert werden noch das Virus übertragen können.

\subsection{Ausbreitungsdynamik}
\label{sec:ausbreitung}
\begin{SCtable}[0.95\textwidth][!ht]
	\centering
	\begin{tabular}{cl}
		\hline \noalign{\smallskip}
		Symbol & Bedeutung\\
		\hline \noalign{\smallskip}
		$N_k$ & Zahl aller je Infizierten in Gruppe $k$ \\
		$S_k$ & Zahl der Infizierbaren in Gruppe $k$ \\
		$I_k$ & Zahl der effektiv Infektiösen in Gruppe $k$\\
		$Q_k$ & Zahl der extern Infizierten in Gruppe $k$ \\
		$D_k$ & Zahl der entdeckten Infizierten in Gruppe $k$\\	
		$D^0_k$ & Zahl der bisher Infizierten in Gruppe $k$, die irgendwann entdeckt werden\\	
		$P_k$ & Zahl der nur durch Impfung erfolgreich Geschützten in Gruppe $k$\\
		\noalign{\smallskip} \hline
	\end{tabular}
	\caption{Kohorten.}
	\label{tab:zustandsgroessen}
\end{SCtable}
Um die Dynamik der Ausbreitung zu beschreiben, werden zunächst die in Tabelle~\ref{tab:zustandsgroessen} aufgeführten Kohorten verwendet. 
\begin{align}
\dot{N}_k(t) &= \frac{S_k(t)}{G_k} \sum\limits_{l=0}^{g-1} \kappa_{lk}(t) \, I_l(t) + Q_k(t)
\label{eq:nt:abs} \\
\dot{D}_k(t) &= \lambda_k(t) \, \frac{d}{dt} N_k(t-\tau^d_k(t)) 
\label{eq:dt:abs} \\
\dot{P}_k(t) &= \varepsilon_k(t') \frac{S_k(t')}{G_k-V_k(t')} \dot{V}_k(t') \; , \quad t' = t-\tau^p_k.
\label{eq:dp:abs}
\end{align}
Die Summe in Gl.~\eqref{eq:nt:abs} gibt die Gesamtzahl von Personen in Gruppe $k$ an, die pro Tag in kritischen Kontakt mit Infektiösen aus allen Gruppen treten. Infiziert wird von diesen der Anteil der Infizierbaren (erster Quotient). Das Modell für die Entdeckung ist schlicht. $\frac{d}{dt} N_k(t-\tau^d_k(t))$ ist die Zahl der Personen in Gruppe $k$, die vor $\tau^d_k(t)$ Tagen neu infiziert wurden. Genau nach $\tau^d_k(t)$ Tagen entscheidet sich also, ob Infizierte mit der Wahrscheinlichkeit $\lambda_k(t)$ durch Tests erkannt werden oder von da an nie wieder. Wir setzen voraus, dass die Entdeckungsdauer zwischen Beginn und Ende der infektiösen Zeit liegt:
\begin{align}
\tau^s_k \le \tau^d_k(t) \le \tau^e_k  \; . 
\end{align}
Für $\tau^d_k = \tau^s_k$ gibt eine entdeckte Person das Virus nicht mehr weiter, d.h. noch kürzere Entdeckungsdauern hätten keinen weiteren Effekt und wir können sie durch $\tau^d_k=\tau^s_k$ erfassen. Eine Entdeckung nach der infektiösen Zeit, z.B. durch Antikörpertests, mag von wissenschaftlichem Interesse sein -- die Ausbreitung beeinflusst sie nicht mehr. Wir erfassen diese Fälle durch $\tau^d_k = \tau^e_k$. 

Gl.~\eqref{eq:dp:abs} beschreibt, wie sich die Zahl der nur durch Impfung Immunisierten entwickelt. Erklärungsbedürftig ist sicher zweierlei, die eigentümliche Definition von $P_k$ und der Quotient. Die Menge der nur durch Impfung Immunisierten wird eingeführt, um eine Überlappung mit Genesenen zu vermeiden. Der Quotient repräsentiert die bedingte Wahrscheinlichkeit, unter den Ungeimpften einen noch nie Infizierten zu treffen. Hier geht ein, dass in der Praxis einer Impfung kein Antikörpertest vorausgeschickt wird. Ein gewisser Schwachpunkt des Modells ist, dass die Wirkzeit $\tau^p_k$ nicht zeitabhängig angelegt ist. Dies wäre z.B. realistischer, wenn sich die Anteile von Impfstoffen mit unterschiedlichen Wirkzeiten verschieben. Die numerischen Implikationen wären aber erheblich. Und da wir die unsichere Wirkzeit ohnehin durch Unter- und Obergrenzen abschätzen, lohnt der Aufwand nicht.    

In Gl.~\eqref{eq:nt:abs} bis \eqref{eq:dp:abs} gelten die folgenden Zusammenhänge:
\begin{align}
S_k(t) &= G_k - N_k(t)-P_k(t) \label{eq:infizierbar}\\
D_k(t) &= D^0_k(t-\tau^d_k(t)) \label{eq:entdeckt} \\
I_l(t) &= \int_{t-\tau^e_l}^{t-\tau^s_l} \dot{N}_l(t') \, dt' -
\int_{t-\tau^e_l}^{t-\tau^d_l(t)} \dot{D}^0_l(t') \, dt' \; .
\label{eq:infektioes:1} 
\end{align}
In unserem Modell bilden die Infizierbaren das Komplement zu den jemals Infizierten und erfolgreich Geimpften, d.h. Angehörige der beiden letzten Gruppen können sich später nicht mehr anstecken oder das Virus übertragen. 

Gl.~\eqref{eq:entdeckt} definiert die zunächst etwas abstrakte Zahl $D^0_k(t)$ der Personen, die zur Zeit $t$ infiziert sind und bereits entdeckt wurden oder erst noch entdeckt werden. Wir benötigen sie zur Formulierung der effektiv Infektiösen in Gl.~\eqref{eq:infektioes:1}.
Das erste Integral erstreckt sich über alle Personen, die vor $\tau^e_l$ bis $\tau^s_l$ Tagen neu infiziert wurden, jetzt also Viren produzieren.  Man beachte, dass Infizierte in dieser Phase als gleichbleibend infektiös modelliert werden. 
Im zweiten Integral werden diejenigen abgezogen, die nach der Infektion bis heute entdeckt und in Quarantäne genommen wurden. 

Um das endgültige Modell formulieren zu können, müssen wir noch die Zeit der zukünftigen Entdeckung $t^d_k$ einführen: 
\begin{align}
t' &= t^d_k(t) \Leftrightarrow t = t' - \tau^d_k(t') \; .
\label{eq:entdeckungszeit} 
\end{align}     
Für konstante Entdeckungsdauer gilt einfach 
$ t^d_k(t) = t + \tau^d_k$ . 
Damit schreibt sich unser Dynamik-Modell als retardierte Differentialgleichung in den drei Größen $N_k$, $D^0_k$ und $P_k$:
\begin{align}
\dot{N}_k (t) &= Q_k(t) + \left( 1 - \frac{N_k(t)+P_k(t)}{G_k} \right)\sum\limits_{l=0}^{g-1}\kappa_{lk}(t) \, \times \nonumber \\
& \hspace{3em} \left[
N_l(t-\tau^s_l) - N_l(t-\tau^e_l) - D^0_l(t-\tau^d_l(t)) + D^0_l(t-\tau^e_l) \right] \label{eq:zwei:Nt} \\
\dot{D}^0_k(t) &= \lambda_k\left( t^d_k(t) \right) \dot{N}_k(t)
\label{eq:zwei:D0t} \\
\dot{P}_k(t) &= \varepsilon_k(t') \, \frac{G_k-N_k(t')-P_k(t')}{G_k-V_k(t')} \, \dot{V}_k(t') \; , \quad t' = t-\tau^p_k. 
\label{eq:zwei:Pt}
\end{align}
Man beachte, dass Gl.~\eqref{eq:zwei:D0t} zukünftige  Entdeckungsraten- und -dauern verwendet. 

\subsection{Symmetrie der Kontakte und Rechnung mit Anteilen}
\label{sec:symmetrie}
Der Kontakt einer Person aus Gruppe $k$ mit einer Person aus Gruppe $l$ ist ebenso ein umgekehrter Kontakt. Daher gilt:
\begin{align}
G_l \, \kappa_{lk} = G_k \, \kappa_{kl} \;.
\label{eq:kappa}
\end{align}
Gehen wir nun zu Anteilen an der Gesamtgruppe über und benennen diese Anteile mit kleinen Buchstaben, dann vereinfachen sich Gleichungen~\eqref{eq:zwei:Nt} bis \eqref{eq:zwei:Pt} nochmals:
\begin{align}
\dot{n}_k (t) &= q_k(t) + \left[ 1 - n_k(t) - p_k(t) \right] \sum\limits_{l=0}^{g-1}\kappa_{kl}(t) \, \times \nonumber \\
& \hspace{3em} \left[
n_l(t-\tau^s_l) - n_l(t-\tau^e_l) - d^0_l(t-\tau^d_l(t)) + d^0_l(t-\tau^e_l) \right] \label{eq:skal:nt} \\
\dot{d}^0_k(t) &= \lambda_k\left( t^d_k(t) \right) \dot{n}_k(t) 
\label{eq:skal:d0t} \\
\dot{p}_k(t) &= 
\varepsilon_k(t') \, \frac{1-n_k(t')-p_k(t')}{1-v_k(t')} \, \dot{v}_k(t') \; , \quad t' = t-\tau^p_k. \label{eq:skal:pt} \\
n_k(t) &= \frac{N_k(t)}{G_k}\, , 
\;  q_k(t) = \frac{Q_k(t)}{G_k}\, ,
\;  d^0_k(t) = \frac{D^0_k(t)}{G_k}, 
\; p_k(t) = \frac{P_k(t)}{G_k}\, ,
\; v_k(t) =  \frac{V_k(t)}{G_k}\; .
\label{eq:skal}
\end{align}
Wir weisen darauf hin, dass wegen Gl.~\eqref{eq:kappa} die Indizes von $\kappa$ vertauscht sind.

\subsection{Bezug zu gemeldeten Größen}
\label{sec:bezug}
\begin{SCtable}[0.95\textwidth][!ht]
	\centering
	\begin{tabular}{cl}
		\hline \noalign{\smallskip}
		Symbol & Bedeutung\\
		\hline \noalign{\smallskip}
		$R_k$ & Kumulierte gemeldete Fälle in Gruppe $k$ ({\em reported})\\
		$M_k$ & Kumulierte Verstorbene in Gruppe $k$ ({\em mortus})\\
		\noalign{\smallskip} \hline
	\end{tabular}
	\caption{Meldegrößen.}
	\label{tab:meldegroessen}
\end{SCtable}
Um die Modellparameter zu kalibrieren, müssen simulierte und gemeldete Größen in Beziehung gesetzt werden. Die Erstimpfungen $V_k(t)$ erscheinen bereits als Eingabe des dynamischen Modells \eqref{eq:skal:nt} - \eqref{eq:skal}. Die gemeldeten kumulierten Fall- und Verstorbenenzahlen ergeben sich dann wie folgt, s. auch Tabelle~\ref{tab:parameter}.
\begin{align}
R_k(t) &= G_k\; d^0_k(t-\tau^r_k(t)) \label{eq:reported} \\
\dot{M}_k(t) &= G_k \; \rho^m_k(t)\; \dot{n}_k(t-\tau^m_k(t)) \label{eq:dead} \; .
\end{align}
Die kumulierte Zahl gemeldeter Fälle ergibt sich einfach durch Verschiebung um die Meldezeit $\tau^r_k$. Bei der kumulierten Zahl der Verstorbenen ist das nicht so einfach möglich, da sich die Sterberate über die Zeit ändern kann, z.B. durch Verschiebung der betroffenen Bevölkerungsteile zu Jüngeren hin. Daher müssen wir zu den Tageswerten übergehen, d.h. eine Differentialgleichung lösen. $\tau^m_k$ ist die mittlere Sterbezeit. Man beachte, dass wir die Zahl der Toten proportional zur Zahl der wirklich Infizierten ansetzen, d.h. dass alle Covid-19-Toten auch als solche erkannt werden. 

\subsection{Anfangsbedingungen}
\label{sec:anfangsbedingungen}
Die Differentialgleichungen \eqref{eq:skal:nt} - \eqref{eq:skal:d0t} und \eqref{eq:dead} sind retardiert. Genauer hängen die rechten Seiten von den Infizierten $n_k$ und den erfolgreich Geimpften $p_k$ in der Vergangenheit ab. Aus diesem Grunde müssen wir für diese Größen nicht nur einzelne Startwerte vorgeben, sondern Verläufe vor Simulationsbeginn. Wir nehmen exponentielle Verläufe an, die sich jeweils durch zwei neue Parameter beschreiben lassen, s. auch Tabelle~\ref{tab:parameter}.
\begin{align}
    n_k(t) &= \frac{N_k^0}{G_k} \, \mbox{e}^{\alpha^N_k \, \left(t-t_0 \right)} \, , \quad 
    p_k(t) = \frac{P_k^0}{G_k} \, \mbox{e}^{\alpha^P_k \, \left(t-t_0 \right)}  \; , \quad t \le t_0 \label{eq:start:history} \; .
\end{align}
Die eigentlichen Startwerte lauten dann
\begin{align}
    n_k(t_0) &= \frac{N^0_k}{G_k} \, , \quad
    d^0_k(t_0) = \lambda_k(t_0) \, n_k(t_0) \, , \quad
    p_k(t_0) = \frac{P^0_k}{G_k} \, , \quad
    M_k(t_0) = M^0_k \label{eq:start:t0} \, ,
\end{align}
d.h. außer den Parametern der exponentiellen Verläufe brauchen wir nur noch den Startwert $M^0_k$ der Sterbefälle als weiteren Parameter.

\subsection{Numerische Umsetzung}
\label{sec:numerische}
Wir diskretisieren Gleichungen~\eqref{eq:skal:nt} bis \eqref{eq:skal:pt} so , dass wir ein explizites Verfahren 2. Ordnung erhalten. Dazu definieren wir
\begin{align}
\Delta t &> 0 \quad \mbox{(konstanter Zeitschritt)} \label{eq:model:dt} \\
t_j &= j\, \Delta t  \label{eq:modell:ti} \\
i_l(t) &= n_l(t - \tau^s_l) - n_l(t-\tau^e_l) - d^0_l (t - \tau^d_l(t)) +d^0_l(t - \tau^e_l) \label{eq:modell:il} \\
\sigma_k(t) &= \frac{\Delta t}{2} \sum\limits_{l=0}^{g-1} 
\kappa_{kl}(t) \, i_l(t) \label{eq:modell:sigma:k} \; .
\end{align} 
Unser Modell benötigt die Zustandsgrößen $n_k$, $d^0_k$ und $p_k$ zu vergangenen Zeiten. Dazu speichern wir diese bis zu einer maximalen Rückschauzeit $j_{\!R} \Delta t$ für Zeiten $t_{j-j_{\!R}}\, , \ldots ,\, t_{j-1}$ in einer Pipeline. Vergangene Werte gewinnen wir dann durch quadratische Interpolation. Ist ein Zeitschritt des numerischen Verfahrens abgearbeitet, werden die neuen Werte in die Pipeline geschoben und die ältesten Werte fallen hinaus. 

Die Evolutionsgleichung~\eqref{eq:skal:pt} für erfolgreich Geimpfte kann in 2. Ordnung wie folgt approximiert werden:
\begin{align}
\frac{p_k(t_{j+1})-p_k(t_j)}{\Delta t} &\approx
\varepsilon_k(t') \, \frac{1-n_k(t')-p_k(t')}{1-v_k(t')} \, \dot{v}_k(t') \; , \quad t' = t_{j\!+\!\frac 1 2}-\tau^p_k 
    \label{eq:modell:diskret:impfung:approx} \\             
\Rightarrow p_k(t_{j+1}) &\approx  
p_k(t_j) + \Delta t  \; 
\varepsilon_k(t') \, \frac{1-n_k(t')-p_k(t')}{1-v_k(t')} \, \dot{v}_k(t')\;  .  
    \label{eq:modell:diskret:impfung}
\end{align}
Mit $p_k( t_{j\!+\!\frac 1 2}) \approx \frac{ p_k(t_{j})+p_k(t_{j+1})}{2}$ lautet eine Approximation 2. Ordnung von Gl.~\eqref{eq:skal:nt}:
\begin{align}
\frac{n_k(t_{j+1}) - n_k(t_{j})}{\Delta t} \approx q_k(t_{j\!+\!\frac 1 2}) + 
\left[ 1- \frac{n_k(t_{j+1}) + n_k(t_{j})}{2} - p_k(t_{j+ \frac 1 2}) \right] 
\frac{2}{\Delta t} \sigma_k(t_{j+ \frac 1 2}) \label{eq:modell:diskret} \\
\Rightarrow n_k(t_{j+1}) \approx \frac{\left[ 1-\sigma(t_{j\!+\!\frac 1 2}) \right] n_k(t_j) + 2 \left[ 1- p_k(t_{j\!+\!\frac 1 2}) \right] \sigma_k(t_{j\!+\!\frac 1 2}) + \Delta t \, q_k(t_{j\!+\!\frac 1 2})}
{1+\sigma_k({t_{j\!+\!\frac 1 2}})} .\label{eq:modell:diskret:n}
\end{align}
Mit bekanntem $n_k(t_{j+1})$ finden wir dann in gleicher Approximationsordnung:
\begin{align}
d^0_k(t_{j+1}) &\approx
d^0(t_j) + \lambda_k\big( t^d_k(t_{j\!+\!\frac 1 2}) \big)
\left[ n_k(t_{j+1}) - n_k(t_j) \right] \; . 
\label{eq:modell:diskret:d0} 
\end{align} 

\section{Anpassung der Modellparameter}
\label{sec:anpassung}
Um möglichst unabhängig von falschen Grundannahmen zu bleiben, füttern wir unser Modell nur mit den vom RKI erfassten Fall-, Sterbe- und Impfzahlen, aber sonst mit keinen weiteren Literaturwerten, s. Tabelle~\ref{tab:meldedaten}.
Bis auf zwei Impfparameter -- Wirksamkeit und Wirkbeginn -- passen wir alle übrigen Modellparameter an diese Meldedaten an, insbesondere zwei Kontaktraten und eine Entdeckungsrate pro Woche sowie alle Zeitkonstanten des individuellen Krankheitsverlaufs. Wir betonen, dass bei der Parameteranpassung nicht regularisiert wird. Da die Anpassung dennoch konvergiert, bedeutet dies, dass die Parameter lokal eindeutig bestimmt sind. 

\subsection{Datenbasis}
\label{sec:datenbasis}

\begin{SCtable}[0.95\textwidth][!ht]
	\centering
	\begin{tabular}{lp{0.6\textwidth}}
		\hline \noalign{\smallskip}
		Größe & Quelle \\
		\noalign{\smallskip} \hline \noalign{\smallskip}
		Einwohner &   \href{https://www.destatis.de/DE/Themen/Laender-Regionen/Regionales/Gemeindeverzeichnis/Administrativ/04-kreise.html}{Statistisches Bundesamt} \\
		gemeldete Neuinfektionen & 
		\href{https://raw.githubusercontent.com/KITmetricslab/covid19-forecast-hub-de/master/data-truth/RKI/by_age/truth_RKI-Cumulative Cases by Age_Germany.csv}{Robert Koch Institut, aufbereitet am Karlsruher Institut für Technologie durch German and Polish COVID-19 Forecast Hub} \\
		gemeldete Sterbefälle & 
		\href{https://raw.githubusercontent.com/KITmetricslab/covid19-forecast-hub-de/master/data-truth/RKI/by_age/truth_RKI-Cumulative Deaths by Age_Germany.csv}{RKI, KIT} \\
		gemeldete Erstimpfungen & \href{https://www.rki.de/DE/Content/InfAZ/N/Neuartiges_Coronavirus/Daten/Impfquotenmonitoring.xlsx?__blob=publicationFile}{RKI} \\
		\noalign{\smallskip} \hline
	\end{tabular}
	\caption{Datenquellen.}
	\label{tab:meldedaten}
\end{SCtable}

\subsection{Impfparameter}
\label{sec:impfparameter}
Für die beiden Impfparameter bestimmen wir nun plausible Referenzwerte und zwei Wertepaare, die die Impfwirkung nach oben und nach unten abschätzen. Unser Modell vereinfacht die reale Entwicklung des Impfschutzes. Während er sich tatsächlich nach der zweiten Impfung nochmals steigert, springt er in unserem Modell nur einmal. Ferner setzen wir voraus, dass der Schutz vor Erkrankung und der Schutz davor, das Virus zu übertragen, miteinander einhergehen.  

Laut \href{https://www.rki.de/SharedDocs/FAQ/COVID-Impfen/FAQ_Liste_Wirksamkeit.html}{RKI} setzt ein erster Impfschutz von 60\% bis 70\% bereits 10-14 Tage nach der Erstimpfung ein -- unabhängig vom Impfstoff. Durch die Zweitimpfung erreicht die Wirksamkeit dann bei mRNA-Impfstoffen 4 bis 8 Wochen nach der Erstimpfung 95\% und bei AstraZeneca spätestens 14 Wochen nach der Erstimpfung 80\% \cite{der2021epidemiologisches}. Wir leiten daraus folgende Szenarien ab. Eine Wirksamkeit von $\varepsilon=60\%$ und ein Wirkbeginn von $\tau^p = 14$ d liefern eine untere Abschätzung für den Impfeffekt. Eine obere Abschätzung ergibt sich für $\varepsilon=100\%$ und $\tau^p = 10$ d. Als Referenzszenario wählen wir $\varepsilon=91\%$ und $\tau^p = 14$ d. Die 91\% ergeben sich als mittlere finale Wirksamkeit von mRNA- und Vektorimpfstoffen, gewichtet gemäß der \href{https://www.rki.de/DE/Content/InfAZ/N/Neuartiges_Coronavirus/Daten/Impfquotenmonitoring.xlsx?__blob=publicationFile}{Verabreichung in Deutschland} (Stand 10.6.2021) . Die zwei Wochen sind geschätzt: Der Wirkbeginn ist etwas später als der mittlere Wirkbeginn der Erstimpfung angesetzt, um zu berücksichtigen, dass der volle Schutz erst nach der Zweitimpfung vorliegt.

\subsection{Kontaktraten}
\label{sec:anpassung:kontakt}
Wir passen pro Woche zwei Kontaktraten an, eine für Wochentage und eine für das Wochenende. Über die Osterzeit erweitern wir das Wochenende von Karfreitag bis Ostermontag. Da sich geänderte Kontaktraten erst verzögert um die Meldezeit $\tau^r$ auf die Meldezahlen auswirken, können wir die Kontaktraten der letzten beiden Wochen nicht an Meldedaten anpassen, die auch nur bis zum Ende des Simulationszeitraums reichen. Deswegen setzten wir die Kontaktraten der letzten beiden Wochen jeweils durch die entsprechende Rate der Vorwoche fort. Die angepasste Kontaktrate für Deutschland ist in Abbildung~\ref{fig:deutschland:kontakte} dargestellt.

\subsection{Entdeckungsraten}
\label{sec:anpassung:entdeckung}
Es wird eine Entdeckungsrate pro Woche angepasst. Für die Zeit vor Simulationsbeginn wird die Rate der ersten Woche fortgesetzt und in den zwei letzten Wochen die Rate der drittletzten Woche. Damit die Kurve der Neuinfektionen überall geformt werden kann, erfolgen die Wechsel der Entdeckungsrate an Donnerstagen. Dem verlängerten Osterwochenende wird noch eine weitere eigene Entdeckungsrate zugewiesen. Die angepasste Entdeckungsrate ist in Abbildung~\ref{fig:deutschland:entdeckung} abgebildet.    

\subsection{Maximum-Likelihood-Schätzer}
\label{sec:maximum:likelihood}
Wir schätzen unsere Modellparameter per \href{https://de.wikipedia.org/wiki/Maximum-Likelihood-Methode}{Maximum-Likelihood-Methode}. Es seien $\vec{x} \in \mathbb{R}^p$ der Vektor aller unbekannten Parameter, $T = \left\{ t_s,\ldots,t_e \right\}$ die $q$ Tage des Simulationszeitraums und $T' = \left\{ t_f,\ldots,t_e \right\}$ die $q'$ Tage, an denen wir Sterbezahlen anpassen. Zu Parametern $\vec{x}$ simulieren wir Fallmeldungen $\vec{r}(\vec{x})$ und Sterbezahlen $\vec{m}(\vec{x})$. Jeder Eintrag von $\vec{r}$ gehört zum Beginn eines Tages $i \in T$ und jeder Eintrag von $\vec{m}$ zum Beginn eines Tages $i \in T'$. $\hat{r}$ und $\hat{m}$ seien die zugehörigen Meldewerte. Wir nehmen an, dass die Messfehler von $\hat{r}$ und $\hat{m}$ jeweils unabhängig identisch normalverteilt sind mit Standardabweichungen $\sigma_r$ bzw. $\sigma_m$. Dann ergibt sich der Maximum-Likelihood-Schätzer durch Lösen des nichtlinearen gewichteten Kleinste-Quadrate-Problems
\begin{align}
    \min\limits_{\vec{x} \in \mathbb{R}^p} 
    \sum\limits_{i=1}^q \left( \frac{r_i(\vec{x}) - \hat{r}_i}{\sigma_r}\right)^2 +  
    \sum\limits_{i=1}^{q'} \left( \frac{m_i(\vec{x}) - \hat{m}_i}{\sigma_m}\right)^2 \; .
\end{align}
Um die Lösung zu beschleunigen, verwenden wir in allen Rechnungen \href{https://de.wikipedia.org/wiki/Automatisches_Differenzieren}{Automatisches Differenzieren}. Dies liefert insbesondere die in einem Newton-Verfahren benötigten Jacobi-Matrizen
    $D_{\vec{x}}\,\vec{r}$ und $D_{\vec{x}} \, \vec{m}$ .

Ein Problem liegt darin, dass wir die Messfehler nicht kennen. Deshalb betten wir die Parameteranpassung in eine äußere Schleife ein, in der wir $\sigma_r$ und $\sigma_m$ so lange durch die empirischen Standardabweichung ersetzen, bis die Werte konvergieren:
\begin{align}
\sigma_r &= \sqrt{\frac{1}{q} \sum\limits_{i=1}^q \left( r_i(\vec{x})-\hat{r}_i \right)^2} \, , \quad 
\sigma_m = \sqrt{\frac{1}{q'} \sum\limits_{i=1}^{q'} \left( m_i(\vec{x})-\hat{m}_i \right)^2} \label{eq:sigma} \; .
\end{align}
Mit Hilfe der Jacobi-Matrizen bestimmen wir anschließend die Kovarianzmatrix der identifizierten Parameter:
\begin{align}
    C &= \mbox{cov}(\vec{x}) = \left( M^t M \right)^{-1}\; \mbox{mit} \;
    M = \left[\begin{array}{c}
    \sigma_r^{-1} D_{\vec{x}} \vec{r} \\
    \sigma_m^{-1} D_{\vec{x}} \vec{m} \\    
    \end{array} \right] \; .
\end{align}
$\sqrt{c_{ii}}$ liefert uns schließlich eine Schätzung für die Standardabweichung $\sigma_i$ des $i$-ten Parameters $x_i$. Die in Tabelle~\ref{tab:parameter} grün gedruckten Parameterintervalle haben die Form $x_i \pm 3\, \sigma_i$. Die Fehlerschläuche in den nachfolgenden Abbildungen haben, je nach Angabe, entweder einen Radius von $\sigma_i$ oder $3 \, \sigma_i$, d.h. sie stellen 68,3\%- oder 99,7\%-Konfidenzintervalle dar.

\section{Ergebnisse}
\label{sec:ergebnisse}
Wir präsentieren die Ergebnisse der Parameteranpassung für Deutschland und vier ausgewählte Bundesländer: Hessen, Niedersachsen, Rheinland-Pfalz und Thüringen.
Der Zeitraum für Simulation und Anpassung erstreckt sich vom 25.01. bis 17.05.2021, enthält also die dritte Corona-Welle. Wir zeigen, dass die Meldedaten nicht nur im Wochenmittel, sondern tagesgenau reproduziert werden können, setzen die wechselnden Kontakt- und Entdeckungsraten mit realen Ereignissen wie Ostern oder Schulbeginn in Beziehung und folgern schließlich, dass vor allem die Massentests an Schulen die dritte Welle gebrochen haben.        

\subsection{Fall- und Sterbezahlen}
\label{sec:ergebnis:faelle}
\begin{SCfigure}[0.95\textwidth][!ht]
\includegraphics[width=0.95\textwidth]{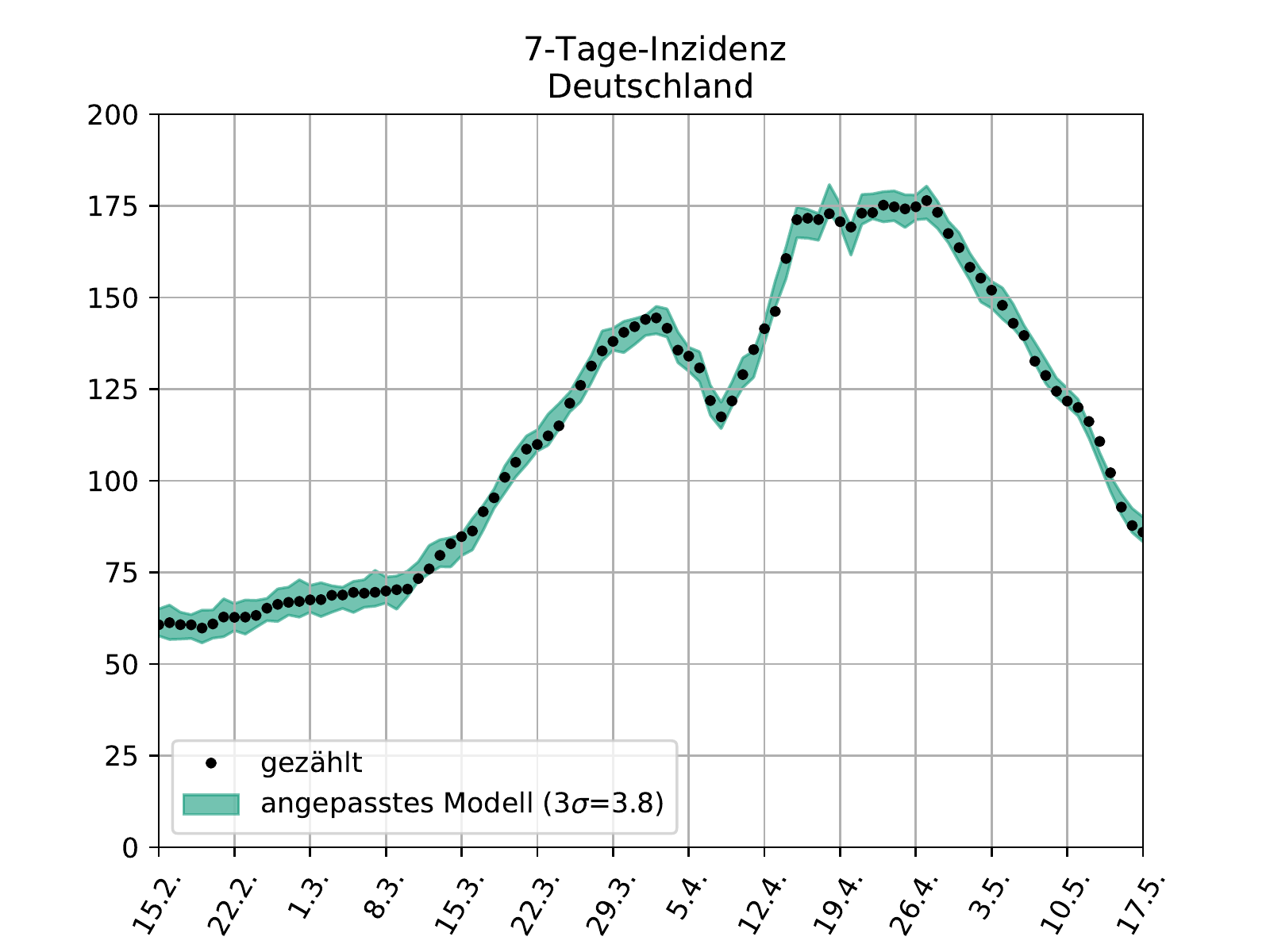}
\caption{Modellanpassung.\\ 7-Tage-Inzidenz laut \href{https://raw.githubusercontent.com/KITmetricslab/covid19-forecast-hub-de/master/data-truth/RKI/by_age/truth_RKI-Cumulative Cases by Age_Germany.csv}{RKI} und mit angepasstem Modell simuliert.}
\label{fig:deutschland:anpassung}
\end{SCfigure}

Abbildung~\ref{fig:deutschland:anpassung} zeigt die gemessene und simulierte 7-Tage-Inzidenz für Deutschland vor und während der dritten Welle. Offenbar gibt das angepasste Modell den wahren Verlauf sehr genau wieder, insbesondere die Zeitpunkte, an denen sich die Steigung abrupt ändert.  

\begin{SCfigure}[][!htb]
\includegraphics[width=0.95\textwidth]{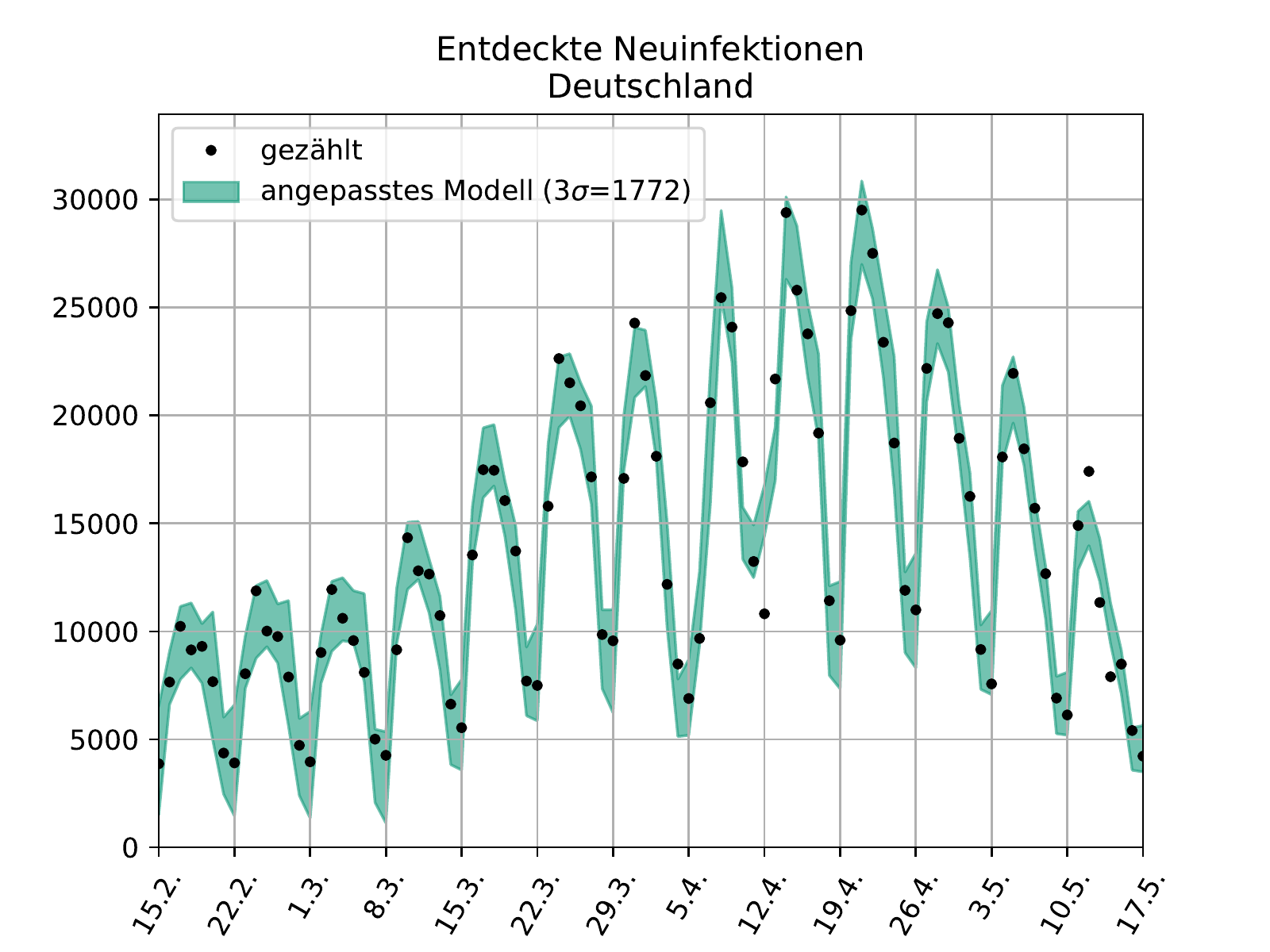}
\caption{Tagesgenaue Anpassung des Modells an gemeldete Neuinfektionen.}
\label{fig:deutschland:faelle}
\end{SCfigure}

Die Anpassung von zwei Kontaktraten und einer Entdeckungsrate pro Woche wird erst möglich, weil wir eine weit verbreitete Grundannahme in Frage stellen. Üblicherweise werden die starken Schwankungen der Fallzahlen im Wochenverlauf damit \href{https://www.rki.de/SharedDocs/FAQ/NCOV2019/FAQ_Liste_Fallzahlen_Meldungen.html}{erklärt}, dass Menschen am Wochenende seltener zum Arzt gehen und  Gesundheitsämter verzögert melden. Die gängige Vorstellung ist also, dass lediglich die gemeldeten Fallzahlen mäandern, während sich die Infektion selbst deutlich gleichmäßiger in der Gesellschaft ausbreitet. In dieser Vorstellung liefern gemessene Tageswerte also keinen Informationsgewinn gegenüber wöchentlichen Mittelwerten. Unsere Analyse zeigt jedoch, dass sich die Schwankungen im Wochenverlauf gut durch verminderte Kontakte am Wochenende erklären lassen, also einem realen Geschehen folgen, s. Abbildung~\ref{fig:deutschland:faelle} und \ref{fig:deutschland:kontakte}. 
Nur weil nun tatsächlich sieben mal mehr nutzbare Messungen zur Verfügung stehen, können wir überhaupt Kontakt- und Entdeckungsraten gleichzeitig anpassen. 

\begin{SCfigure}[][!htb]
\includegraphics[width=0.95\textwidth]{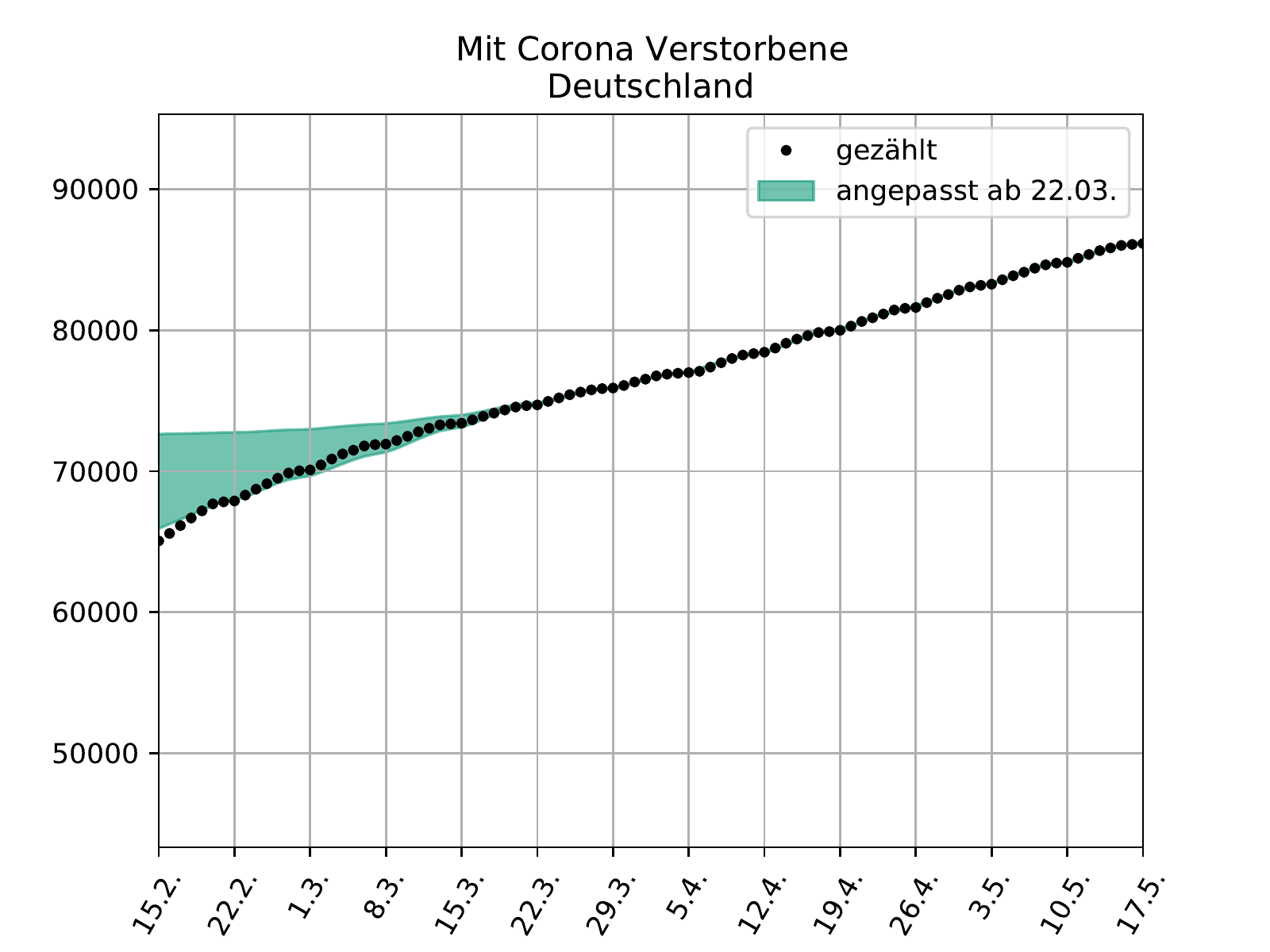}
\caption{Verstorbene laut \href{https://raw.githubusercontent.com/KITmetricslab/covid19-forecast-hub-de/master/data-truth/RKI/by_age/truth_RKI-Cumulative Deaths by Age_Germany.csv}{RKI} und mit angepasstem Modell simuliert. Die Anpassung beginnt am 22.03., da zuvor die Annahme einer konstanten Sterberate nicht erfüllt ist. Im weiteren Verlauf sind Messung und Fit nicht zu unterscheiden.} 
\label{fig:deutschland:verstorbene}
\end{SCfigure}
Umgekehrt erlauben wir uns aber immer noch deutlich weniger Parameter (51) als Meldewerte (168). D.h. die gute Übereinstimmung ist nicht das selbstverständliche Ergebnis einer Überanpassung, sondern zeigt, dass unser Modell die Struktur des Problems erfasst. Neben den gemeldeten Fällen passen wir ab dem 22.03. auch noch die Verstorbenen an. Das Ergebnis ist in Abbildung~\ref{fig:deutschland:verstorbene} dargestellt. Im Anpassungszeitraum sind Modell und Realität optisch nicht zu trennen. Davor gibt es allerdings größere Abweichungen. Dies erklärt sich daraus, dass wir eine konstante Sterberate annehmen. Tatsächlich ist die Sterberate aber seit dem Winter deutlich gefallen. Dies liegt daran, dass die besonders Gefährdeten mittlerweile weitgehend geimpft sind. Da es uns in dieser Studie aber um die Ausbreitung der Epidemie geht und nicht um eine genaue Vorhersage der Sterbefälle, ist die Fehlanpassung im Winter unproblematisch. Dennoch ist es wichtig, die Sterbefälle in der Anpassung zu berücksichtigen: Wenn die Todeszahlen weniger stark steigen als die gemeldeten Fälle, dann ist daran bei konstanter Sterberate zu erkennen, wie stark die Entdeckungsrate gestiegen ist.

\subsection{Kontakt- und Entdeckungsraten}
\label{sec:ergebnis:kontakte}
\begin{SCfigure}[][!ht]
\includegraphics[width=\textwidth]{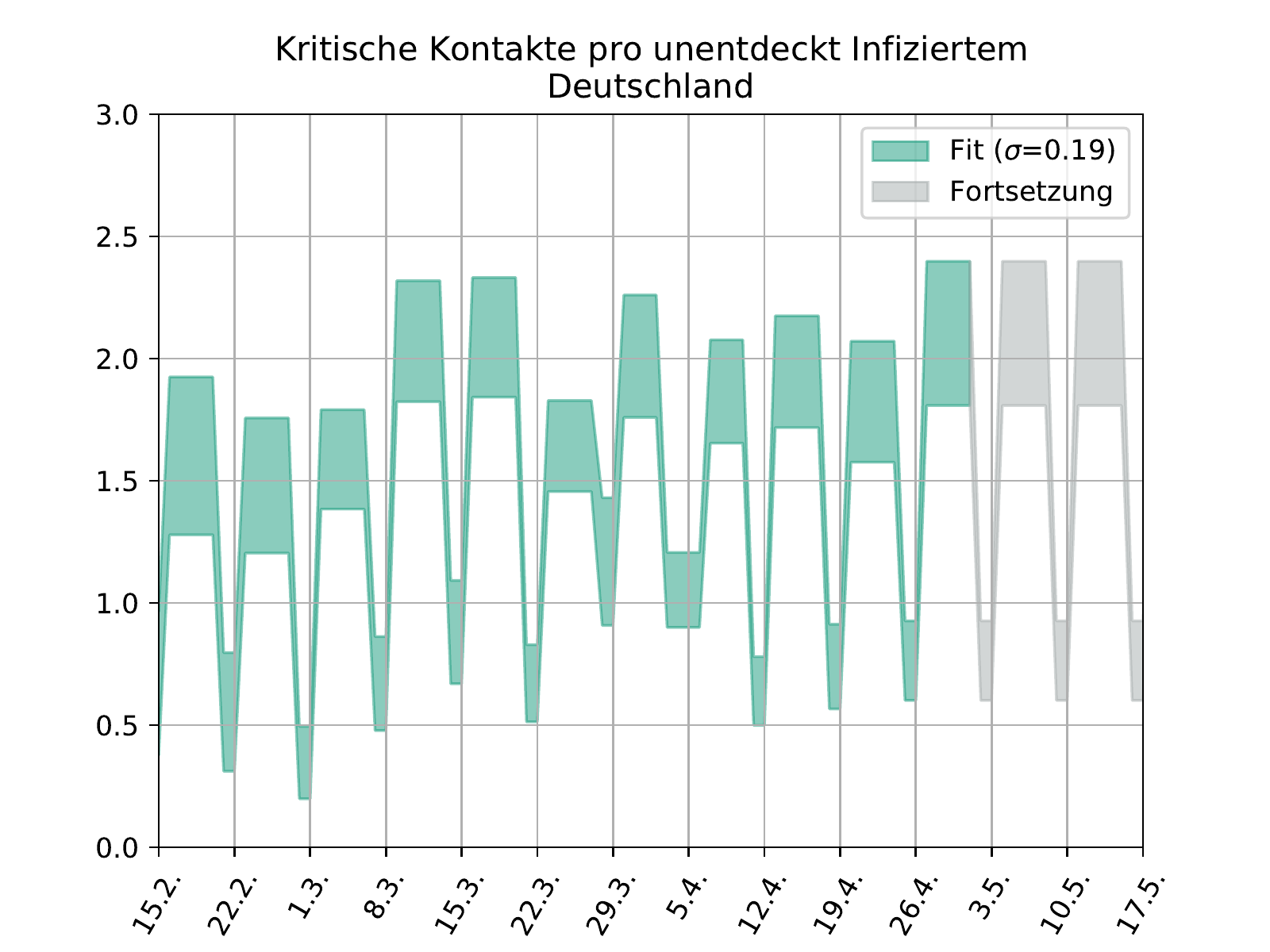}
\caption{Rekonstruierte Basisreproduktionsrate für Deutschland, genauer
$\boldsymbol{\kappa(t) \left(\tau^e - \tau^s \right)}$. Da sich eine geänderte Kontaktrate erst verzögert in den Meldezahlen widerspiegelt, werden die letzten Raten nicht identifiziert, sondern wochenperiodisch fortgesetzt. Markante Daten: Ankündigung einer Osterruhe (22.03.), Ostern (02.04.-05.04.), Bundesnotbremse (24.04.).}
\label{fig:deutschland:kontakte}
\end{SCfigure}
Abbildung~\ref{fig:deutschland:kontakte} zeigt die rekonstruierte Kontaktrate für Deutschland. Genauer ist das Produkt aus Kontaktrate und Dauer der infektiösen Phase dargestellt, was ungefähr der Basisreproduktionsrate entspricht: 
\begin{align}
 R_0(t) &\approx \kappa(t) \left( \tau^e - \tau^s \right)\; .
 \label{eq:basisreproduktion}
\end{align}
Bei einem Wert von z.B. 2 gibt ein unentdeckter Infizierter in einer nicht immunisierten Umgebung das Virus an 2 Personen weiter. 
Da sich eine geänderte Kontaktrate erst verzögert in den gemessenen Fallzahlen zeigt, lassen sich die letzten Werte nicht identifizieren und werden extrapoliert.

Abbildung~\ref{fig:deutschland:kontakte} lässt sich sehr schön im Lichte realer Ereignisse interpretieren. Zunächst fällt auf, dass die Kontaktrate am Wochenende systematisch und gleichmäßig niedriger ausfällt als unter der Woche. D.h. Beruf, Schule oder Geschäfte scheinen sich kritischer auszuwirken als das familiäre Umfeld. Öffnen Gastronomie und kulturelle Einrichtungen erst wieder in größerem Umfang, dann könnte dieser Effekt allerdings weniger deutlich ausfallen.     

Gut sichtbar ist ferner der Anstieg der Kontaktrate seit Ende Februar infolge der sich durchsetzenden britischen Variante. Wir betonen hier noch einmal, dass die Kontaktrate die Häufigkeit {\em kritischer} Kontakte beschreibt. Selbst bei gleichem Kontaktverhalten steigt sie also, wenn sich ansteckungsfähigere Virusvarianten durchsetzen.  

Ab dem 22.03. sinkt die Kontaktrate dann wieder. Dieser Einschnitt fällt mit der Ankündigung einer Osterruhe durch die Regierung zusammen. Es scheint sich also um einen rein psychologischen Effekt zu handeln. Am Wochenende vor Ostern fiel der Wochenendrückgang dafür schwächer aus als an jedem anderen Wochenende. Möglicherweise versuchten die Menschen hier, noch schnell anstehende Erledigungen und Kontakte vorzuziehen. Ostern selbst war die Kontaktrate vergleichsweise niedrig und in den drei folgenden Wochen stabilisierte sie sich bei einem Wert unter 2, also deutlich unter dem Wert von vor dem 22.03. Die Bundesnotbremse wurde am 24.04. wirksam. In der Folge ist die Kontaktrate -- jedenfalls nach unserer Berechnung -- eher noch gestiegen. Möglicherweise haben andere Effekte, z.B. Schulöffnungen nach den Ferien, die Bundesnotbremse überlagert. In jedem Fall scheint ihre Wirkung gemessen an den Einschränkungen eher gering gewesen zu sein. 

\begin{SCfigure}[][!ht]
\includegraphics[width=0.95\textwidth]{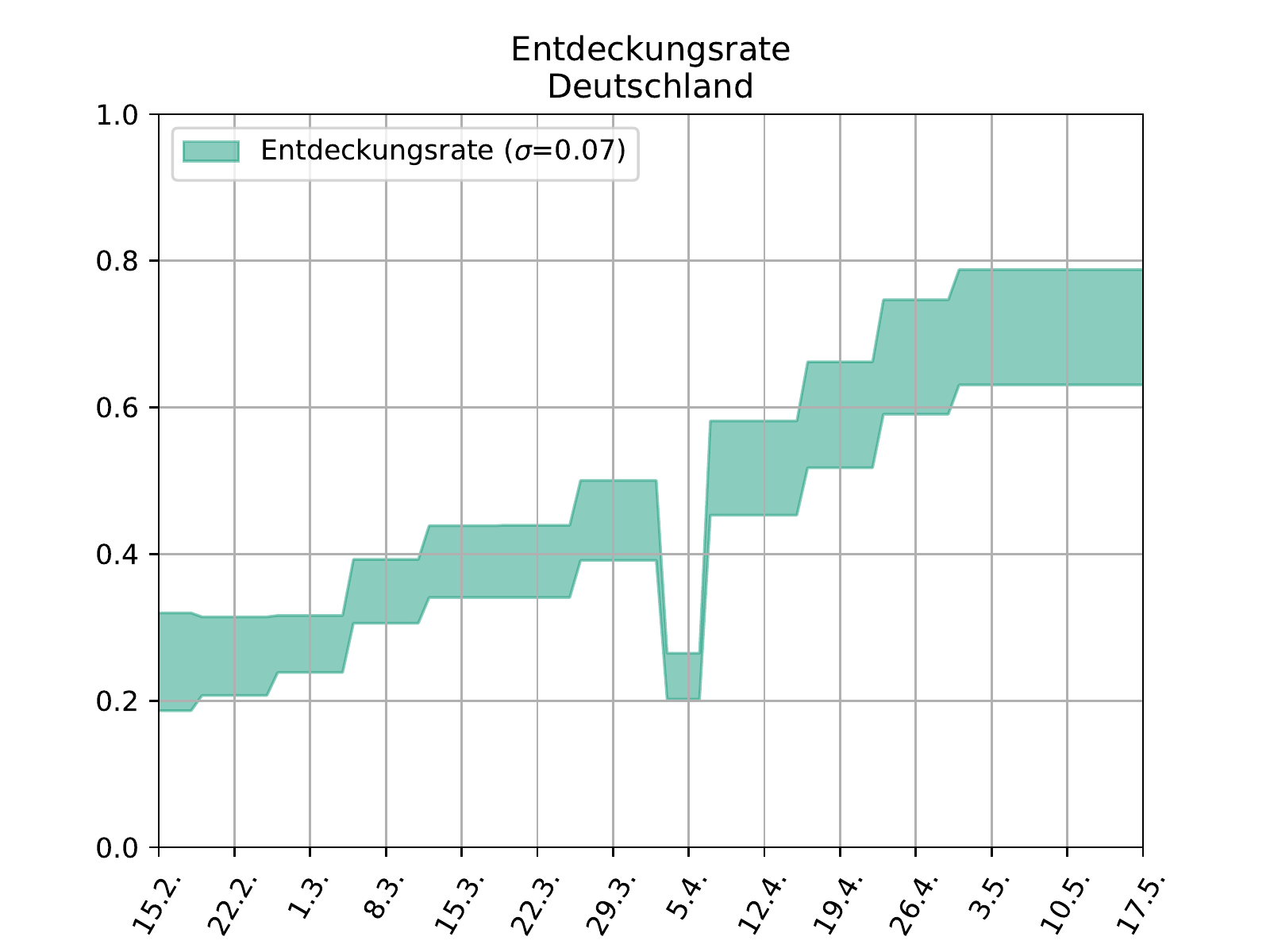}
\caption{Rekonstruierte Entdeckungsrate für Deutschland. Erster sprunghafter Anstieg mit Einführung der \href{https://www.bundesgesundheitsministerium.de/coronavirus/nationale-teststrategie/faq-schnelltests.html}{Schnelltests für alle am 08.03.2021}. Einbruch um Ostern. Finale Entdeckungsrate um 71\% deckt sich mit Ergebnis der  \href{https://elisa-luebeck.de/einjaehrige-corona-studie-zu-antikoerpern-mit-rund-3000-probandinnen-und-probanden-zeigt-dunkelziffer-ist-deutlich-zurueckgegangen/}{ELISA-Studie}.}   
\label{fig:deutschland:entdeckung}
\end{SCfigure}
Wir wenden uns nun Abbildung~\ref{fig:deutschland:entdeckung} zu. Eines der wichtigsten Ergebnisse dieser Arbeit ist der Umstand, dass es uns gelungen ist, die Entdeckungsrate alleine aus den Meldedaten des RKI zu rekonstruieren. Wichtige Ereignisse, denen man einen deutlichen Einfluss auf die Entdeckungsrate zutrauen würde, spiegeln sich auch tatsächlich im rekonstruierten Verlauf wider. Seit Einführung der \href{https://www.bundesgesundheitsministerium.de/coronavirus/nationale-teststrategie/faq-schnelltests.html}{Schnelltests für alle am 08.03.} steigt die Rate monoton von 25\% auf 71\% -- mit einem Einbruch um Ostern. Der finale Wert entspricht erstaunlich genau der Dunkelziffer von 30\%, die in der \href{https://elisa-luebeck.de/einjaehrige-corona-studie-zu-antikoerpern-mit-rund-3000-probandinnen-und-probanden-zeigt-dunkelziffer-ist-deutlich-zurueckgegangen/}{ELISA-Studie} experimentell gefunden wurde. 

\subsection{Zeitkonstanten des Krankheitsverlaufs}
\label{sec:krankheitsverlauf}
\begin{SCfigure}[][!htb]
\includegraphics[width=0.95\textwidth]{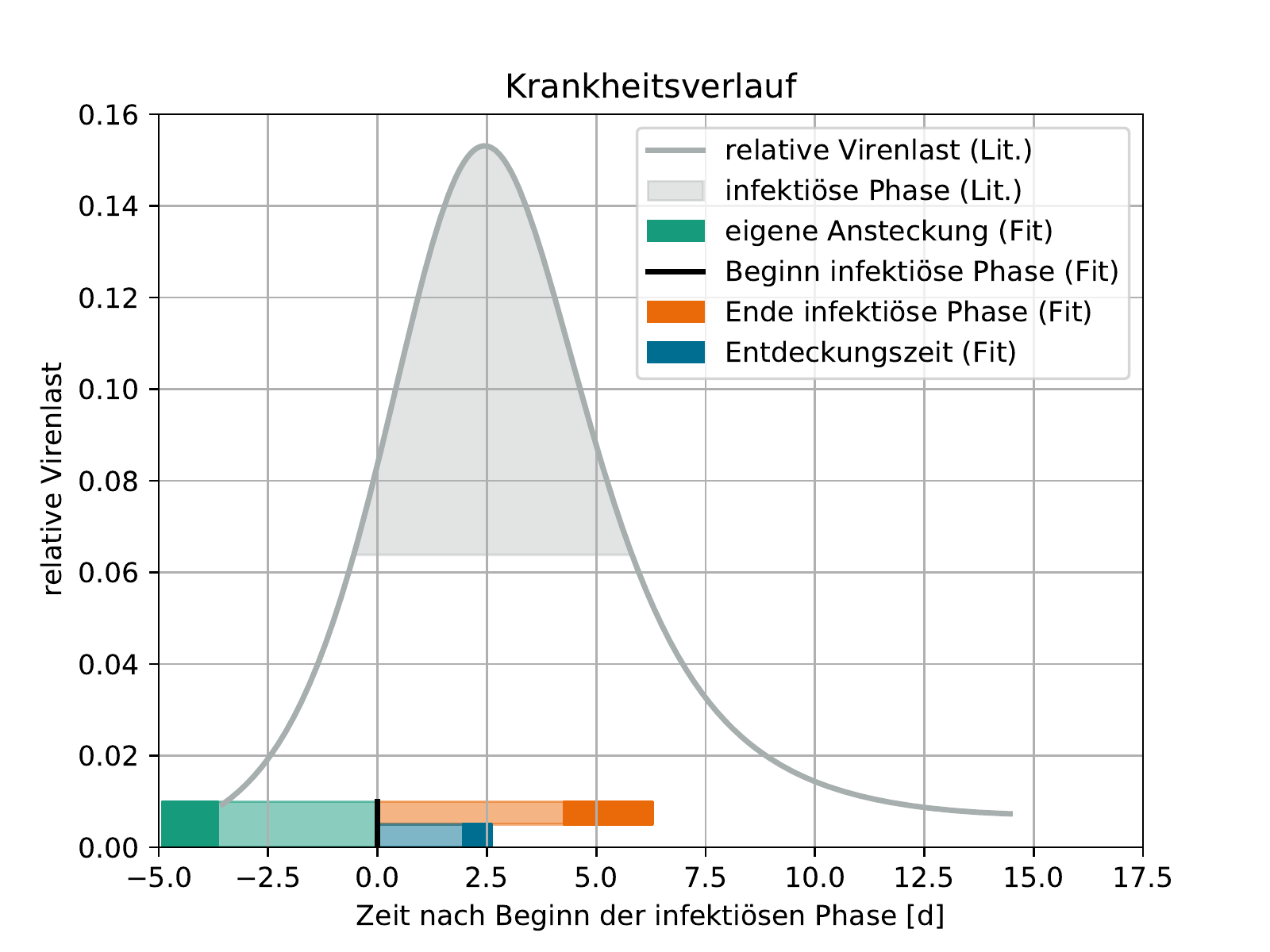}
\caption{Gefundene Zeitkonstanten des Krankheitsverlaufs im Vergleich zur Virenlast.}
\label{fig:krankheit}
\end{SCfigure}
In diesem Abschnitt vergleichen wir die rekonstruierten Zeitkonstanten des individuellen Krankheitsverlaufs mit dem Verlauf der Virenlast, den wir in einem älteren Bericht an Literaturwerte angepasst haben. Daraus leiten wir Thesen zum ungenutzten Potential des Maßnahmenpakets aus Tests, Nachverfolgung und Isolation ab.  

In einem früheren Bericht \cite{mohring21tests} haben wir den zeitlichen Verlauf der Virenlast an Literaturwerte \cite{he2020author} angepasst. Nach \cite{he2020author} erreicht die Virenlast ihr Maximum, wenn auch erste Symptome auftreten.
Zu diesem Zeitpunkt sei bereits 44\% der Virenlast gestreut. 5 Tage zuvor seien 1\% und 3 Tage zuvor 9\% der Virenlast produziert. Die angepasste Kurve ist in Abbildung~\ref{fig:krankheit} grau dargestellt. 

Die Zeitachse und die farbigen Balken beziehen sich dagegen auf die kastenförmige Infektiosität, deren Parameter wir in dieser Arbeit an die Meldedaten angepasst haben und die in Tabelle~\ref{tab:parameter} aufgelistet sind. Die dort angegebene Streuung entspricht 3 $\sigma$ bzw. einem 99,7\%-Konfidenzintervall.
Der grüne Balken gibt die Latenzzeit $\tau_s = 4,3\pm0,6$ d an, der orange Balken die Dauer der infektiösen Phase $\tau^e - \tau^s = 5,3\pm 1$ d. 

Auch in \cite{he2020author} wird ein Beginn der infektiösen Phase genannt, und zwar 3 Tage vor dem Maximum der Infektiosität. Der Bereich, in dem der dortige Grenzwert überschritten wird, haben wir hellgrau eingefärbt und mit {\em infektiöse Phase, Lit.} beschrieben. Der Beginn der infektiösen Phase in \cite{he2020author} und von unserer kastenförmigen Infektiosität kann man nicht gleichsetzen, da die natürliche Infektiosität langsam einsetzt, während unsere von gar nicht infektiös auf voll infektiös springt. Daher haben wir zur Vergleichbarkeit die beiden infektiösen Phasen zentriert. Offenbar passen unsere angepassten Zeitkonstanten recht gut zum Verlauf der Virenlast.

Als nächstes werden wir Thesen zum ungenutzten Potential des Maßnahmenpakets Testung formulieren. Hierfür ist ein dritter Parameter zentral -- die mittlere Zeit ab Infektion, nach der ein Entdeckter isoliert wird: $\tau^d = 6,6 \pm 0,3$ d. Nach unserem Modell beträgt die bis zur Entdeckung gestreute relative Virenlast 
\begin{align}
    \nu^d = \frac{\tau^d - \tau^s}{\tau^e-\tau^s} \approx 0,43\; .
\end{align}
Dies entspricht ziemlich genau den 44\% aus \cite{he2020author} bei Auftritt von Symptomen. Insgesamt passen die Daten also zu folgendem Szenario. Infizierte werden erst entdeckt, wenn sie Symptome zeigen, und nicht Entdeckte streuen Viren solange, bis die Virenlast natürlicherweise nachlässt. Dies führt zu folgenden Thesen:
\begin{enumerate}
    \item Anlasslose Antigen-Schnelltests können Infektiöse nicht aufspüren, bevor sich Symptome zeigen.
    \item In der dritten Welle beruhte der Erfolg von Tests vor allem darauf, dass Entdeckte direkt nach dem Auftreten von Symptomen isoliert wurden.
    \item Nachverfolgung trug dagegen kaum zum Erfolg bei. 
\end{enumerate}
These 1 deckt sich mit Aussagen von \href{https://www.ndr.de/nachrichten/info/Drosten-Schnelltests-sind-wohl-weniger-zuverlaessig-als-gedacht,coronavirusupdate178.html}{Drosten}. Um Infizierte in einer früheren Krankheitsphase zu entdecken, könnten bei Reihenuntersuchungen z.B. \href{https://www.rki.de/SharedDocs/FAQ/NCOV2019/FAQ_Liste_Diagnostik.html}{gepoolte PCR-Tests} eingesetzt werden, die früher anschlagen als Antigen-Tests.

Zu These 2. Unsere Parameteranpassung liefert, dass die Zeitspanne, über die ein Infizierter ungehindert andere ansteckt, bei Entdeckten um 57\% gegenüber Unentdeckten reduziert ist. Zumindest in der Mechanik unseres Modells ist dies der wesentliche Hebel von Tests. Umgekehrt heißt das aber auch, dass Infizierte, die {\em nicht} durch Tests aufgespürt werden, sich meist auch {\em nicht} sofort eigenständig isolieren, sobald Symptome auftreten. Denn Symptome treten ja im Gros der Verläufe auf, laut \cite{buitrago2020occurrence} bei ca. 80\%.

Zu These 3. Die identifizierte Zeit $\tau^d$ bis zur Entdeckung ist in unserem Modell ein effektiver Parameter. Das Verhältnis $\nu^d = \frac{\tau^d - \tau^s}{\tau^e-\tau^s}$ legt fest, wie stark die Weitergabe des Virus bei Entdeckten effektiv gemindert wird. Hätte Nachverfolgung einen großen Einfluss gehabt, dann müsste $\tau^d$ deutlich kleiner sein als die Zeit bis zum Auftritt von Symptomen. Denn dann wären auch Personen, die durch einen Infizierten zuvor angesteckt wurden, in der Regel rechtzeitig isoliert worden. Netto hätte das dann den Effekt gehabt, als hätte der Infizierte früher aufgehört, andere anzustecken. Tatsächlich entspricht der identifizierte Wert von $\tau^d$ aber dem Symptombeginn.       
Um die Nachverfolgung effektiver zu machen, könnten aktuelle Fälle 
zuerst verfolgt werden, auf Kosten unbearbeiteter Altfälle, deren Verfolgung ohnehin zu spät käme. 

Insgesamt können wir an dieser Stelle aber nur von Thesen sprechen. Unser Modell von Ansteckung und Testung ist zu einfach, als dass unsere Schlüsse ohne Prüfung durch andere Disziplinen als Belege gelten könnten.



\subsection{Schlüsselrolle von Schnelltests an Schulen}
\label{sec:schlüsselrolle}

\begin{SCfigure}[][!htb]
\includegraphics[width=0.95\textwidth]{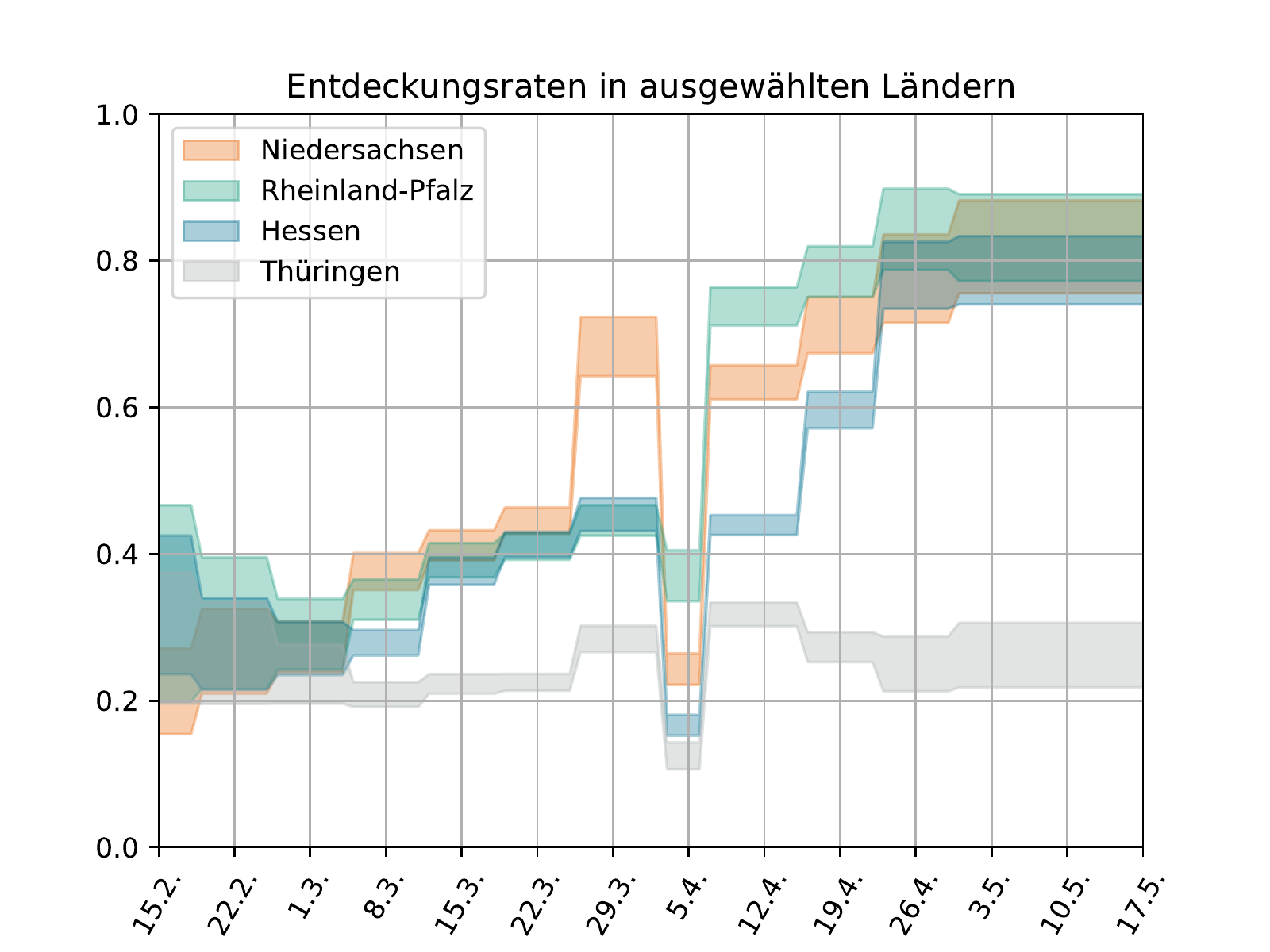}
\caption{Schnelltests an Schulen steigern Entdeckungsrate deutlich. Testbeginn: Niedersachsen 22.03. (Probewoche) bzw. 12.04. (Ferienende), Rheinland-Pfalz 07.04. (Ferienende), Hessen 19.04 (Ferienende). Schultests in Thüringen meist nur vom 12.04. (Ferienende) bis 24.04. (Bundesnotbremse).}
\label{fig:schultests}
\end{SCfigure}

Aus Abbildung~\ref{fig:deutschland:entdeckung} geht hervor, dass die Entdeckungsrate in Deutschland seit der zweiten Märzwoche kontinuierlich angestiegen ist, also seitdem \href{https://www.bundesgesundheitsministerium.de/coronavirus/nationale-teststrategie/faq-schnelltests.html}{Selbsttests für alle} verfügbar sind. Lediglich über Ostern ist die Entdeckungsrate kurz eingebrochen, um nach den unterschiedlich endenden Schulferien umso steiler anzusteigen.
In diesem Abschnitt liefern wir Hinweise darauf, dass die Selbsttests an Schulen in besonderem Maße zur Steigerung der Entdeckungsrate beigetragen haben dürften. 

Dazu wiederholen wir die Anpassungen, die wir bisher nur für Deutschland als Ganzes durchgeführt haben, für vier einzelne Bundesländer. Lediglich die Parameter des individuellen Krankheitsverlaufs werden von Deutschland übernommen, da diese Werte recht sensibel auf statistische Streuungen reagieren -- und die fallen für die einzelnen Bundesländer natürlich größer aus als für ganz Deutschland. 

Drei der vier Bundesländer wählen wir danach aus, dass ihre Osterferien zu unterschiedlichen Zeiten endeten. 
\href{https://www.mk.niedersachsen.de/startseite/aktuelles/presseinformationen/selbsttests-an-schulen-laufen-an-erste-400-000-tests-ab-heute-in-der-auslieferung-testwoche-startet-montag-198555.html}{Niedersachsen} führte bereits ab dem 22.03. eine erste Testwoche durch. Ab dem 12.04., also nach Ende der dortigen Osterferien, wurden die Schultests dann fortgesetzt. In  \href{https://corona.rlp.de/fileadmin/bm/Bildung/Corona/Konzept_Selbsttests_an_Schulen.pdf}{Rheinland-Pfalz} gab es keine vorgezogene Probewoche. Dafür begann die Schule bereits am 07.04. In \href{https://kultusministerium.hessen.de/schulsystem/umgang-mit-corona-an-schulen/fuer-eltern/elternbriefe/schul-und-unterrichtsbetrieb-ab-dem-19-april-2021}{Hessen} schließlich setzten die Schultests erst mit dem 19.04. ein, dem dortigen Ende der Osterferien.

In Abbildung~\ref{fig:schultests} sind die rekonstruierten Entdeckungsraten der drei Bundesländer dargestellt. Es fällt auf, dass die Werte vor dem 22.03. und nach dem 26.04. ähnlich sind, dazwischen aber stark abweichen, und zwar genau so, wie es die unterschiedlichen Startzeitpunkte der Schultests nahelegen: In Niedersachsen schnellt die Entdeckungsrate bereits vor Ostern hoch, bricht über die Feiertage ein und erreicht erst in der Woche nach dem 12.04. wieder das Niveau von vor Ostern. In Rheinland-Pfalz erfolgt der große Sprung unmittelbar nach Ostern, in Hessen erst zwei Wochen später.  

Man beachte, dass die Entdeckungsraten alleine durch Anpassung an die gezählten Neuinfektionen und Todesfälle bestimmt werden. Der Algorithmus hat keinen Zugriff auf Schulferien und die Feiertage werden nur dadurch berücksichtigt, dass dort das wöchentliche Zeitraster ergänzt ist. Ansonsten fallen die vorgegebenen Sprungstellen auf Donnerstage. Dies führt zum numerischen Effekt, dass Sprünge an Montagen auf zwei Intervalle verteilt werden und dürfte z.B. den Zwischenanstieg in Hessen vor Ferienende erklären. Im Rahmen der verfügbaren Auflösung werden die Startzeitpunkte der Schultests also gut getroffen. 

\subsection{Das Test-Inzidenz-Dilemma}
\label{sec:dilemma}
Wenden wir uns nun Thüringen zu. Hier fällt in Abbildung~\ref{fig:schultests} zunächst auf, dass das Niveau der Entdeckungsrate insgesamt viel niedriger liegt als bei den übrigen Ländern. Dies kann zum Teil mit einem systematischen Anpassungsfehler zusammenhängen: Wir nehmen für alle Bundesländer die für Deutschland gefundene Sterberate an. Tatsächlich aber hat Thüringen das zweithöchste \href{https://de.statista.com/statistik/daten/studie/1093993/umfrage/durchschnittsalter-der-bevoelkerung-in-deutschland-nach-bundeslaendern/}{Durchschnittsalter} in Deutschland und damit sicher eine erhöhte Sterberate. Dies würde bedeuten, dass die tatsächliche Entdeckungsrate Thüringens insgesamt etwas höher liegt als dargestellt. Der relative zeitliche Verlauf bliebe davon aber unbeeinflusst. Insbesondere trifft sicher zu, dass die Entdeckungsrate nach einem kurzen Anstieg direkt nach Ostern sofort wieder abfällt -- im Gegensatz zu allen drei anderen Ländern. 

Auch dies korreliert direkt mit Schnelltests an Schulen. Diese fanden in Thüringen im Wesentlichen nur zwischen dem Ende der Osterferien (12.04.) und der erneuten Schließung der meisten Schulen im Rahmen der \href{https://www.bundesgesundheitsministerium.de/service/gesetze-und-verordnungen/guv-19-lp/4-bevschg-faq.html}{Bundesnotbremse} (24.04.) statt.
Auch die für Thüringen gefundene Basisreproduktionsrate (Abbildung~\ref{fig:rheinland-pfalz:thüringen}) passt zu diesem Szenario. Auf niedrige Werte zwischen Karfreitag und Ferienende folgen hohe Werte nach Schulöffnung und wieder niedrige Werte nach Einsetzen der Bundesnotbremse. 
Im Gegensatz zu Deutschland und den anderen Bundesländern knickt die Inzidenz erst eine Woche später steil nach unten ab, nachdem der bereits eingeleitete Abstieg durch eine Woche mit leichtem Anstieg unterbrochen ist.   

Es könnte also sein, dass Thüringen ein Beispiel für das Test-Inzidenz-Dilemma darstellt: Vermehrte Tests treiben die Entdeckungsrate und damit auch die Inzidenz nach oben, inzidenzgebundene Kontaktbeschränkungen wie die Bundesnotbremse greifen, Schulen werden geschlossen, es werden weniger Tests durchgeführt und die positive Wirkung einer hohen Entdeckungsrate kann sich nicht entfalten. Am Ende müssen alleine harte Kontaktbeschränkungen die Infektionszahlen senken.
Einen interessanten Einblick in das Test-Inzidenz-Dilemma liefert später auch Abbildung~\ref{fig:deutschland:maßnahmen:inzidenz}.

\subsection{Bundesländer auf einen Blick}
\label{sec:bundesländer}
\begin{SCfigure}[][p]
\includegraphics[width=0.49\textwidth]{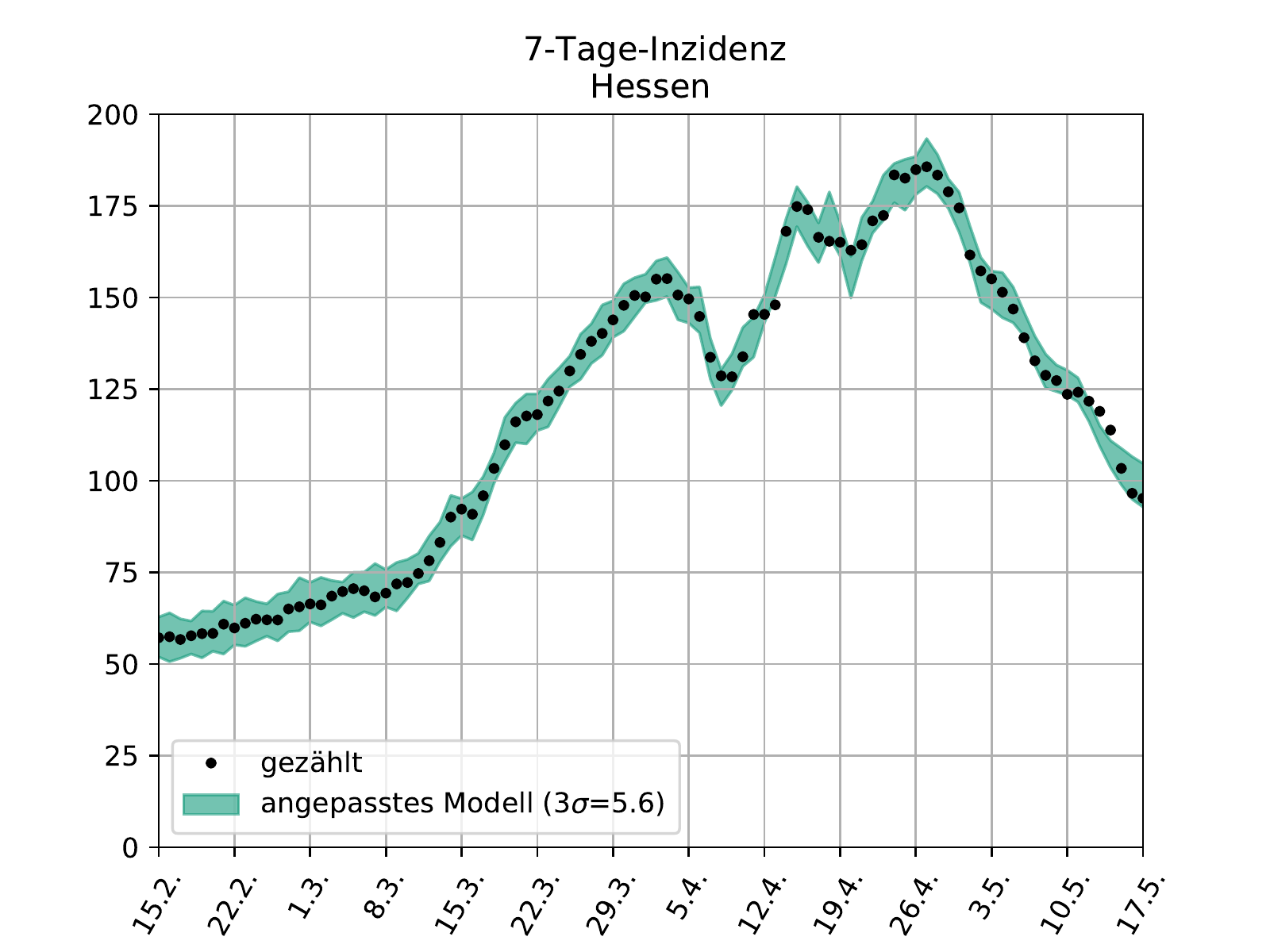}
\hfill
\includegraphics[width=0.49\textwidth]{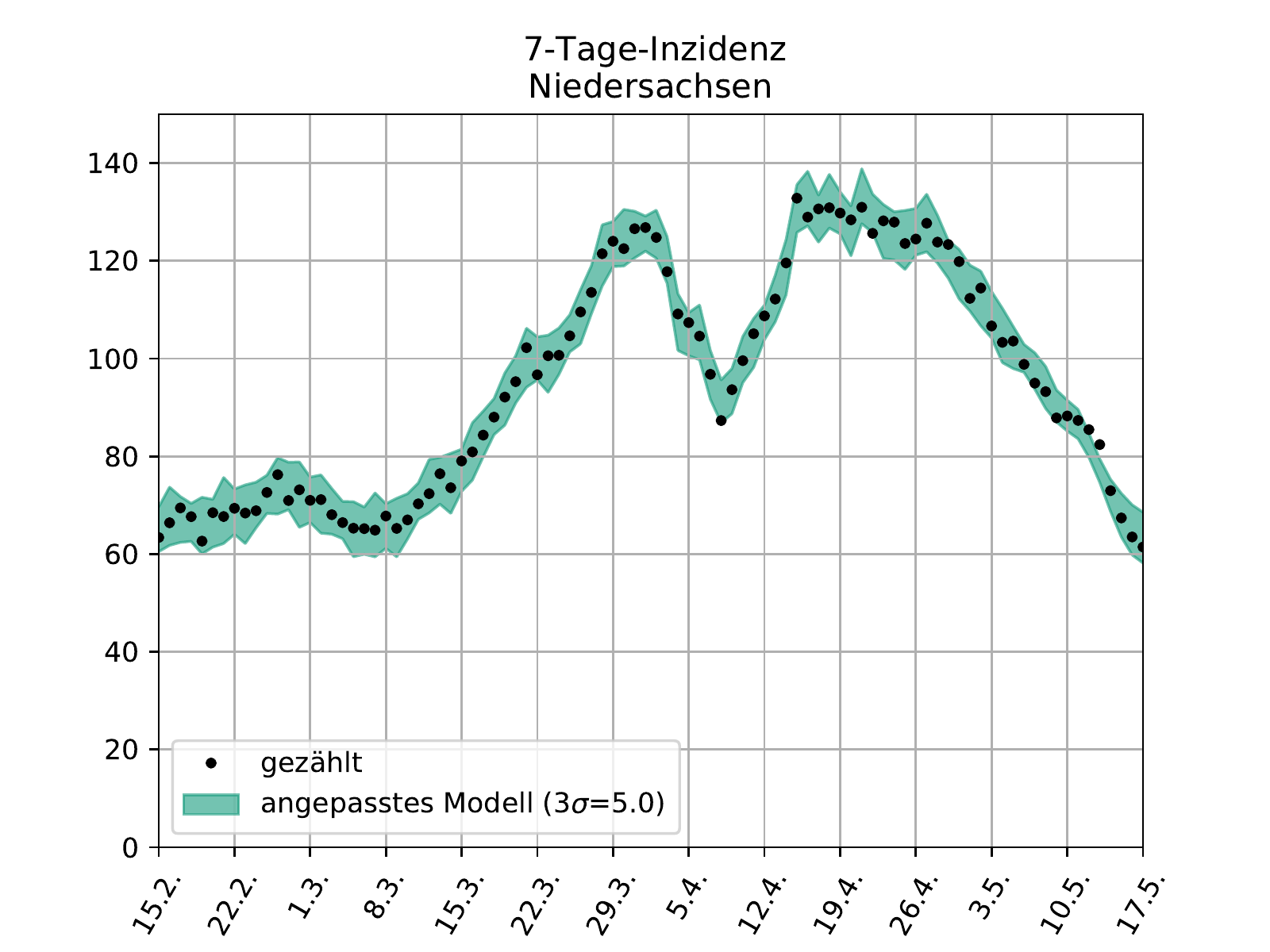}\\
\includegraphics[width=0.49\textwidth]{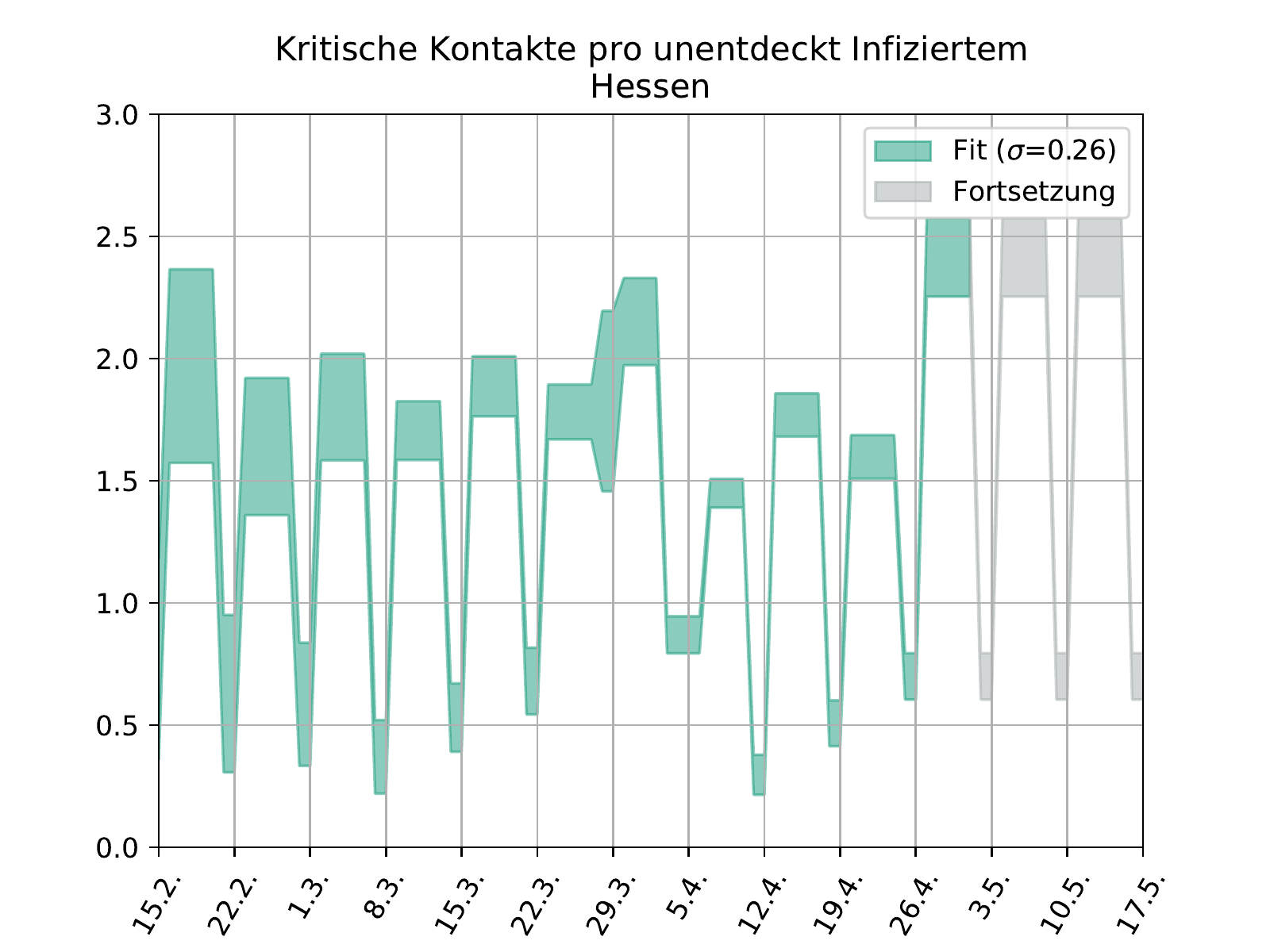}
\hfill
\includegraphics[width=0.49\textwidth]{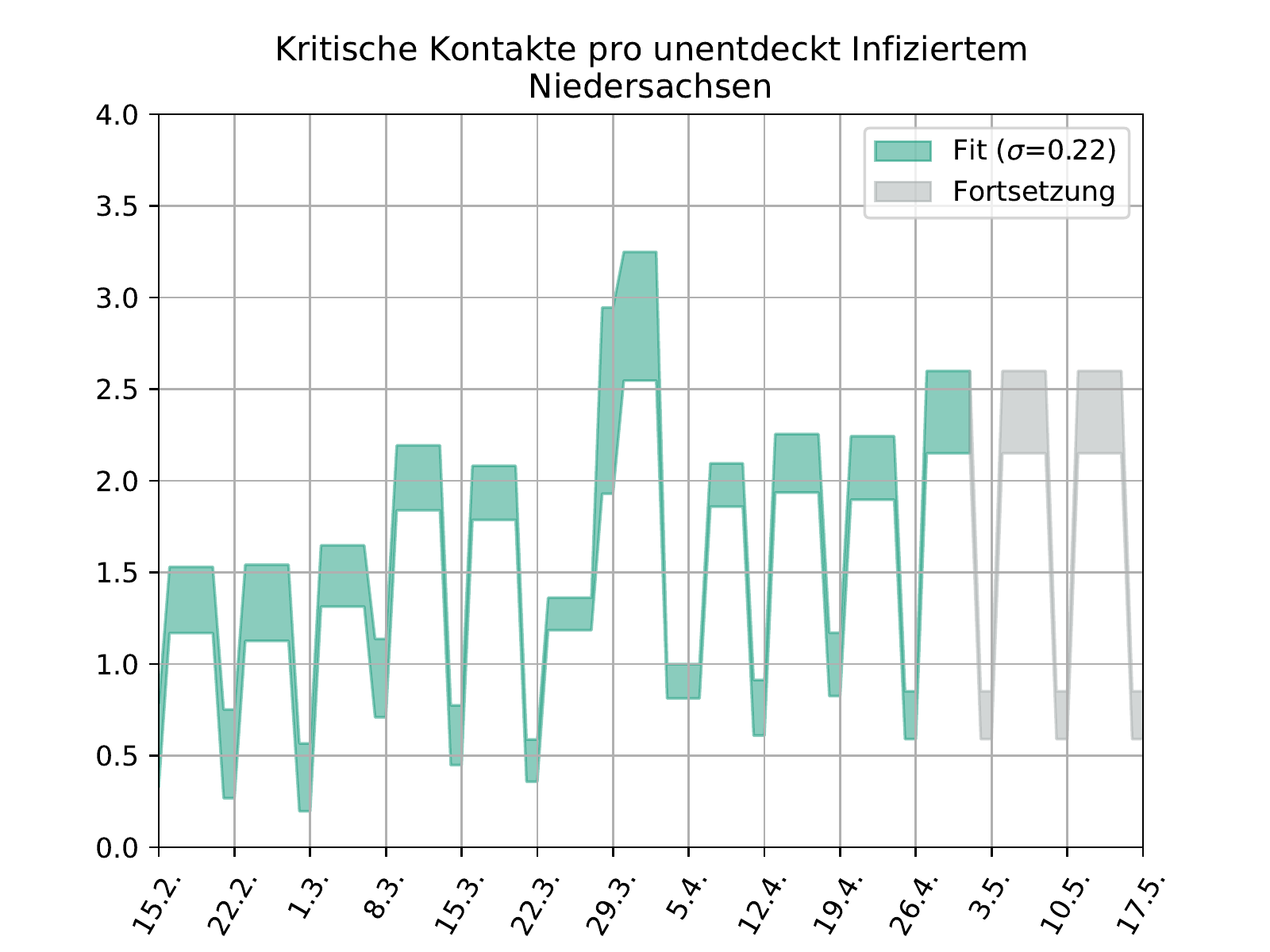}\\
\includegraphics[width=0.49\textwidth]{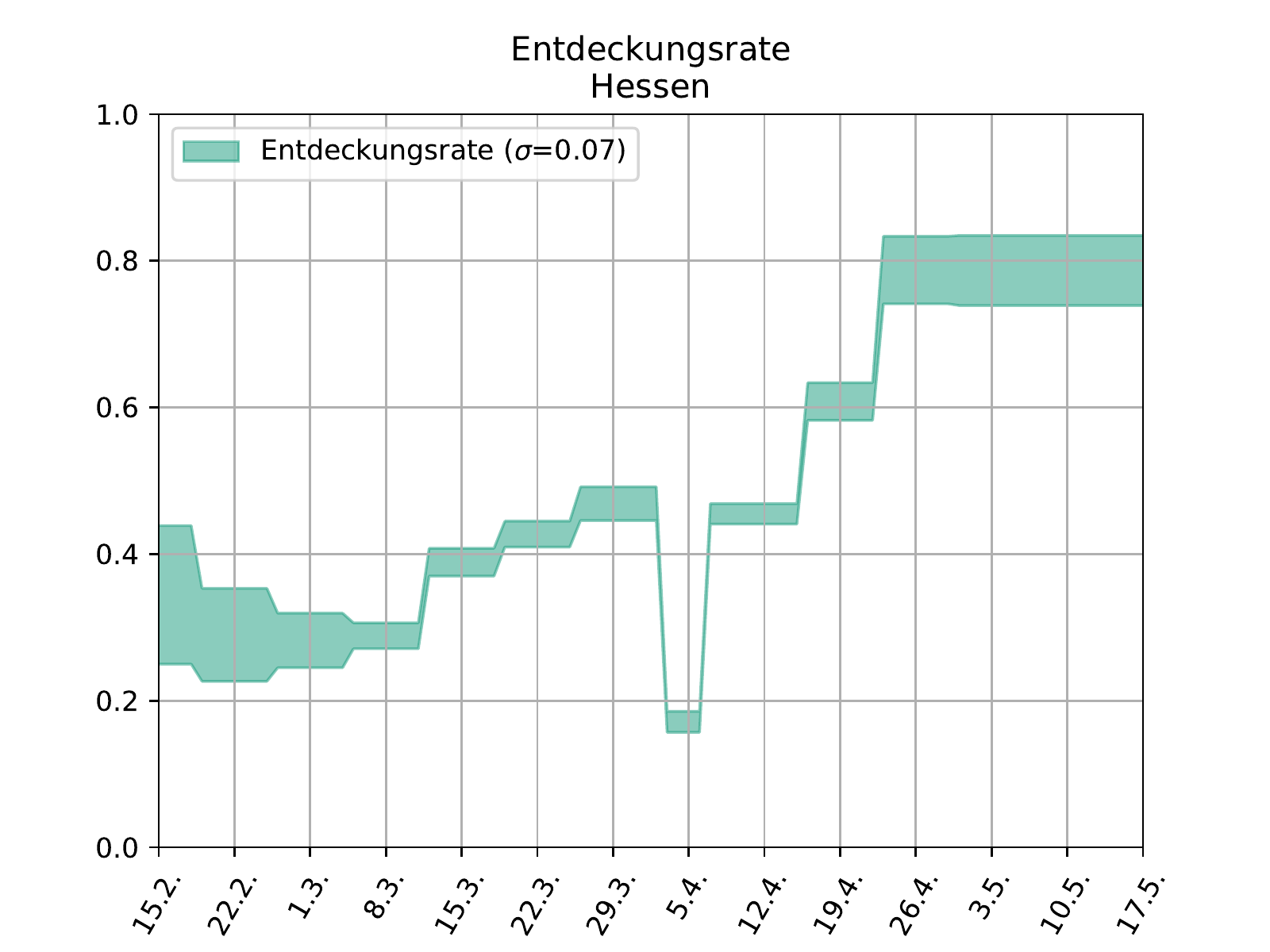}
\hfill
\includegraphics[width=0.49\textwidth]{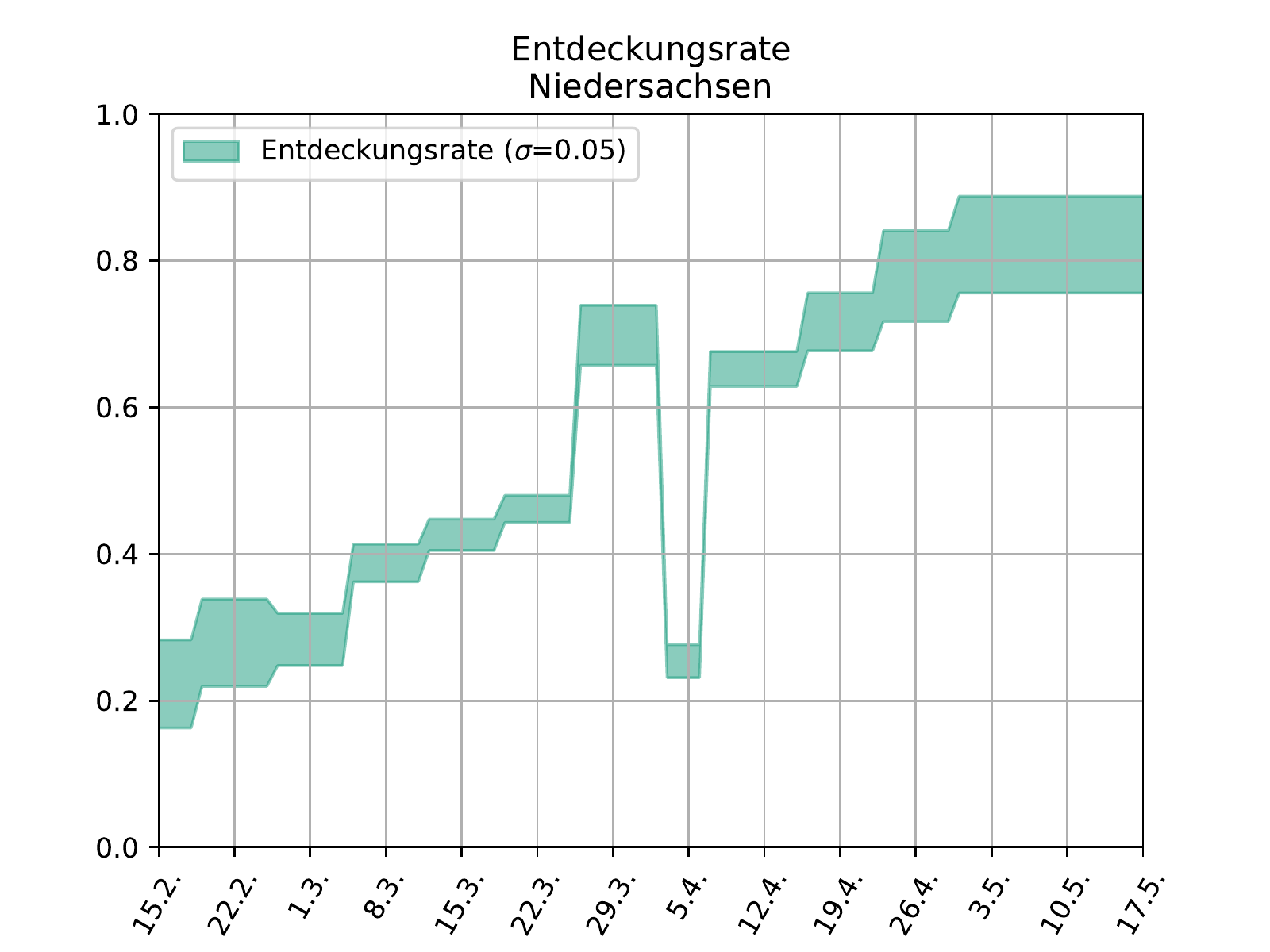}
\caption{7-Tage-Inzidenz, Basisreproduktionsrate (genauer  $\boldsymbol{\kappa(t) \left(\tau^e - \tau^s \right)}$) und Entdeckungsrate für Hessen (links) und Niedersachsen (rechts). Wichtige Daten: 8.3. Recht auf Schnelltest, 22.3. Ankündigung Osterruhe und Probewoche in Niedersachsen, 2.4. Karfreitag, 12.4. erster Schultag in Niedersachsen, 19.4. erster Schultag in Hessen, 24.04. Bundesnotbremse wird wirksam. Die letzten Kontaktraten können nicht aus Meldungen des Simulationszeitraums gefittet werden und sind fortgesetzt. Bemerkenswert: starker Anstieg der Kontakte in der Woche vor Ostern, besonders in Niedersachsen.} 
\label{fig:hessen:niedersachsen}
\end{SCfigure}

\begin{SCfigure}[][p]
\includegraphics[width=0.49\textwidth]{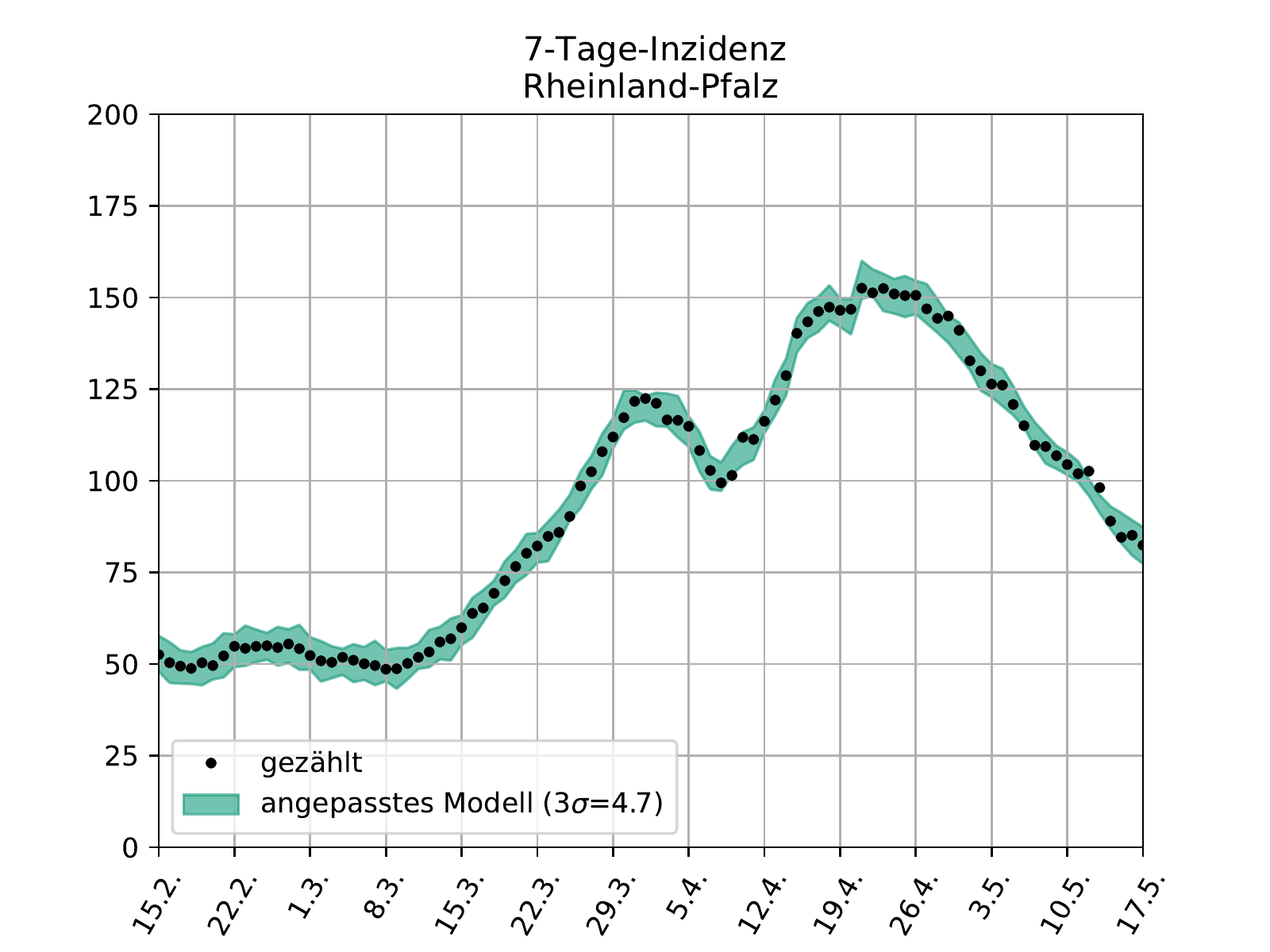}
\hfill
\includegraphics[width=0.49\textwidth]{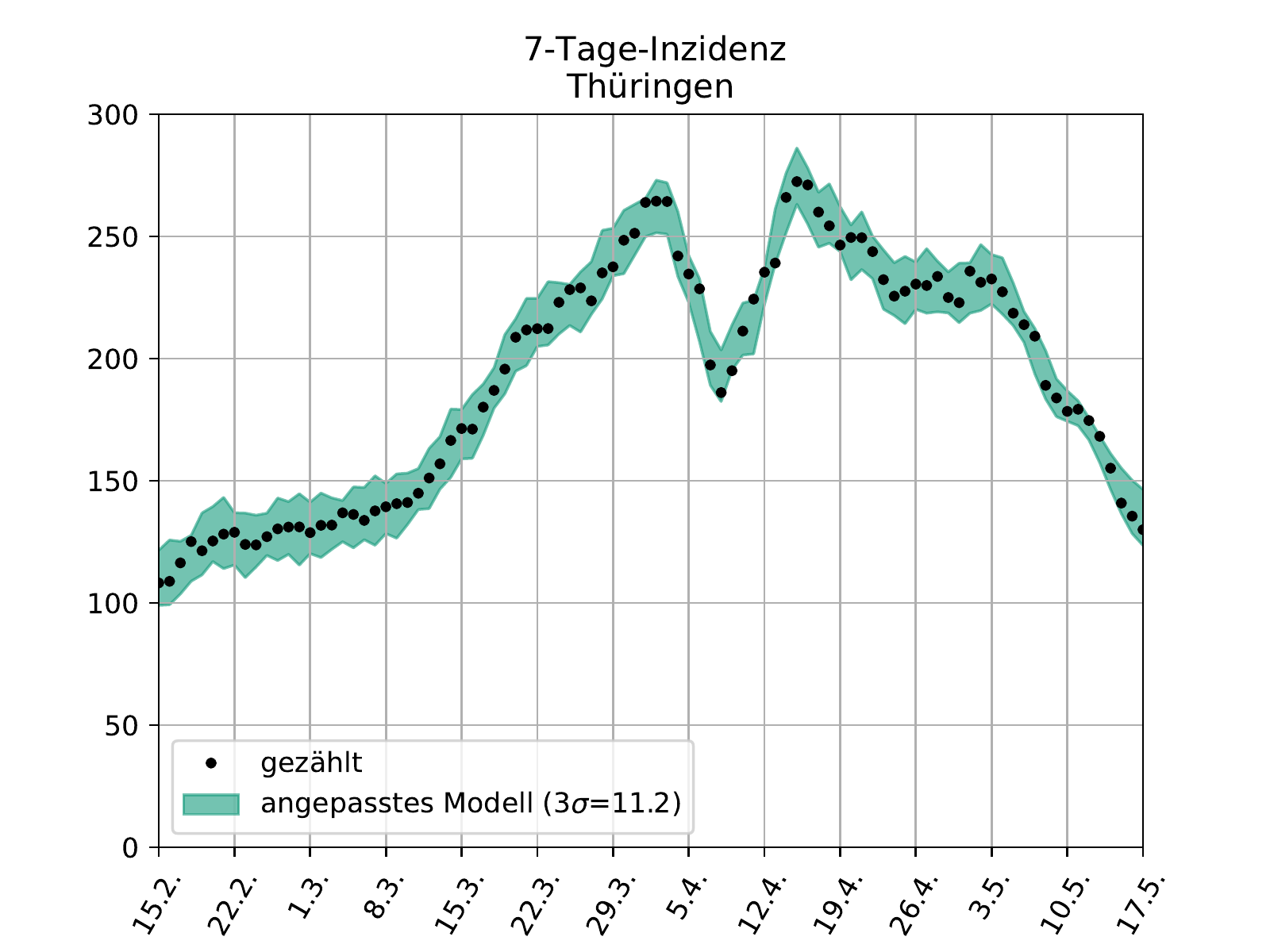}\\
\includegraphics[width=0.49\textwidth]{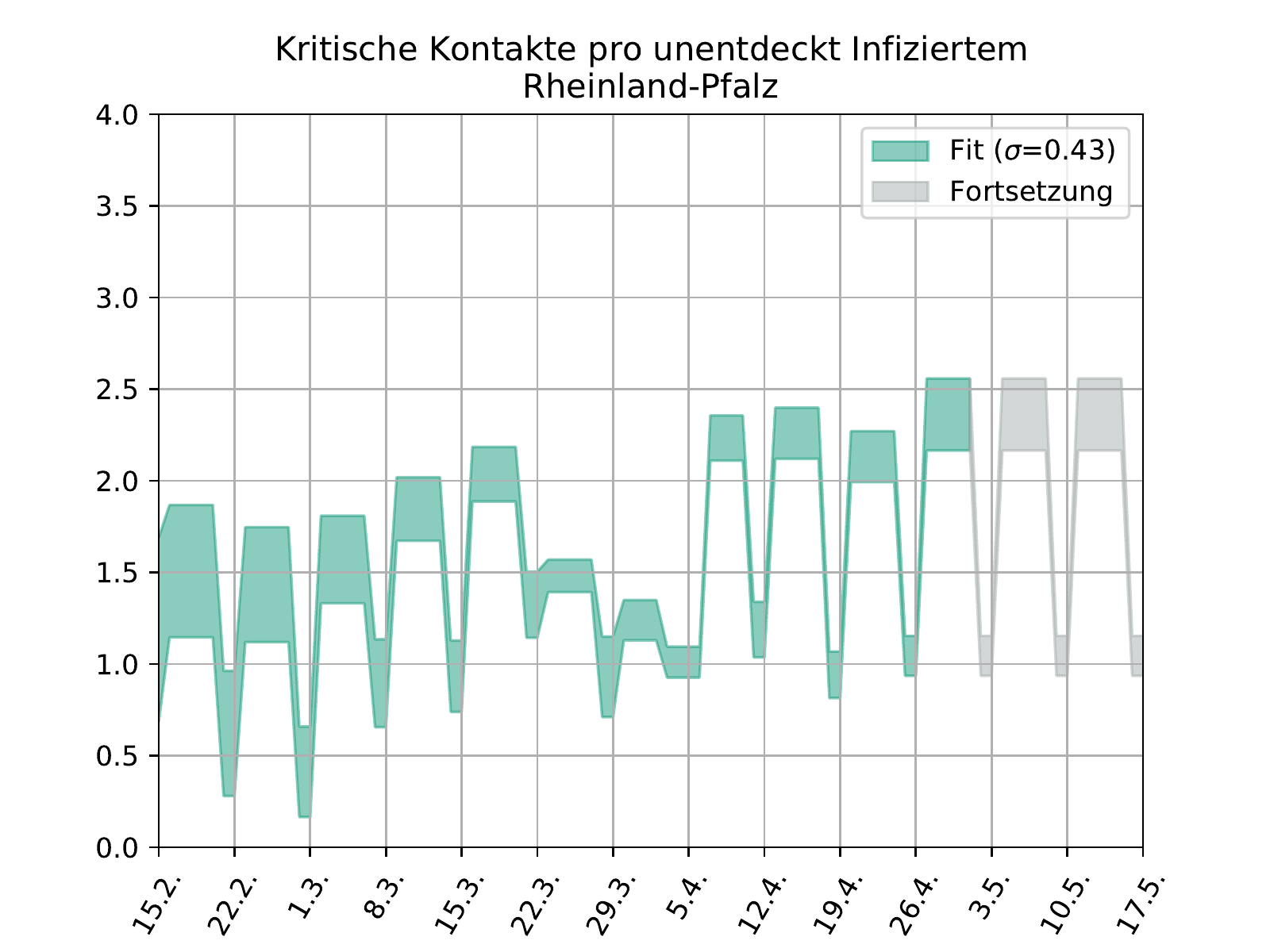}
\hfill
\includegraphics[width=0.49\textwidth]{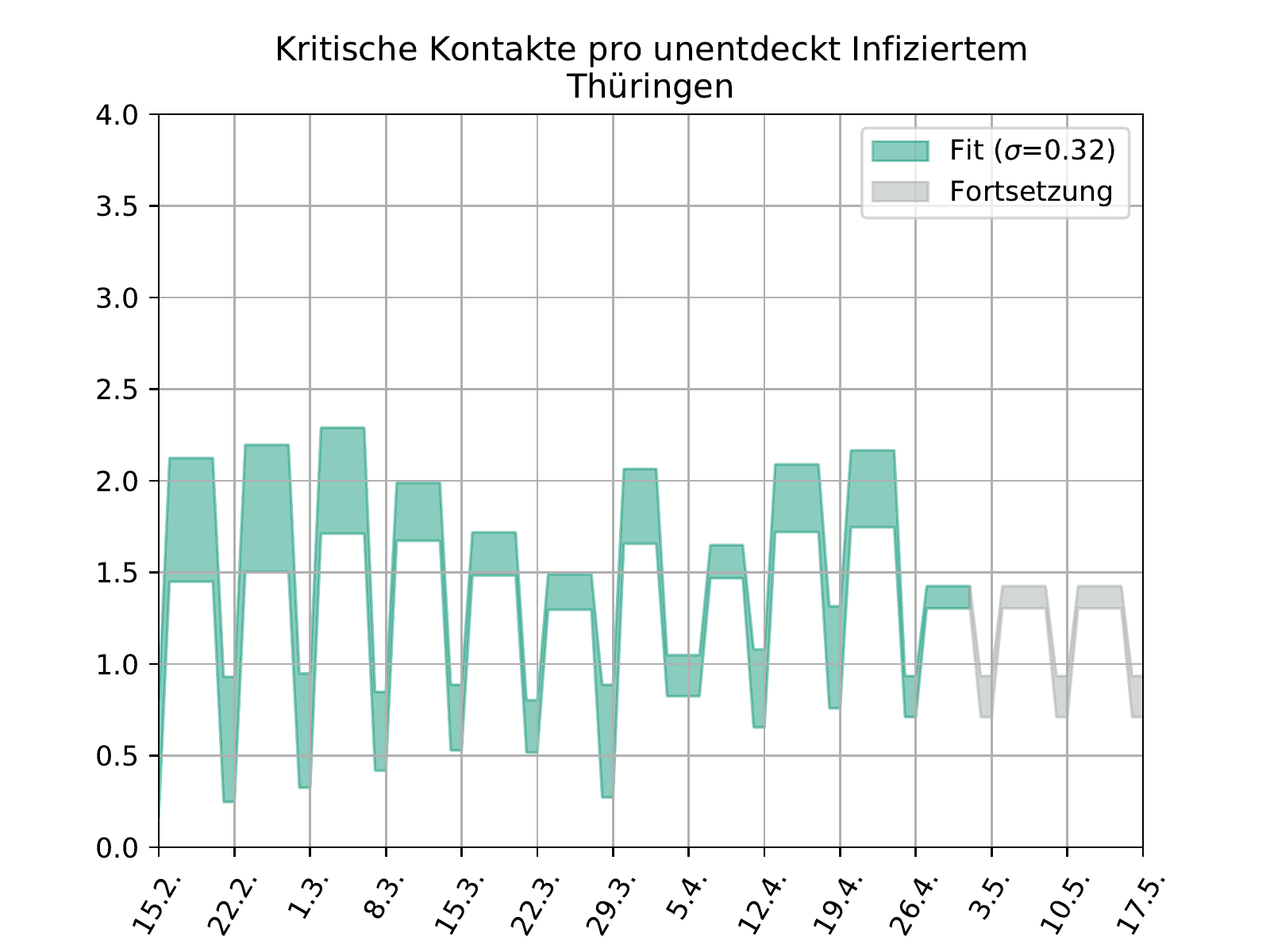}\\
\includegraphics[width=0.49\textwidth]{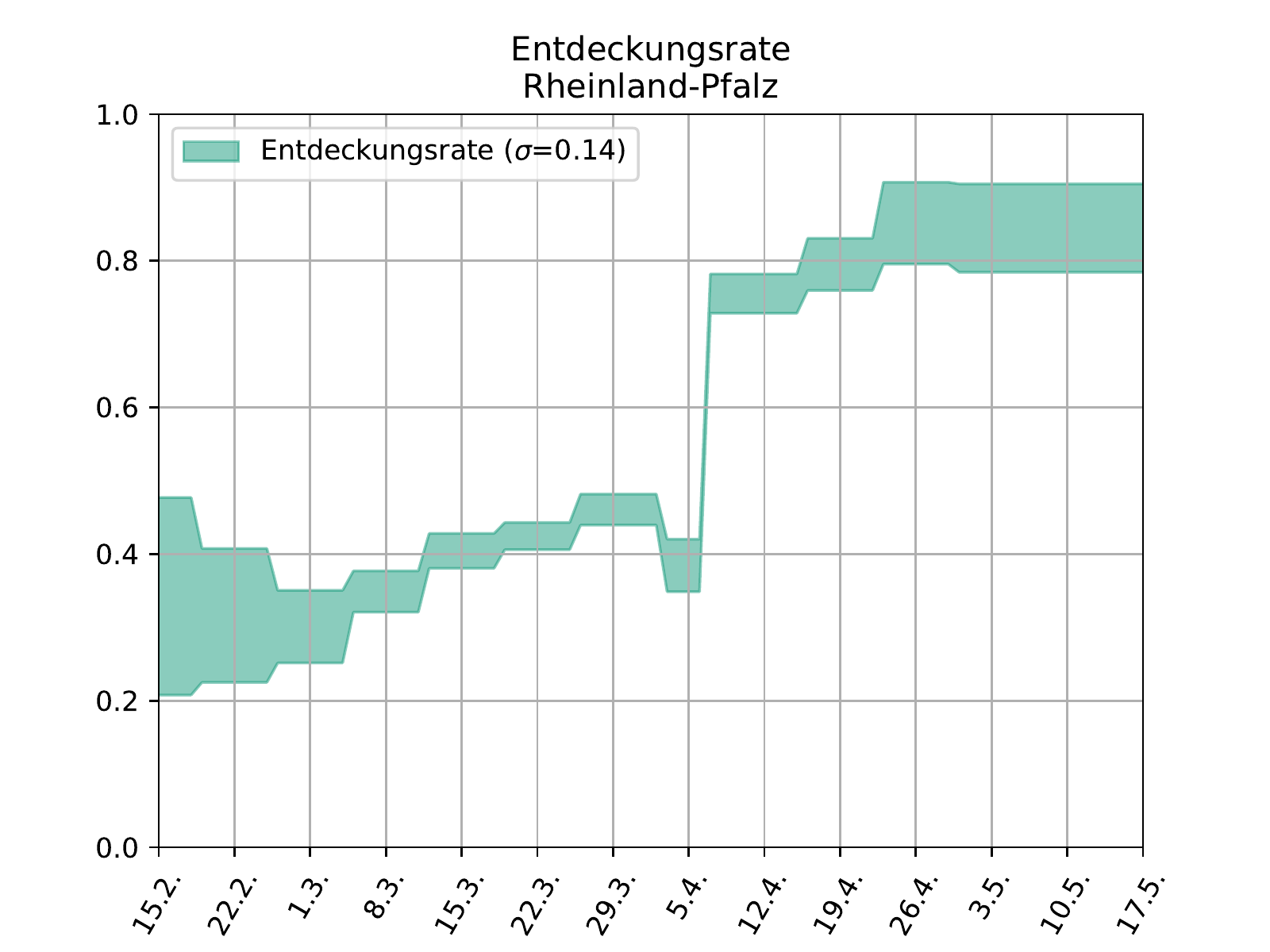}
\hfill
\includegraphics[width=0.49\textwidth]{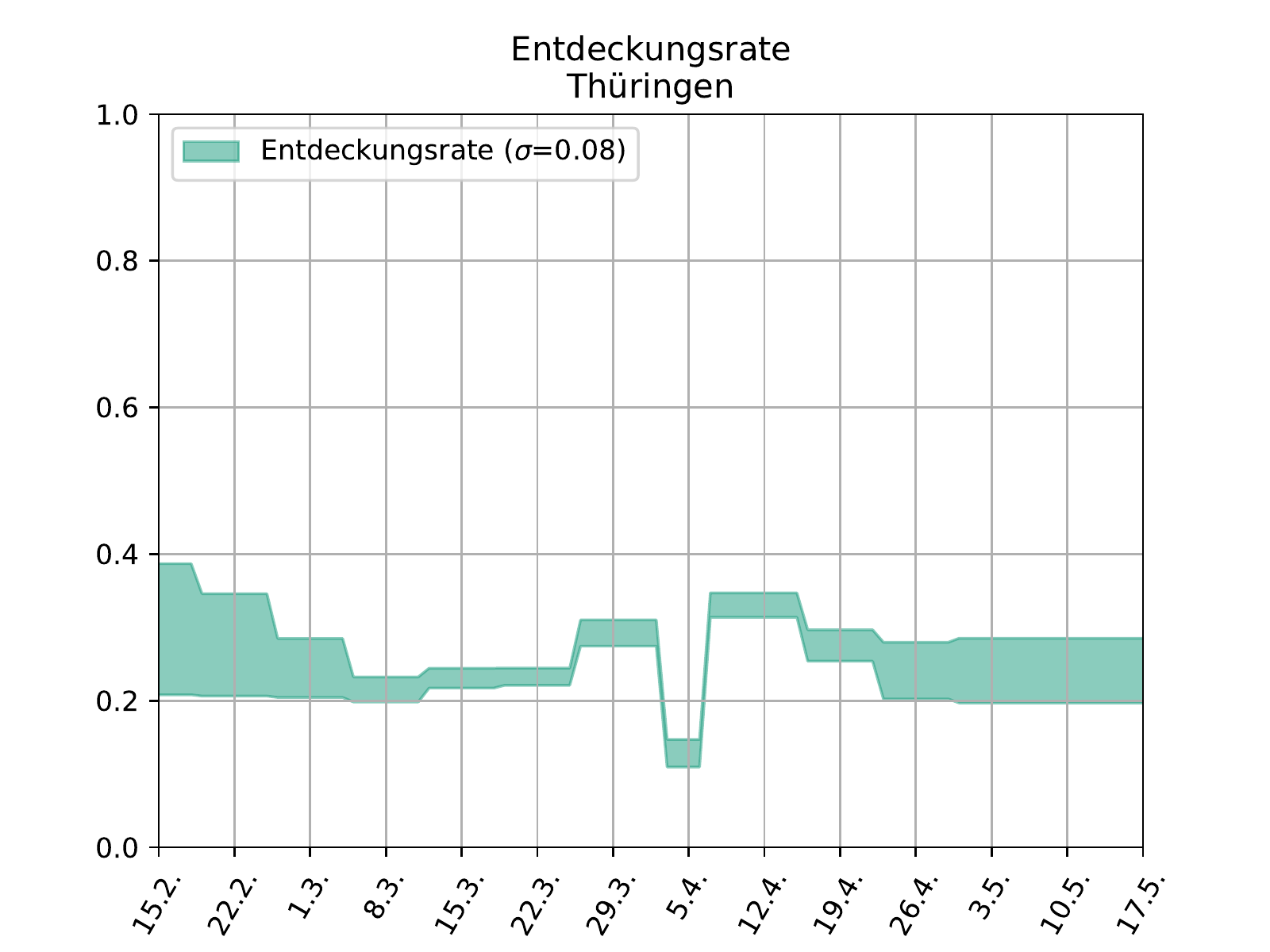}
\caption{7-Tage-Inzidenz, Basisreproduktionsrate (genauer  $\boldsymbol{\kappa(t) \left(\tau^e - \tau^s \right)}$) und Entdeckungsrate für Rheinland-Pfalz (links) und Thüringen (rechts). Wichtige Daten: 8.3. Recht auf Schnelltest, 22.3. Ankündigung Osterruhe, 2.4. Karfreitag, 7.4. erster Schultag in Rheinland-Pfalz, 12.4. erster Schultag in Thüringen, 24.04. Bundesnotbremse wird wirksam. Die letzten Kontaktraten können nicht aus Meldungen des Simulationszeitraums gefittet werden und sind fortgesetzt. Bemerkenswert: Entdeckungsrate steigt in RLP nach Ostern massiv, was mit der Einführung von Schultests und dem Aufbau neuer Testzentren korreliert. In Thüringen erreichen die Kontaktraten ihr erstes Maximum bereits Anfang März, in RLP erst vor dem 22.3., möglicherweise ein Hinweis auf den Weg der britischen Variante. In Thüringen wirkt die Bundesnotbremse und die Entdeckungsrate bleibt niedrig.} 
\label{fig:rheinland-pfalz:thüringen}
\end{SCfigure}

In den Abbildungen~\ref{fig:hessen:niedersachsen} und \ref{fig:rheinland-pfalz:thüringen} sind nun noch einmal für alle vier untersuchten Bundesländer die angepassten Inzidenzen, Kontakt- und Entdeckungsraten dargestellt.

\newpage

\subsection{Beiträge der Maßnahmen}
\label{sec:beiträge}
Wir kommen nun zum zentralen Ergebnis der Arbeit: Welchen Anteil haben die Maßnahmenbündel {\em Testen}, {\em Impfen} und {\em Kontaktbeschränkungen} an der Eindämmung der dritten Ausbreitungswelle in Deutschland? 

\begin{SCfigure}[][p]
\includegraphics[width=0.95\textwidth]{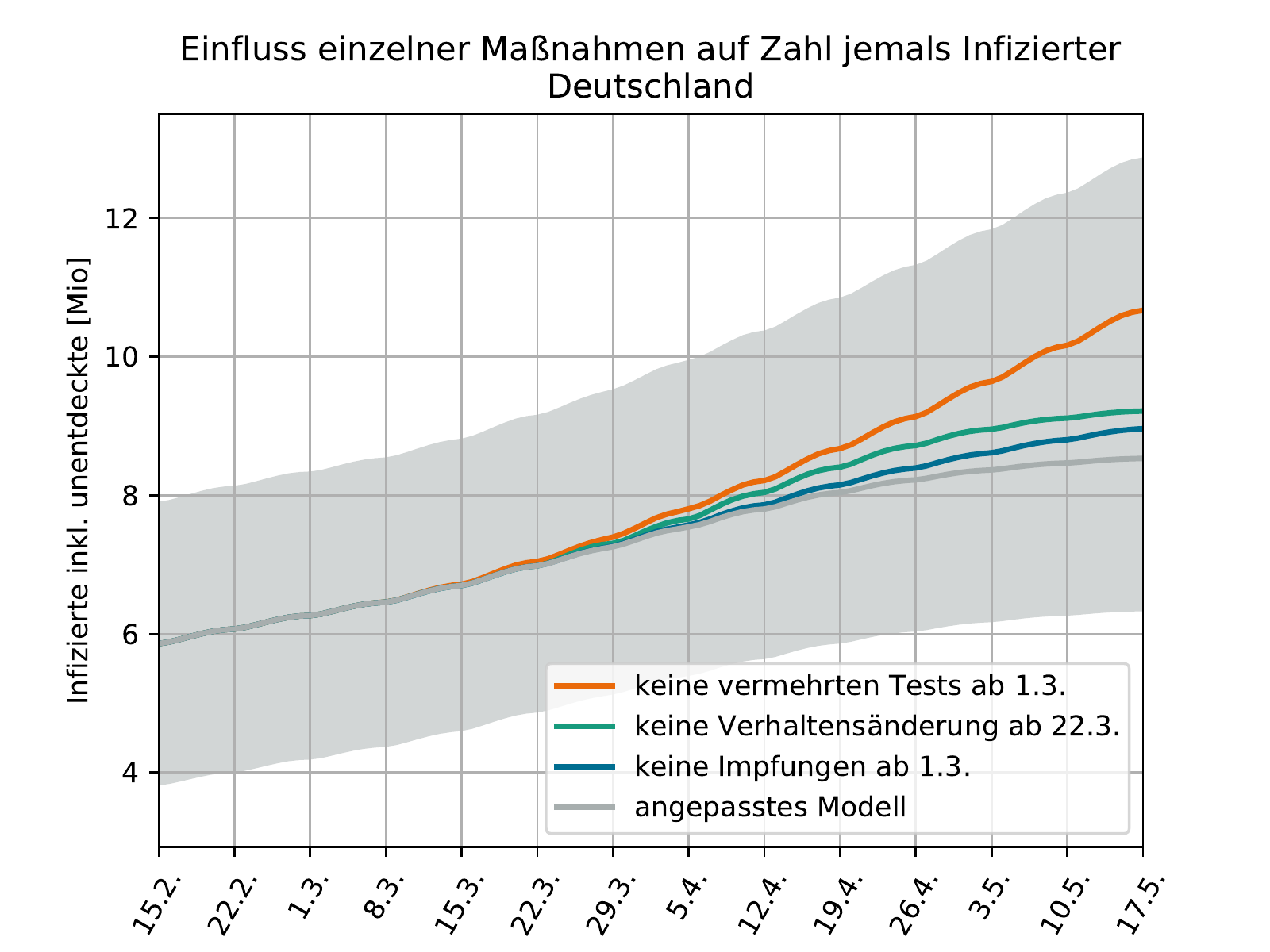}
\caption{So hätte sich die Zahl der Infizierten entwickelt, wenn je eine Maßnahme ausgeblieben wäre. Dargestellt ist die berechnete Zahl aller jemals Infizierten inklusive der Unentdeckten. Der graue Bereich ist der $\boldsymbol{\sigma}$-Fehlerschlauch der angepassten Lösung (68\% Konfidenz), erweitert auf alle Kurven. Der große Schätzfehler resultiert vor allem aus der ungewissen Zahl von Genesenen zu Beginn der Simulation, die alle Kurven gemeinsam nach oben oder unten verschiebt.} 
\label{fig:deutschland:maßnahmen}
\end{SCfigure}

\begin{SCfigure}[][p]
\includegraphics[width=0.95\textwidth]{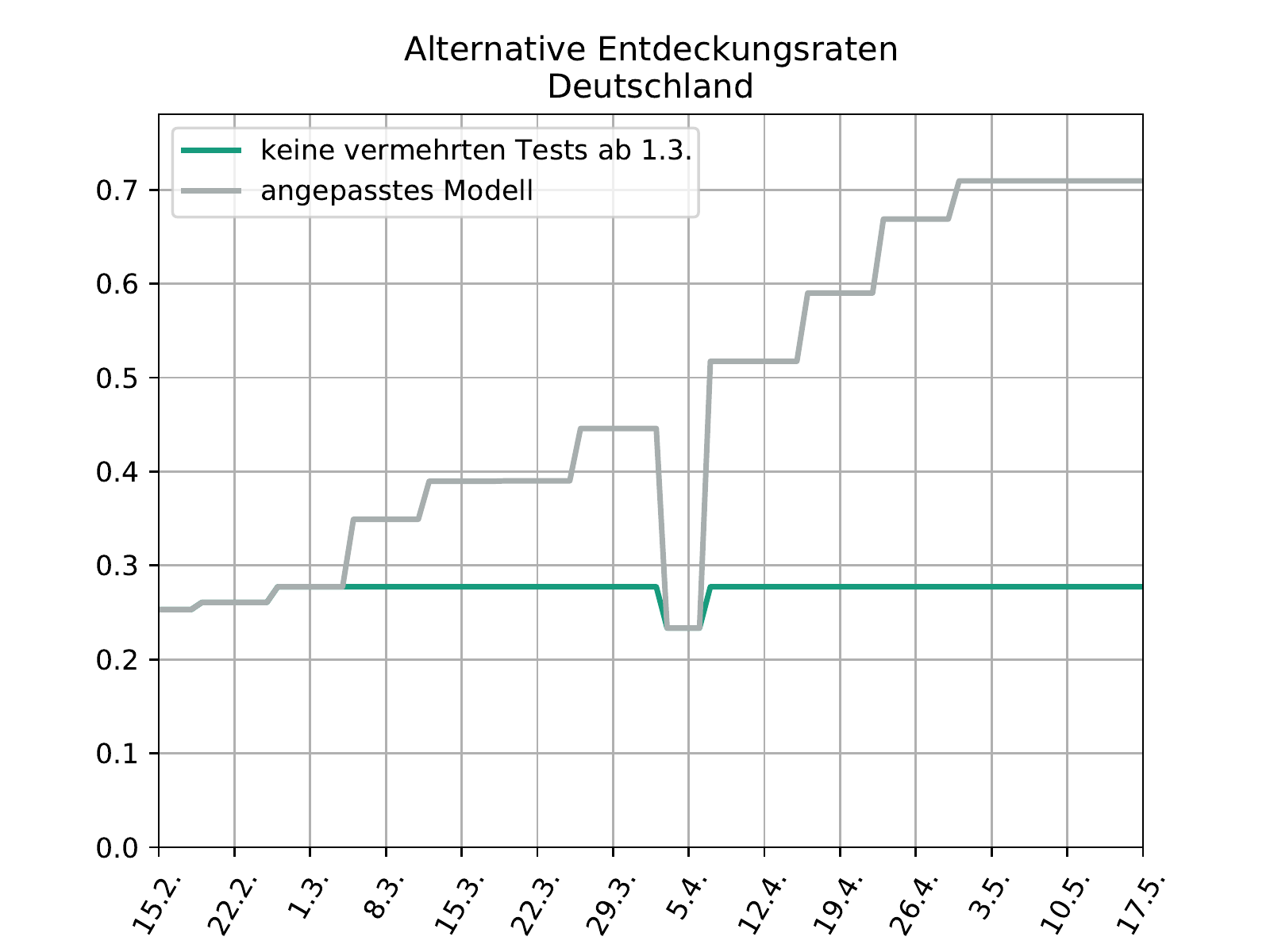}
\caption{Entdeckungsrate ohne Verbesserung gegenüber 01.03. und angepasstes Modell zum Vergleich.} 
\label{fig:deutschland:alternativentdeckungen}
\end{SCfigure}

\begin{SCfigure}[][p]
\includegraphics[width=0.95\textwidth]{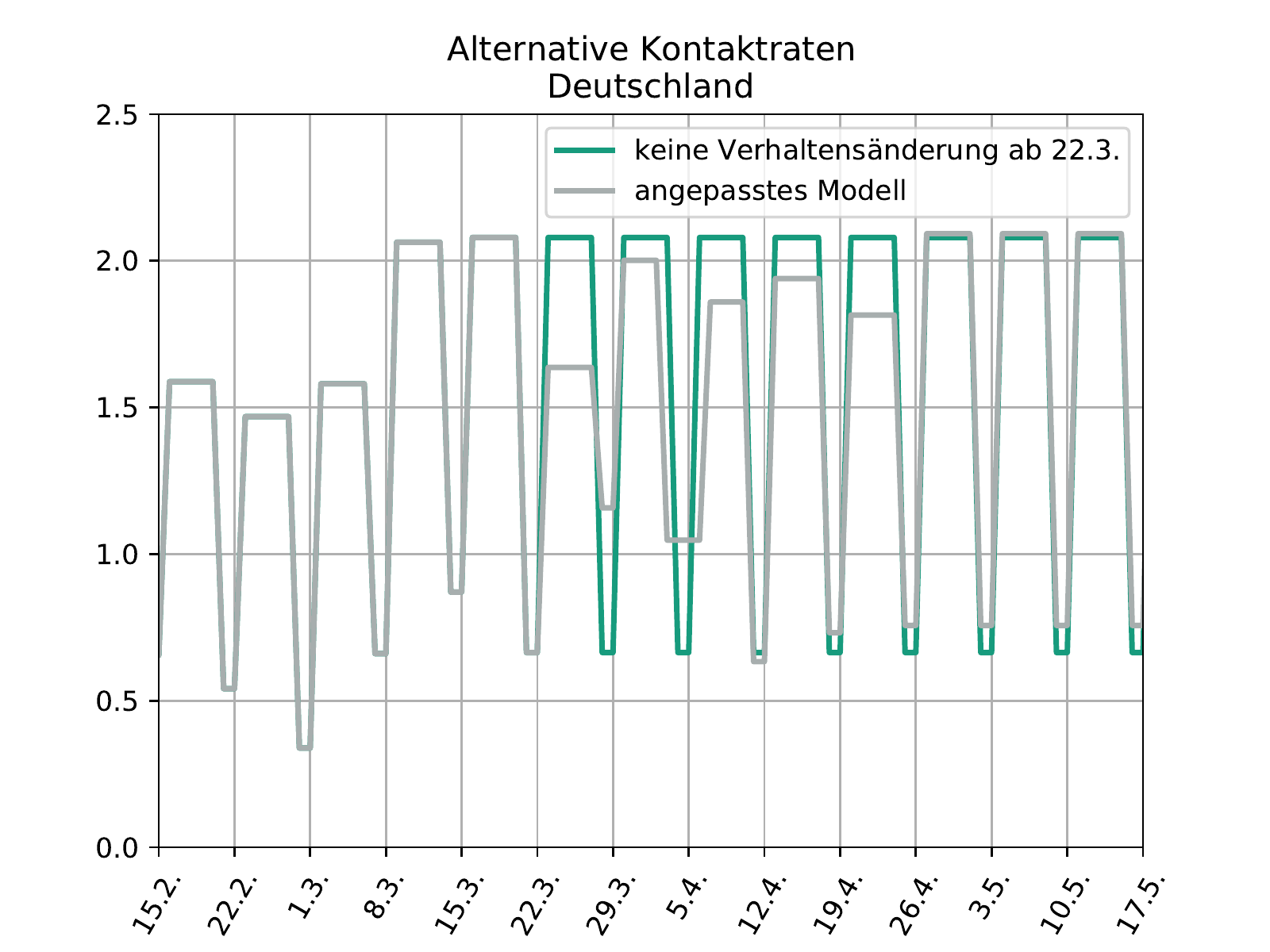}
\caption{Kritische Kontakte pro Infiziertem $\boldsymbol{\left[ \kappa \left( \tau^e - \tau^s \right)\right]}$,  fortgesetzt wie in Woche vor 22.03. und angepasstes Modell zum Vergleich. Bei einem Wert von 2 gibt ein unentdeckter Infizierter in einer nicht immunisierten Umgebung das Virus an 2 Personen weiter.} 
\label{fig:deutschland:alternativkontakte}
\end{SCfigure}

\begin{SCfigure}[][p]
\includegraphics[width=0.95\textwidth]{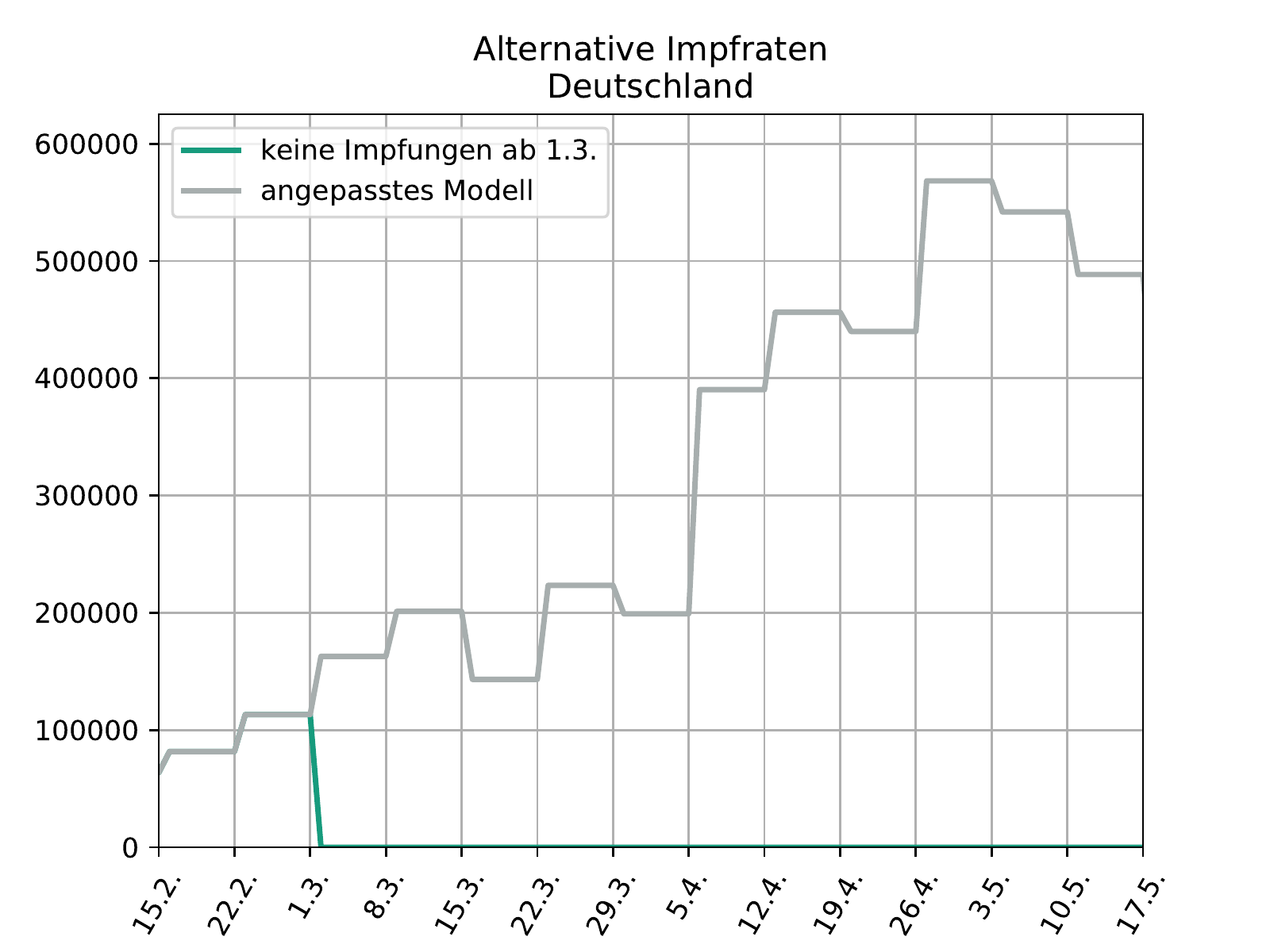}
\caption{Tägliche Erstimpfungen bei Impfstopp ab 01.03. und  \href{https://www.rki.de/DE/Content/InfAZ/N/Neuartiges_Coronavirus/Daten/Impfquotenmonitoring.xlsx?__blob=publicationFile}{reale Wochenmittel} zum Vergleich.}
\label{fig:deutschland:alternativimpfung}
\end{SCfigure}

In Abschnitt~\ref{sec:ergebnis:kontakte} haben wir bereits die Kontakt- und Entdeckungsraten identifiziert, die während der dritten Welle wirksam waren. Daraus ergibt sich aber noch nicht, welcher Effekt dominant ist.  
Um dies zu klären, rechnen wir drei weitere Szenarien durch. In jedem Szenario wird nun eine der drei Raten für Impfen, Kontakte und Tests so eingefroren, wie es dem Wegfall der zugehörigen Maßnahme entspräche. Anschließend wird geprüft, wie stark sich die Fallzahlen gegenüber dem realen Szenario erhöhen. Die Maßnahme, deren Wegfall zur größten Erhöhung führt, gilt uns dann als die mit der größten Wirkung. 
Das Ergebnis ist in Abbildung~\ref{fig:deutschland:maßnahmen} dargestellt.
Die Abbildungen~\ref{fig:deutschland:alternativentdeckungen}--\ref{fig:deutschland:alternativimpfung} beschreiben die gewählten Alternativszenarien im Detail.

Um die drei Maßnahmen quantitativ vergleichen zu können, führen wir folgende Maßzahl für den Wegfall einer Maßnahme ein:
\begin{align}
    \mu_{m} &= 100 \left[ \frac{N_m(17.05.2021)-N_m(01.03.2021)}
    {N(17.05.2021)-N(01.03.2021)} - 1 \right] \; ,
    \label{eq:maßzahl}
\end{align}
wobei $N(t)$ die Zahl aller jemals bis zur Zeit $t$ infizierten Personen im angepassten Modell darstellt und $N_m(t)$ die entsprechende Zahl im Alternativszenario mit weggelassener Maßnahme $m$. Die Maßzahl gibt also an, um welchen Prozentsatz sich der Anstieg der Fälle im betrachteten Zeitraum erhöht hätte, wenn eine Maßnahme ausgeblieben wäre.

Um auszuschließen, dass der Befund bei geringfügig anderen Parametern ganz anders aussieht, variieren wir die unsicheren Impfparameter innerhalb plausibler Grenzen. Eine Wirksamkeit von $\varepsilon = 100\%$ und ein früher Wirkbeginn bereits nach $\tau^p = 10$ Tagen schätzen den Impfeffekt nach oben ab. $\varepsilon = 60\%$ und $\tau^p = 14$ Tage liefern eine untere Grenze.

\begin{SCtable}[0.95\textwidth][b]
	\centering
   \begin{tabular}{lrrrr}
		\hline \noalign{\smallskip}
		Fall & Abb. & \parbox{0.15\textwidth}{Kein Impfen ab 1.3.}  
		& \parbox{0.17\textwidth}{Verhalten wie vor 22.3.}  
		& \parbox{0.11\textwidth}{Tests wie vor 1.3.} \\
		\noalign{\smallskip} \hline \noalign{\smallskip}
        Deutschland                   & \ref{fig:deutschland:maßnahmen} & 19 & 30 & 94 \\
        $\varepsilon=100$,$\tau^p=10$  & \ref{fig:deutschland:maßnahmen:10:100} & 28 & 27 & 92 \\
         $\varepsilon=60$,$\tau^p=14$ & \ref{fig:deutschland:maßnahmen:14:60} & 11 & 37 & 99 \\
         Wechsel mittwochs            & \ref{fig:deutschland:maßnahmen:mittwochs} & 20 & 18 & 95 \\
         Hessen                       & \ref{fig:hessen:maßnahmen} & 18 & 6 & 95 \\
         Niedersachsen                & \ref{fig:niedersachsen:maßnahmen} & 18 & -16 & 223 \\
         Rheinland-Pfalz              & \ref{fig:rheinland-pfalz:maßnahmen} & 20 & 69 & 222 \\
         Thüringen                    & \ref{fig:thüringen:maßnahmen} & 21 & 97 & 7 \\
		\noalign{\smallskip}
		\hline
	\end{tabular}
	\caption{Maßzahl $\mu$ für den Wegfall einer einzelnen Maßnahme.}
	\label{tab:maßzahl}
\end{SCtable}

\begin{SCfigure}[][!ht]
\includegraphics[width=0.7\textwidth]{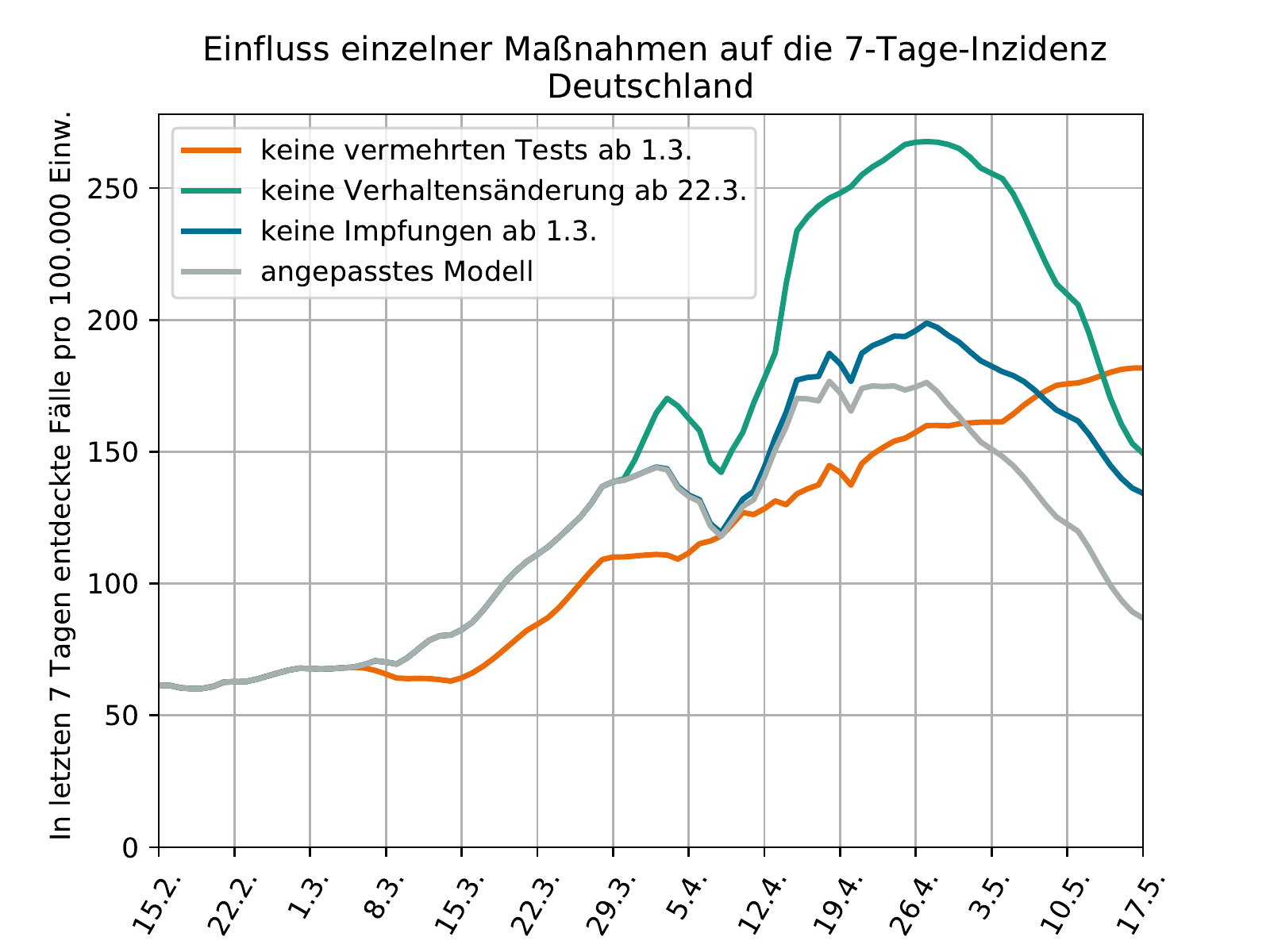}
\includegraphics[width=0.7\textwidth]{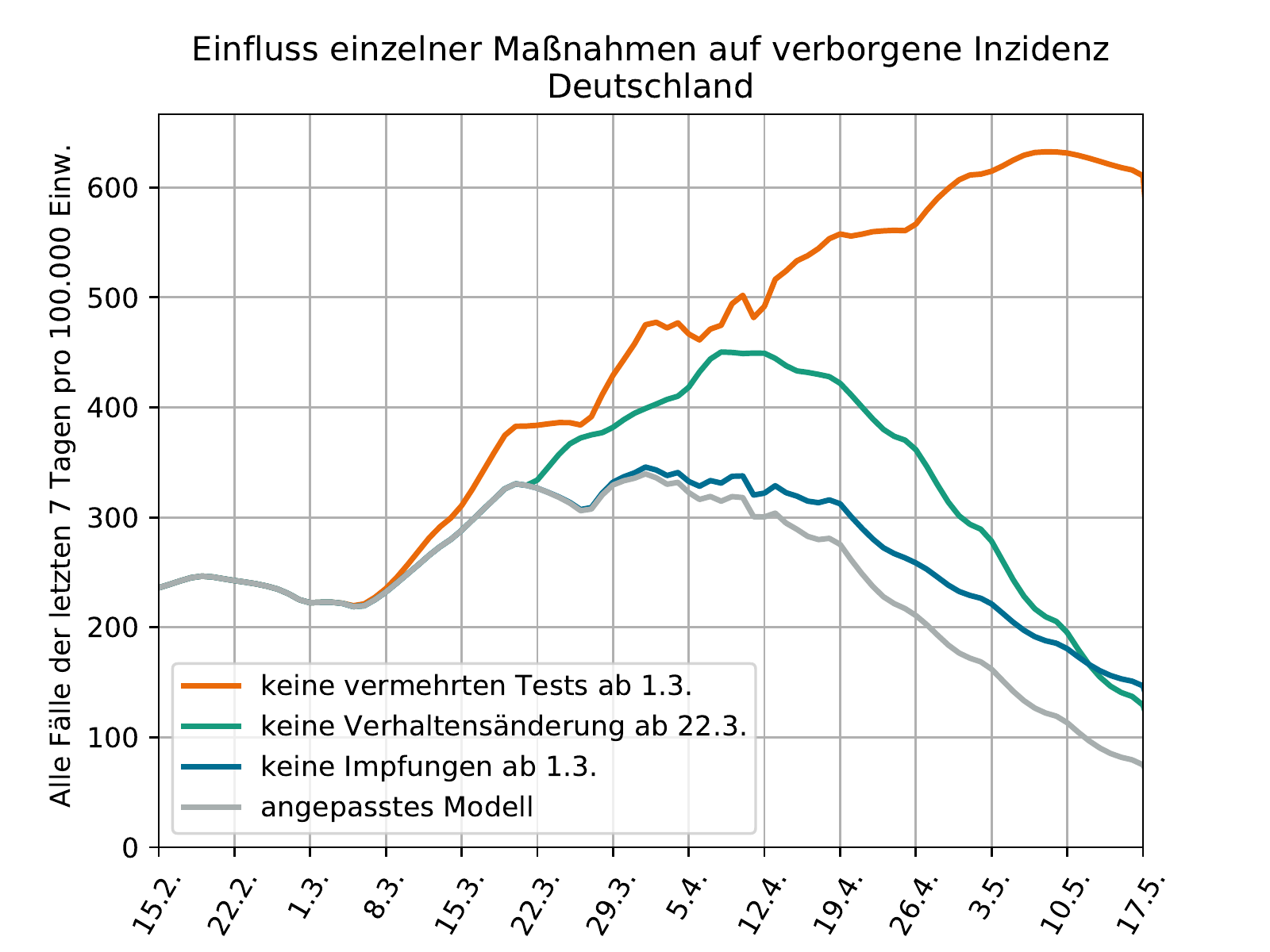}
\caption{In der Realität ist nur die fallbezogene Inzidenz zugänglich (oben). In der Simulation kann man auch die wahre Inzidenz bestimmen, die unentdeckte Infizierte berücksichtigt (unten). Während bezogen auf die fallbezogene Inzidenz das Weglassen zusätzlicher Schnelltests bis Anfang Mai als am wenigsten kritisch gewirkt hätte, offenbart die wahre Inzidenz, wie verheerend weggelassene Schnelltests tatsächlich gewesen wären (Test-Inzidenz-Dilemma). }
\label{fig:deutschland:maßnahmen:inzidenz}
\end{SCfigure}

\begin{SCfigure}[][!ht]
\includegraphics[width=0.7\textwidth]{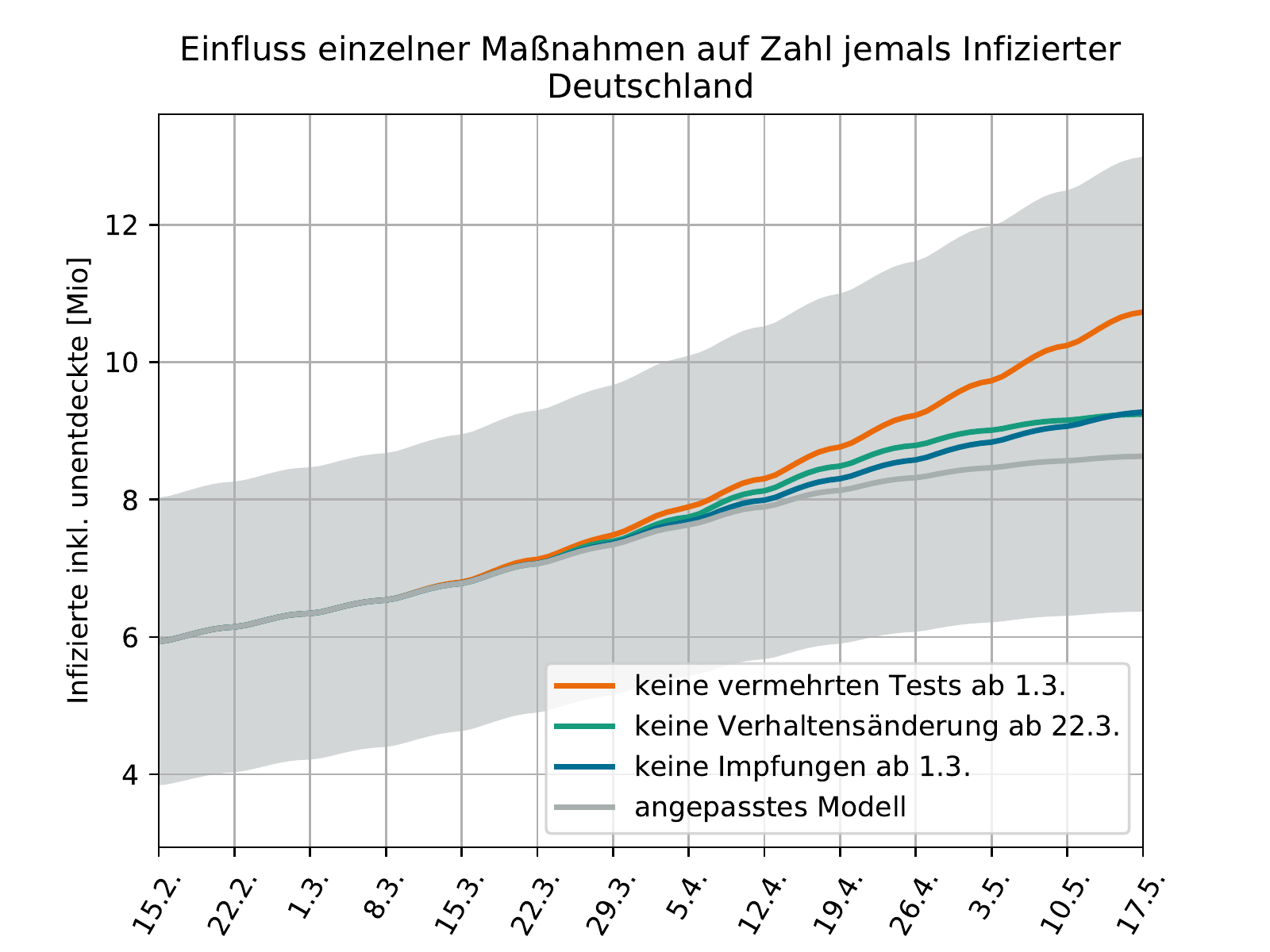}
\caption{Maßnahmenvergleich bei starker Impfwirkung ($\boldsymbol{\varepsilon}$=100\%, $\boldsymbol{\tau^p}$ = 10 d).}
\label{fig:deutschland:maßnahmen:10:100}
\end{SCfigure}

\begin{SCfigure}[][p]
\includegraphics[width=0.7\textwidth]{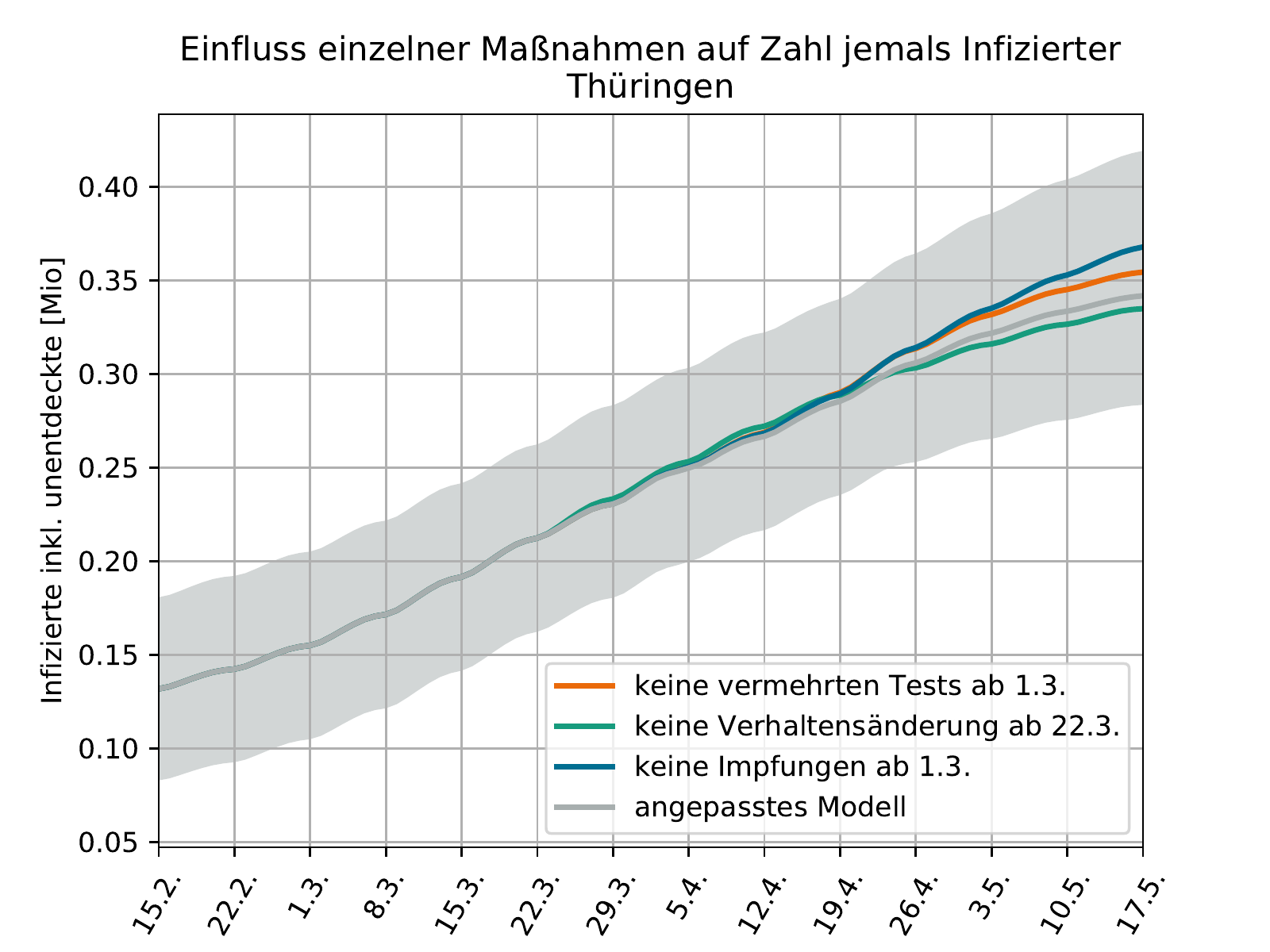}
\caption{Maßnahmenvergleich für Thüringen.}
\label{fig:thüringen:maßnahmen}
\end{SCfigure}

Ferner ändern wir eine numerische Stellgröße, die eher willkürlich gewählt ist, nämlich den Wochentag, an dem die stückweise konstant modellierten Entdeck\-ungsraten springen, von Donnerstag auf Mittwoch. Schließlich führen wir die gleichen Rechnungen auch noch für die bereits in Abschnitt~\ref{sec:schlüsselrolle} untersuchten Bundesländer durch. 
Für die verschiedenen Fälle sind die Maßzahlen $\mu$ in Tabelle~\ref{tab:maßzahl} aufgelistet. Wie sich die Fallzahlen in diesen Szenarien entwickelt hätten, zeigen die Abbildungen~\ref{fig:deutschland:maßnahmen}, \ref{fig:deutschland:maßnahmen:inzidenz} und \ref{fig:deutschland:maßnahmen:10:100} bis \ref{fig:thüringen:maßnahmen} am Ende des Dokuments.

In allen Fällen -- außer für Thüringen -- ergibt sich der gleiche Befund:
\textbf{
\begin{itemize}
    \item Kein Effekt ist gegenüber den anderen vernachlässigbar.
    \item Das Maßnahmenpaket {\em Testen-Nachverfolgen-Isolieren} hat gegenüber {\em Kontaktbeschränkung} und {\em Impfen} den stärksten Einfluss.
    \item Vom 01.03. bis 17.05.2021 wäre der potentielle Anstieg der Fallzahlen durch Weglassen von Tests mindestens 2,5 mal so groß gewesen wie bei Wegfall von Impfungen oder schärferen Kontaktbeschränkungen.
\end{itemize}}
Nur in Thüringen hätte der Wegfall von Tests einen untergeordneten Effekt gehabt. Dies ist aber nicht weiter verwunderlich, da Tests dort ohnehin nie das Dunkelfeld wesentlich erhellt haben, s. Abbildung~\ref{fig:schultests}.

Die sehr niedrigen oder gar negativen Kennwerte von Hessen und Niedersachsen für den Fall, dass es keine Verhaltensänderung nach dem 22.03. gegeben hätte, erklären sich aus Abbildung~\ref{fig:hessen:niedersachsen}. Dort treten die höchsten Kontaktraten erst in der Woche vor Ostern auf. Da im Vergleichsszenario die niedrigeren Kontaktraten aus der Woche vor dem 22.03. fortgeschrieben werden, führt ungeändertes Kontaktverhalten hier sogar zu einer Verbesserung der Lage. Es könnte allerdings sein, dass es sich nicht nur um ein fehlkonstruiertes Szenario handelt. Vielleicht ist das Verhalten, auf baldige Einschränkungen noch einmal konzentriert mit erhöhter Aktivität zu reagieren, tatsächlich ein bedenkenswerter negativer Effekt von Kontaktbeschränkungen.      


\newpage 
\section{Fazit}
Durch Anpassung der Kontakt- und Entdeckungsraten eines epidemiologischen Modells an die Meldedaten des RKI haben wir zeigen können, dass Schnelltests gegenüber Kontaktbeschränkungen und Impfen den größten Beitrag beim Brechen der dritten Corona-Welle erbracht haben. Der  Vergleich von Bundesländern mit unterschiedlich endenden Osterferien legt nahe, dass dieser Effekt zu einem Gutteil auf das Konto von Schnelltests an Schulen geht.   

\section{Danksagung}
\label{sec:dank}
Diese Arbeit entstand im Rahmen der Projekte EpideMSE und SEEvacs, die durch das Aktionsprogramm {\em Franhofer vs. Corona} bzw. das {\em Ministerium für Arbeit, Soziales, Transformation und Digitalisierung des Landes Rheinland-Pfalz} gefördert wurden. Wir danken Anita Schöbel und Raimund Wegener für Durchsicht und Anregungen. 

\newpage
\bibliography{literature}
\bibliographystyle{unsrt}

\begin{SCfigure}[][p]
\includegraphics[width=0.7\textwidth]{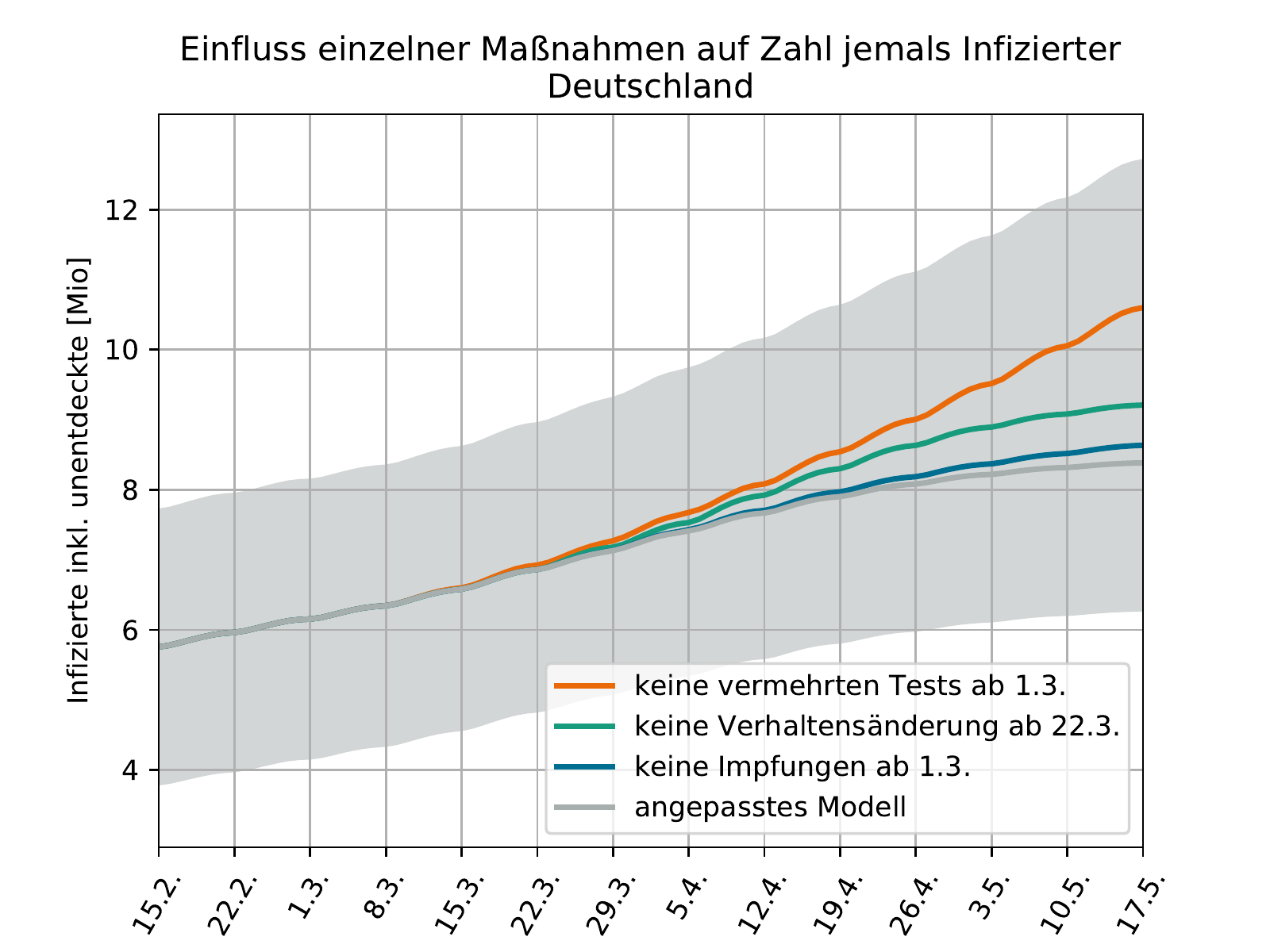}
\caption{Maßnahmenvergleich bei schwacher Impfwirkung ($\boldsymbol{\varepsilon}$=60\%, $\boldsymbol{\tau^p}$ = 14 d).}
\label{fig:deutschland:maßnahmen:14:60}
\end{SCfigure}
\begin{SCfigure}[][p]
\includegraphics[width=0.7\textwidth]{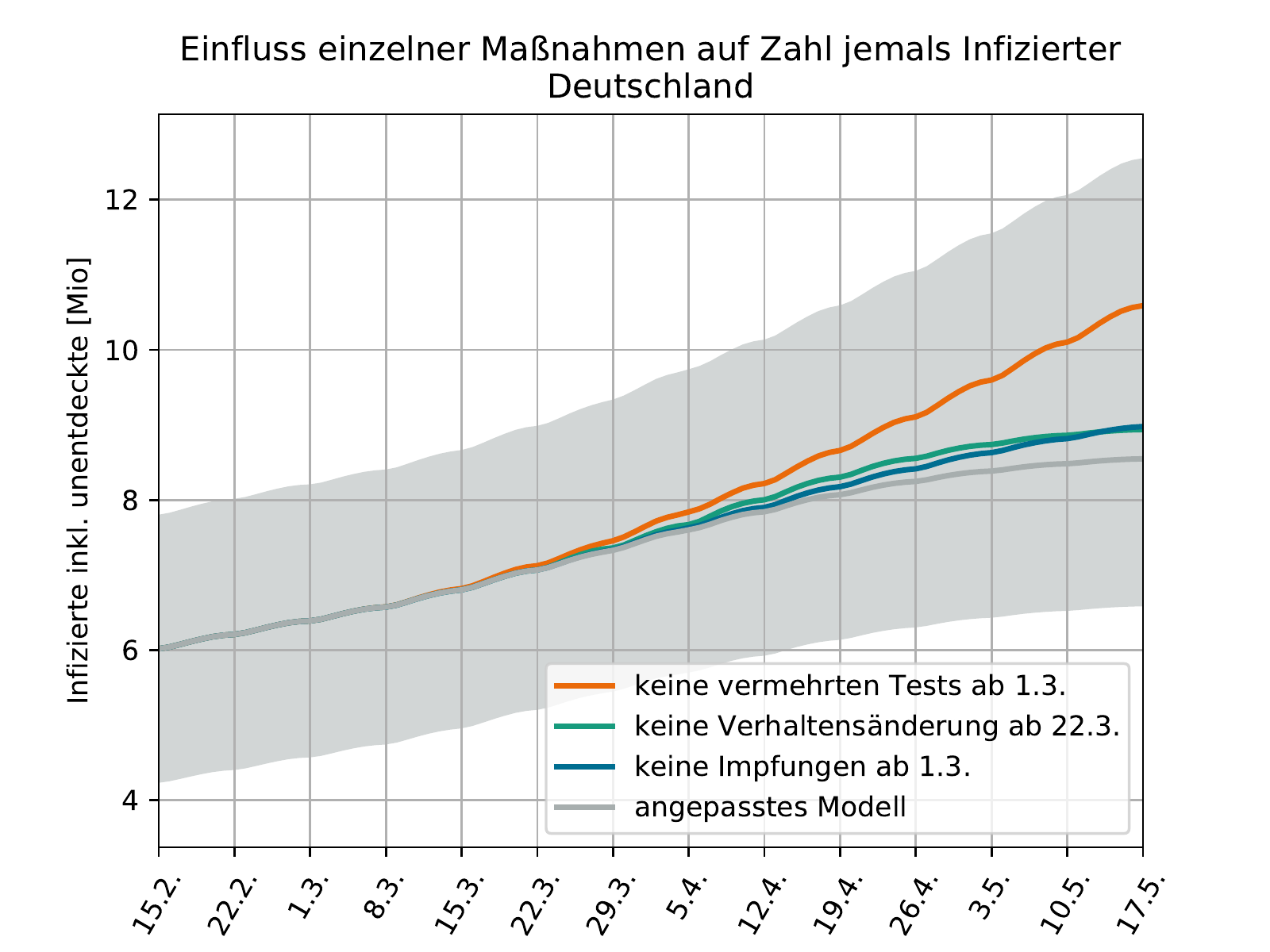}
\caption{Maßnahmenvergleich bei mittwochs springender Entdeckungsrate.}
\label{fig:deutschland:maßnahmen:mittwochs}
\end{SCfigure}
\begin{SCfigure}[][p]
\includegraphics[width=0.7\textwidth]{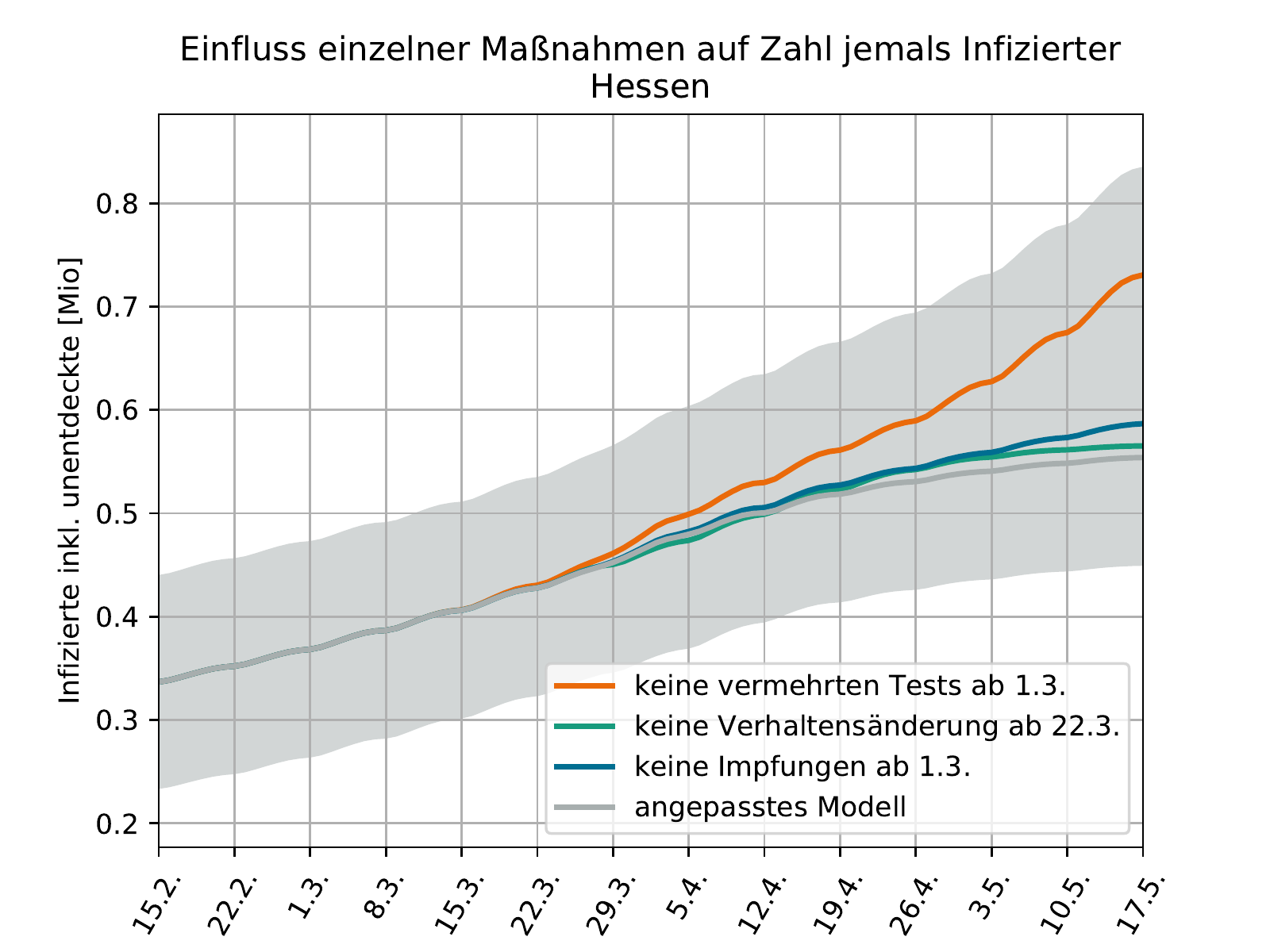}
\caption{Maßnahmenvergleich für Hessen.}
\label{fig:hessen:maßnahmen}
\end{SCfigure}
\begin{SCfigure}[][p]
\includegraphics[width=0.7\textwidth]{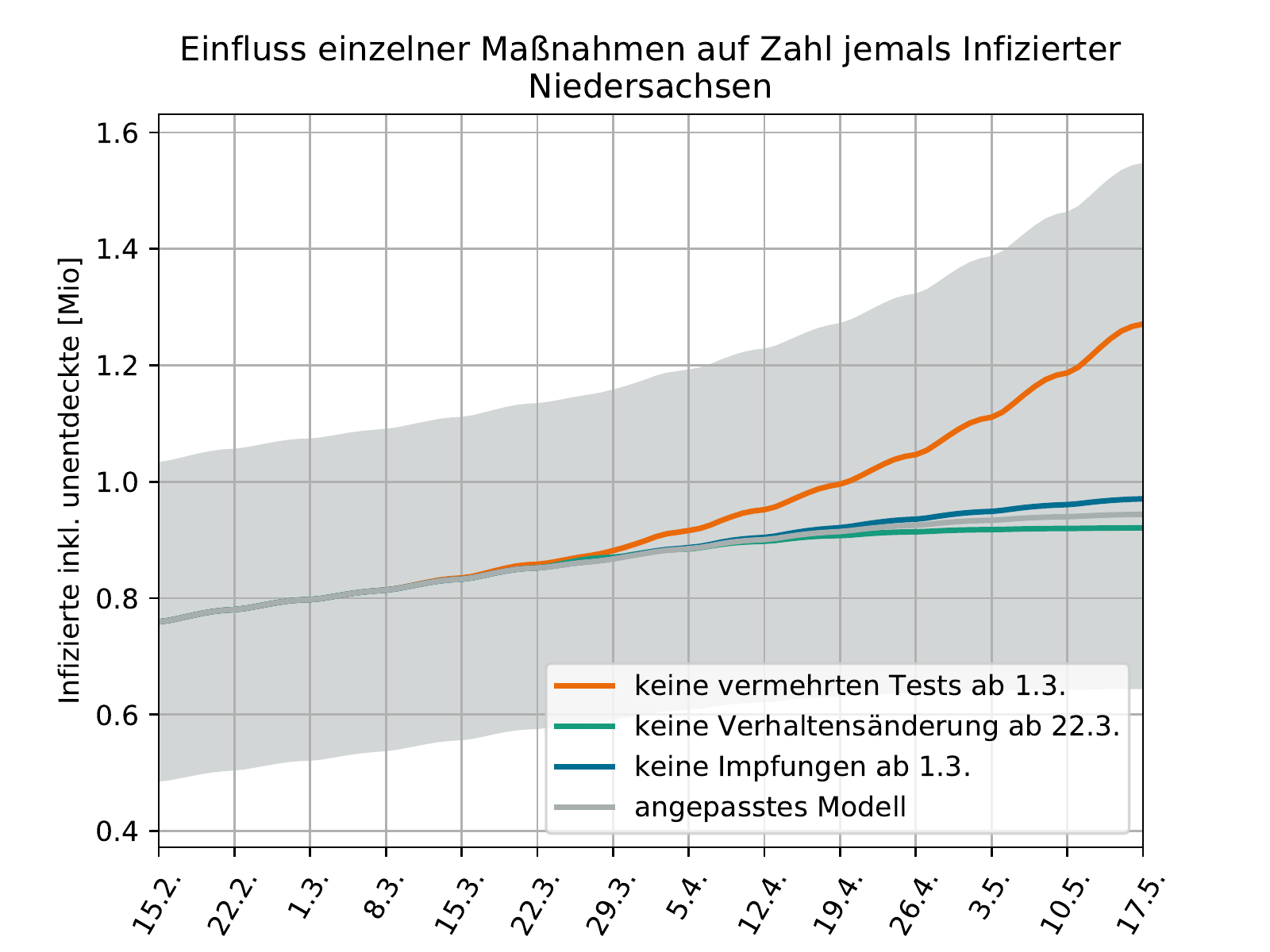}
\caption{Maßnahmenvergleich für Niedersachsen.}
\label{fig:niedersachsen:maßnahmen}
\end{SCfigure}
\begin{SCfigure}[][p]
\includegraphics[width=0.7\textwidth]{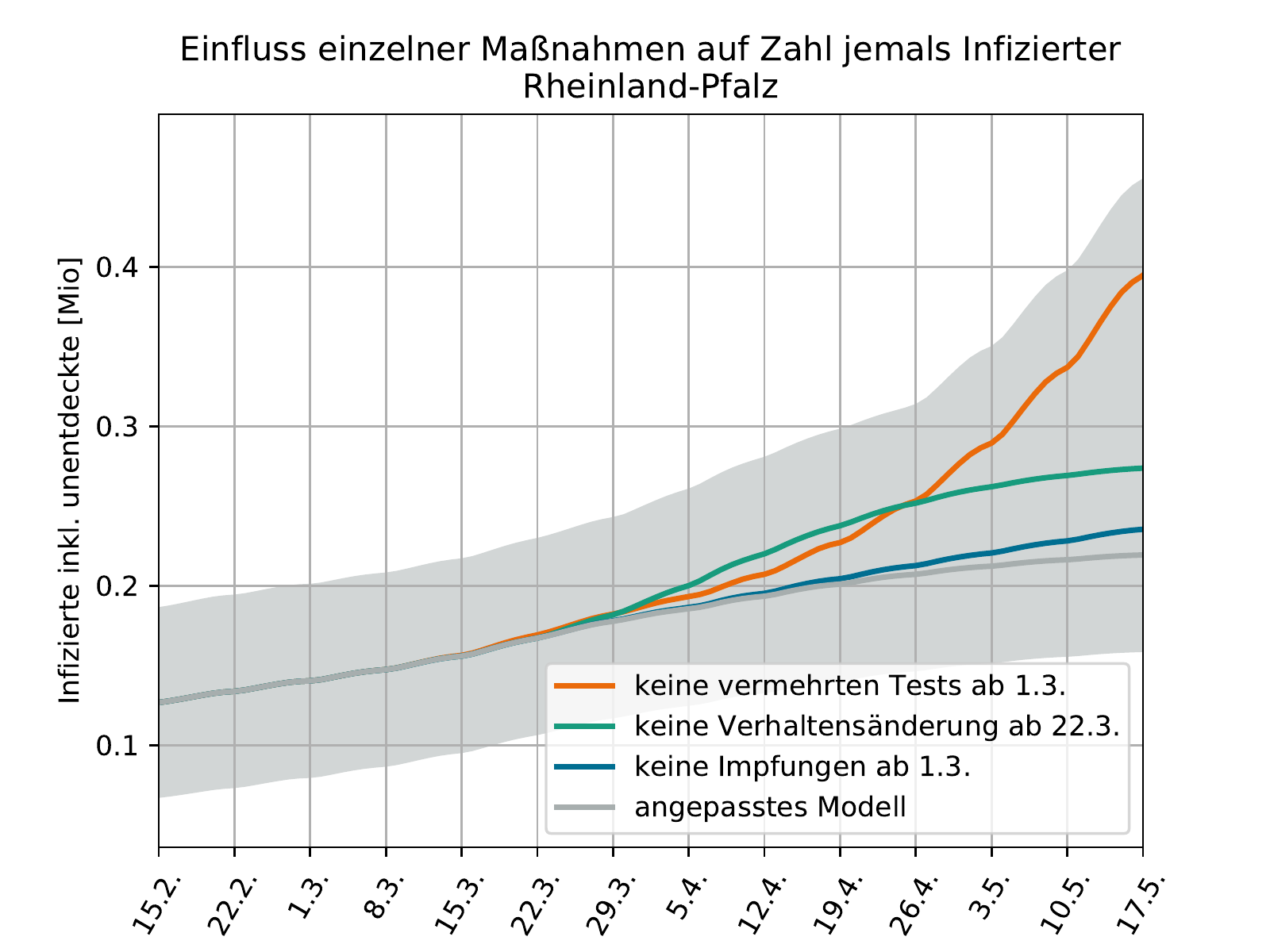}
\caption{Maßnahmenvergleich für Rheinland-Pfalz.}
\label{fig:rheinland-pfalz:maßnahmen}
\end{SCfigure}

\end{document}